\documentclass[a4paper,11pt]{article}
\usepackage{jheppub} 
\usepackage[labelfont=bf]{caption}
\usepackage{lineno}
\usepackage{slashed}
\usepackage{booktabs}
\usepackage[normalem]{ulem}
\usepackage{arydshln}
\usepackage{amssymb}
\usepackage{amsthm}
\usepackage{cancel}
\usepackage{bm}
\usepackage{dsfont}
\usepackage{amsmath}
\usepackage{multicol}
\usepackage{multirow}
\usepackage{xcolor}
\usepackage{color}
\definecolor{myblueold}{RGB}{65,105,225}
\definecolor{myblue}{RGB}{0, 71, 171}
\definecolor{mahogany}{RGB}{156, 0, 0}
\definecolor{mygray}{RGB}{145,145,145}
\definecolor{mybrown}{RGB}{110, 38, 14}
\usepackage{subcaption}

\makeatletter
\def\hlinewd#1{%
	\noalign{\ifnum0=`}\fi\hrule \@height #1 %
	\futurelet\reserved@a\@xhline}
\makeatother

\newcounter{RSQ}

\newcounter{DFQ}

\newcommand{\ar}[1]{\textcolor{mybrown}{(#1)}}

\allowdisplaybreaks

\def\fn{\footnote}
\def\be{\begin{equation}}
\def\ee{\end{equation}}
\def\ba{\begin{alignedat}}
\def\ea{\end{alignedat}}
\def\bea{\begin{eqnarray}}
\def\eea{\end{eqnarray}}
\newcommand{\bs}{\begin{subequations}}
\newcommand{\es}{\end{subequations}}
\usepackage{sectsty}
\sectionfont{\Large}
\subsectionfont{\large}
\subsubsectionfont{\normalsize}

\def\oj{\hspace{1.9mm}}
\def\lj{\hspace{1.6mm}}

\renewcommand{\eqref}[1]{(\ref{#1})}

\newcommand{\ali}[1]{\begin{align}#1\end{align}}

\newcommand{\smallsection}[1]{%
    \ifdim\prevdepth>-1000pt 
        \vspace{0.5cm} 
    \fi
    \noindent\textbf{#1}\\
    \noindent 
}

\def\I{\textcolor{mahogany}{\rm(I)}}
\def\II{\textcolor{mahogany}{\rm(II)}}
\def\III{\textcolor{mahogany}{\rm(III)}}
\def\IV{\textcolor{mahogany}{\rm(IV)}}
\def\V{\textcolor{mahogany}{\rm(V)}}

\def\zz{\textcolor{mygray}{(0)}}
\def\zzzz{\textcolor{mygray}{(00)}}

\def\hn{\textcolor{myblue}{(hn)}}
\def\h{\textcolor{myblue}{(h)}}
\def\sh{\textcolor{myblue}{(sh)}}
\def\hc{\textcolor{myblue}{(hc)}}
\def\p{\textcolor{myblue}{(p)}}
\def\s{\textcolor{myblue}{(s)}}
\def\c{\textcolor{myblue}{(c)}}
\def\sc{\textcolor{myblue}{(sc)}}

\def\hnvar{\textcolor{myblue}{hn}}
\def\hvar{\textcolor{myblue}{h}}
\def\shvar{\textcolor{myblue}{sh}}
\def\svar{\textcolor{myblue}{s}}

\def\scvar{\textcolor{myblue}{sc}}

\title{\boldmath An effective field theory for muon conversion and muon decay-in-orbit}

\author[a,b]{Duarte Fontes,}
\author[a]{Robert Szafron}
\affiliation[a]{Brookhaven National Laboratory, Upton, NY, U.S.A.}
\affiliation[b]{Institute for Theoretical Physics,
Karlsruhe Institute of Technology,
76128 Karlsruhe, Germany}

\date{} 

\emailAdd{duarte.fontes@kit.edu}
\emailAdd{rszafron@bnl.gov}

\abstract{
Muon conversion is one of the best probes of charged lepton flavor violation. The experimental limit is soon expected to improve by four orders of magnitude, thus calling for precise predictions of the shape of the signal spectrum. Equally important are precise predictions for muon decay-in-orbit, the main background for muon conversion. While the calculation of electromagnetic corrections to the two processes above the nuclear scale does not involve significant challenges, it becomes substantially more complex below that scale due to multiple scales, bound-state effects and experimental setup. Here, we present a systematic framework that addresses these challenges by resorting to a series of effective field theories. Combining Heavy Quark Effective Theory (HQET), Non-Relativistic QED (NRQED), potential NRQED, Soft-Collinear Effective Theory I and II, and boosted HQET, we derive a factorization theorem and present the renormalization group equations. Our framework allows for the proper calculation of precise predictions for the rates of the two processes, with crucial implications for the upcoming muon conversion searches. We also provide the most accurate prediction of the signal shape for those searches.  
}

\makeatletter
\gdef\@fpheader{}
\makeatother
\begin{document}

\begin{flushright}
KA-TP-21-2024
\end{flushright}
\vspace*{1cm}

\maketitle
\flushbottom

\section{Introduction}
\label{sec:intro}

Charged lepton flavor violation (CLFV) is highly suppressed in the Standard Model (SM) due to the smallness of neutrino masses~\cite{Petcov:1976ff}.  
If observed, it would provide a clear sign of beyond the SM (BSM) physics (for recent reviews, see refs. \cite{
Kuno:2015tya, Calibbi:2017uvl,Davidson:2022jai,Frau:2024rzt}).
Although searches for CLFV have been carried out in multiple contexts, the most sensitive ones typically involve a muon \cite{Davidson:2022nnl}.
Particularly interesting are processes that involve muons in bound states.
Here, a low-energy muon traveling through matter loses energy until it comes to rest within a target material. This means that it gets bound to a nucleus $N$ of atomic number $Z$, and thus forms a muonic hydrogen, $\mu_{H}$. 
After cascading through progressively lower energy levels in $\mu_{H}$, the muon finally reaches the $1S$ ground state. It then has three possible fates: conversion to an electron, decay-in-orbit (DIO), or nuclear capture. The first two processes, respectively given by $\mu_{H} \to e N$ and $\mu_{H} \to e \nu_{\mu} \bar{\nu}_e N$, 
are such that the final electron is free and the nucleus $N$ provides a source of the electromagnetic field.%
\fn{In DIO, there is a small probability that the final electron is also bound to the nucleus \cite{Aslam:2020pqn}.}
Nuclear capture corresponds to $\mu_{H} \to \nu_{\mu} N'$, where $N'$ is a nucleus with atomic number $Z-1$.

Muon conversion provides one of the most stringent limits on CLFV. It is usually described in terms of the branching ratio, corresponding to the rate of muon conversion normalized to the rate of muon capture,%
\fn{For a recent discussion on an alternative normalization, see ref. \cite{Borrel:2024ylg}.}
\ali{
R_{\mu e} = \dfrac{\Gamma(\mu_{H} \to e N)}{\Gamma(\mu_{H} \to \nu_{\mu} N')}.
}
The current best limit, $R_{\mu e} < 7 \times 10^{-13} $ at $90\%$ confidence level, is reported by the SINDRUM-II experiment \cite{SINDRUMII:2006dvw}.
This limit is expected to be improved by a set of three upcoming experiments: DeeMe at J-PARC  \cite{Natori:2014yba, Teshima:2019orf}, which aims at an accuracy of $10^{-14}$, and Mu2e at Fermilab \cite{Mu2e:2014fns, Diociaiuti:2024stz} and COMET at J-PARC \cite{COMET:2018auw, Fujii:2023vgo}, which expect to reach $10^{-17}$. Planned experiments such as Mu2e-II \cite{Mu2e-II:2022blh}, PRIME \cite{Kuno:2005mm} and the Advanced Muon Facility \cite{CGroup:2022tli} might improve the limit even further.

Such advances on the experimental side should be accompanied by corresponding progress on the theoretical side. Improvements are needed not only for muon conversion, but also for muon DIO near its endpoint (i.e., when the electron energy is close to the muon mass), since it constitutes the only non-reducible background of muon conversion. Several directions have been considered, including improvement of nuclear charge distributions, background estimation, analysis of spin-dependent structures, and dependence on atomic number \cite{Heeck:2021adh, Rule:2021oxe, Haxton:2022piv, Cirigliano:2022ekw, Heeck:2022wer, Borrel:2024ylg, Haxton:2024lyc, Haxton:2024amf}.
A sound comparison between theory and experiment also requires calculating QED corrections, i.e., higher-order effects in the electromagnetic coupling $\alpha$. In the context of muon DIO, some progress has been made in this direction. Ref. \cite{Szafron:2017guu} used collinear factorization to estimate the leading $\mathcal{O}(\alpha)$ corrections enhanced by large logarithmic effects. Furthermore, the hard corrections in the DIO endpoint region have been discussed in ref. \cite{Szafron:2015kja}, while those in the region of intermediate energies have been analyzed in refs. \cite{Szafron:2015mxa, Czarnecki:2014cxa}.

However, a systematic treatment of the QED corrections is still lacking. Crucially, such treatment must ensure the perturbative convergence of the corrections. This is generally spoiled for processes with widely separated scales, like muon conversion and DIO.
Higher-order corrections to these processes receive, in fact, contributions from a heavy scale $\Lambda$, much larger than the nuclear mass $M_N$ (e.g., the electroweak scale or above). This spoils the convergence of perturbation theory because it leads to large logarithms, of the form $\ln(\Lambda/M_N)$. The adequate treatment is to clearly separate the scales, i.e., to isolate single-scale objects out of the multi-scale observable. This, in turn,  can be achieved using an effective field theory (EFT) which results from integrating out all states with masses above $M_N$. The problem of perturbativity is then solved via the usual procedure of matching and renormalization group (RG) running, leading to an RG-improved perturbation theory (see, for example, refs. \cite{Manohar:2018aog, Cohen:2019wxr}).
Additionally, the use of an EFT allows a consistent and, to a large extent, model-independent approach to BSM physics.

An EFT that integrates out states with masses above $M_N$ has been applied to muon conversion in recent works \cite{Rule:2021oxe, Haxton:2022piv, Haxton:2024ecp, Haxton:2024lyc}. If aimed at the energies of the nuclear mass, it provides a proper perturbative description. For energies below those, however, the problem of perturbative convergence shows up again. In other words, such EFT is still plagued by large logarithms: not due to scales above the nuclear mass, but due to scales below it. Moreover, this is caused not simply by one scale below the nuclear mass but by a multiplicity of them. Muon conversion, indeed, is characterized by the following physical scales below the nuclear mass $M_N \sim 25 \; \rm GeV $:
\vspace{-1mm}
\begin{enumerate}
    \item the muon mass, $m_\mu \simeq 105.66 \; \rm MeV$, of the same order as the electron energy, $E_e$,
    \\[-7.6mm]
    \item the muon momentum, $Z \alpha m_\mu  \sim 10 \; \rm MeV$,
    \\[-7.6mm]
    \item the electron mass, $m_e \simeq 0.511 \; \rm MeV$.
\end{enumerate}
\vspace{-1mm}
Since we aim to address QED corrections, we must also consider a quantity $\Delta E$, representing a photon-energy cutoff
that characterizes an inclusive rate. In fact, the experimental setup requires theoretical predictions that contain an arbitrary number of undetected real photons with total energy below $\Delta E$.%
\fn{When including virtual corrections involving massless states, real radiation must also be included to obtain an infrared-finite rate. In the case of muon conversion, then, one should consider not simply $\Gamma[\mu_{H} \to e N]$, but rather $\Gamma[\mu_{H} \to e N] + \Gamma[\mu_{H} \to e N + n \gamma(E_{\gamma} < \Delta E)]$, describing the real emission of $n>0$ photons with total energy below $\Delta E$, which can be understood as the energy resolution of the detector.
An equivalent reasoning applies to muon DIO. }
Even though different scales for $\Delta E$ could be considered, 
we restrict the discussion to the case $\Delta E \sim m_e$. In sum, muon conversion involves a complex hierarchy of scales below the nuclear mass, which can be described as
\ali{
\label{eq:hierarchy}
M_N \gg m_\mu \sim E_e \gg Z \alpha m_{\mu} \gg (Z \alpha)^2 m_{\mu} \sim m_e \sim \Delta E. 
}
Hence, an EFT that simply integrates out states heavier than the nucleus mass still leaves not one but many multi-scale objects, each one of which contains large logarithms. Like $\ln(\Lambda/M_N)$, these spoil the convergence of perturbation theory for higher-order $\alpha$ effects.
Furthermore, a straightforward computation of those effects in processes involving bound states is challenging.

This paper addresses all these problems by providing an EFT framework for bound muons decays, suitable for a systematic computation of QED corrections below the nuclear scale.%
\fn{Our work is complementary to ref. \cite{Haxton:2024lyc}: while that reference provides an EFT description down to the nuclear scale, we focus on scales below it.}
The framework considers the particular cases of muon conversion and muon DIO near its endpoint, and is built by carefully separating all the scales involved in these processes (i.e., by \emph{factorizing} the rates into single-scale objects). 
It exploits Heavy Quark Effective Theory (HQET) \cite{Isgur:1989vq, Isgur:1990yhj, Neubert:1993mb, Manohar:1997qy, Manohar:2000dt}, non-relativistic QED (NRQED) \cite{Caswell:1985ui, Kinoshita:1995mt, Paz:2015uga}, potential NRQED (pNRQED) \cite{Pineda:1997bj, Pineda:1997ie, Brambilla:1999xf, Beneke:1999qg, Beneke:1998jj}, soft-collinear effective theory (SCET) \cite{Bauer:2000ew, Bauer:2000yr, Bauer:2001ct, Bauer:2001yt, Beneke:2002ph, Beneke:2002ni} and boosted HQET (bHQET) \cite{Fleming:2007qr, Fleming:2007xt}. 
These EFTs are examples of what is commonly known as \emph{modern EFTs}, which integrate out modes of fields (instead of integrating out fields altogether). In the context of QED, they have attracted some interest in recent years, e.g., in the context of scattering processes \cite{Hill:2016gdf, Tomalak:2022xup, Tomalak:2022kjd, Hill:2023bfh, Hill:2023acw, Tomalak:2024lme}, heavy quark decays \cite{Beneke:2017vpq, Beneke:2019slt, Beneke:2020vnb, Beneke:2021jhp, Beneke:2022msp} or energy levels for hydrogen-like ions \cite{Czarnecki:2016lzl, Czarnecki:2017kva, Szafron:2019tho, Czarnecki:2020kzi}. Elements of modern EFTs have also been used specifically for muon bound states, albeit without completely separating all the relevant scales involved \cite{Czarnecki:2014cxa, Szafron:2015mxa}.

The modern EFT approach is very powerful: besides providing operator-based, gauge-invariant definitions of each single-scale contribution that appears in the factorization theorem, it highlights the universality of the underlying physics. Indeed, many of the objects that we will encounter in our analysis have been computed before in various other physical contexts.
On the other hand, they
will also be found in several other low-energy processes, so that the EFT framework developed here can also be applied to them, with only minor modifications. One of the aims of this paper is to explore this universality; more specifically, we intend to build a bridge between modern theoretical EFT tools that have been applied predominantly to high-energy processes, on the one hand, and low-energy QED processes at the intensity frontier, on the other.

We emphasize that these low-energy QED processes, despite their apparent simplicity, are notoriously complex due to the presence of multiple scales. Muon conversion and DIO are probably the most challenging ones among the currently searched processes, since they involve both bound-state physics and collinear physics. More specifically, they resemble already complicated exclusive heavy-quark decays (leading to the appearance of ${\rm SCET_{I}}$ and ${\rm SCET_{II}}$), but are further complicated by both the mass of the electron (so that bHQET appears), as well as the perturbative bound-state physics for the non-relativistic muon (leading to NRQED and pNRQED, instead of the simpler HQET used for heavy quarks).

All these EFTs shall then be incorporated in the framework discussed in this paper. As will be seen in detail, this framework involves a sequence of five scales, with an appropriate EFT at each scale. Because the relevant physical scales are the same in muon conversion and DIO near its endpoint, the same sequence of EFTs is adequate to describe both processes.%
\fn{This is not the case for DIO away from the endpoint, as that case would involve different scales.}
On the other hand, the matching coefficients
are in general different in the two processes (even though some of them will be the same, due to the aforementioned universality). Furthermore, the rate of muon DIO near its endpoint is complicated by the presence of the neutrino-antineutrino pair in the final state. Even though the framework presented here holds for both processes, therefore, we perform explicit calculations for the case of conversion, and leave the case of muon DIO near its endpoint for a future publication.

For muon conversion, then, we perform the one-loop matching for the matching coefficients of the five subsequent EFTs. We also derive a factorization theorem, which properly isolates the different scales, and thus leads to a final result for the rate that is factorized into single-scale objects. In addition, we derive the renormalization group equations (RGEs) and evaluate the effect of resummation for the muon conversion rate. It should be clear that, in all of this, we are interested in the QED corrections to the muon conversion rate. We stress that these corrections have two different consequences.

In the first place, they affect the overall normalization (i.e., the absolute value) of the rate. A correct prediction of such normalization is very important for phenomenological purposes, as it allows the experiments to distinguish between different BSM models. While the QED corrections play a non-negligible role in that normalization, other factors (such as nuclear effects) also contribute significantly to it. A proper analysis of those factors is beyond the scope of this work. As a consequence, even though we provide a relevant piece of the normalization by calculating the QED corrections, we leave a complete discussion of the normalization and its theoretical uncertainty for future work.

Secondly, QED corrections affect the shape of the rate for muon conversion, modifying the electron energy spectrum in a non-trivial way. This shape is the main phenomenological focus of our work. In fact, a proper theoretical prediction of the shape of both signal (muon conversion) and background (muon DIO near its endpoint) is crucial for a potential observation of the former in the upcoming experiments. As it turns out, the QED corrections are the only relevant contribution to the shape of the rate (other factors, such as nuclear effects, only affect the normalization). Hence, by calculating the QED corrections to muon conversion, we provide one of the two essential elements for experimental searches of that process. The remaining element concerns QED corrections to the muon DIO near its endpoint --- which, as referred above, we leave to a future work.

The organization of the paper is as follows. After introducing the essentials of muon conversion and DIO in section~\ref{sec:basics}, we describe our treatment of the nucleus in section~\ref{sec:nucleus}. Then, we define the kinematics and the power counting in section~\ref{sec:Kinematics}. After that, and in preparation for the construction of the EFT framework, we present in section~\ref{sec:Framework} a summary of an analysis of the relevant momentum regions. Section~\ref{sec:EFT} is then devoted to describing the sequence of EFTs. For each of them, we provide the Lagrangian, the matching equations, and present the one-loop matching coefficients and their RGEs.
The elaboration of this theoretical building enables us to derive a factorization formula in section \ref{subsec:Factorization-theorem}, with compact formul\ae \, in section \ref{sec:we-love-latin}. Then, after briefly commenting on the dominant nuclear finite-size corrections in section \ref{subsec:FSZ}, we turn to the numerical results in section \ref{sec:Numerical-results}. We present our conclusions in section \ref{sec:Conclusions}, after which we provide technical details about the analysis by momentum regions in appendix~\ref{app:regions}.

\section{Preliminaries}
\label{sec:Preliminaries}


\subsection{Basic aspects}
\label{sec:basics}

Muon conversion can happen either with or without the exchange of a photon between the leptons and the nuclei --- the former being known as photonic conversion, the latter as direct conversion. At energy scales below the muon mass, both photonic conversion and muon DIO reduce to a contact interaction  similar to the one used for direct conversion \cite{Fontes:2025aaa}.
Throughout the paper, therefore, we focus on direct muon conversion. 
The current for this case can be generically written as 
\begin{align}
\label{eq:total-current}
\mathcal{J}
&= -\frac{4 G_F}{\sqrt{2}}  \sum_{X = L,R}
\Bigg\{
%
C_{SX} \, \bar{e} P_X \mu \, \bar{N} N + C_{PX} \bar{e} P_X \mu \, \bar{N} \gamma_5 N
+ C_{VX} \, \bar{e} \gamma^\alpha P_X \mu \, \bar{N} \gamma_\alpha N \nonumber\\
& \hspace{35mm} + C_{AX} \, \bar{e} \gamma^\alpha P_X \mu \, \bar{N} \gamma_\alpha \gamma_5 N + C_{{\rm Der}X} \, \bar{e} \gamma^\alpha P_X \mu \,  (\bar{N} \overleftrightarrow{\partial}_\alpha i \gamma_5 N) \nonumber\\
& \hspace{35mm} + C_{TX} \, \bar{e} \sigma^{\alpha\beta} P_X \mu \,  \bar{N} \sigma_{\alpha\beta} N 
\Bigg\} + \rm h.c. \, ,
\end{align}
where $N$, $\mu$ and $e$ respectively represent the nucleus, muon and electron fields, $G_F$ is the Fermi constant, $P_L$ and $P_R$ are the chirality projection operators and $C_{YX}$ (for $Y \in \{S,P,V,A,{\rm Der}, T\}$) are matching (Wilson) coefficients. 
We focus on the dominant contribution to direct muon conversion, in which the final and initial states of the nucleus are the same --- the so-called coherent conversion. In this case, the terms with $C_{PX}$, $C_{AX}$, $C_{{\rm Der}X}$, and $C_{TX}$ do not contribute at leading power (LP). A generalization to the incoherent case is straightforward.

\subsection{Treatment of the nucleus}
\label{sec:nucleus}

The effective operator in eq.~(\ref{eq:total-current}) belongs to an EFT that takes the nucleus $N$ as a dynamical field; we identify this EFT as nuclear-energy EFT (NEFT) and assume that it is defined above the scale of the muon mass. This approach differs from the one typically used in the literature for muon conversion. In fact, past works either take quarks as dynamical fields --- in which case the EFT is sometimes identified as Weak Effective Theory (WET) --- or nucleons --- in which case the EFT is sometimes identified as Non-Relativistic Effective Theory (NRET) \cite{Kuno:1999jp, Kosmas:2001mv, Cirigliano:2009bz, Davidson:2016edt, Bartolotta:2017mff, Crivellin:2017rmk, Davidson:2018kud}. The use of WET or NRET is undoubtedly more adequate for energies much higher than the typical scales of the muon and the electron in the processes of muon conversion and DIO. In fact, nuclear fields do not exist at such energies, but appear only for distances much larger than the typical size of the nucleus.

The justification for using the NEFT here is a double one: on the one hand, we focus on the coherent conversion, which means that the nucleus state remains unchanged throughout the process  (we neglect recoil corrections). On the other hand, our interest lies precisely on the lower energies --- that is, on the scales of the muon and the electron. More specifically, we focus on the muon and electron involved in muon conversion and DIO, and develop a systematic description of the QED effects between them in the presence of the static electric field of the nucleus.%
\fn{\label{fn:fn1}Here and in what follows, by `QED effects' (or `QED corrections'), we mean higher-order corrections in the electromagnetic coupling $\alpha$, without any dependence on the atomic number $Z$. By contrast, we refer to `nuclear effects' (or `nuclear corrections') as higher-order corrections involving $Z$, such as $Z \alpha$ or $Z^2 \alpha$.}
As mentioned before, it is this description that allows a proper theoretical prediction of the shape of the muon conversion spectrum.
For our purposes, then, the nucleus is essentially a spectator; that is, it is merely the source of the background electric field in which the QED interactions between muon and electron take place. Accordingly, the details of its structure can be ignored, so that the nucleus can be taken as a point-like particle. We will do this in the following, assuming a mass $M_N$ for it.

It should be clear, however, that $M_N$ is a placeholder for a proper nuclear description.%
\fn{$M_N$ might even differ substantially from any real nucleus mass; all that matters here is that it is much larger than the scales of the muon and the electron in muon conversion and DIO, which are the scales that our analysis is focused on.}
Even though we leave that description for future work, we summarize its essential elements~\cite{Kuno:1999jp}. Starting from the low energies, the nucleus must first be matched onto nucleons and, subsequently, the latter must be matched onto quarks. The first transition can be done using chiral perturbation theory. At the LP, this theory implies that the nucleus can be seen as the coherent sum of protons and neutrons, whose effects involve the proton and neutron densities inside the nucleus, respectively. The second transition (between nucleons and quarks) involves non-perturbative physics, and can be done using nucleon form factors. These two transitions would establish a proper connection between the NEFT description of eq.~(\ref{eq:total-current}) and energies at the parton level, described by WET. As usual, if BSM physics lies at much higher energies --- above the electroweak scale --- it must first be matched onto the Standard Model EFT (SMEFT), and this, in turn, to WET.

Regarding our treatment of the nucleus, two obvious corrections could be discussed: corrections to the point-like assumption for the nucleus and corrections due to nuclear excitations.
The first type constitutes finite-size corrections and can be accounted for by including form-factors. This constitutes a trivial generalization in our framework, which we will briefly discuss in section \ref{subsec:FSZ}.
As for the effects involving the transition of the nucleus from its ground state to one of the short-lived excited states (due to interactions with electromagnetic field),  we leave it for future work. A further generalization of our EFT setup is required to address those effects (such as a dispersive treatment of two-photon exchange corrections, as well as inelastic conversion%
\fn{The latter was recently discussed in the context of NRET in ref. \cite{Haxton:2024amf}.}).

In what follows, we treat the nucleus as a fermion. A generalization to an arbitrary spin is trivial. Indeed, as will be seen, the nucleus is described by a HQET field, which naturally exhibit spin symmetry. In other words, the nucleus spin decouples as $M_N \to \infty$, in such a way that light fermions only see the total charge of the nucleus.

\subsection{Kinematics of direct muon conversion and power counting}
\label{sec:Kinematics}

\begin{figure}[t!]
    \centering
        \begin{subfigure}[t]{0.4\textwidth}
        \centering
        \includegraphics[width=0.71\textwidth]{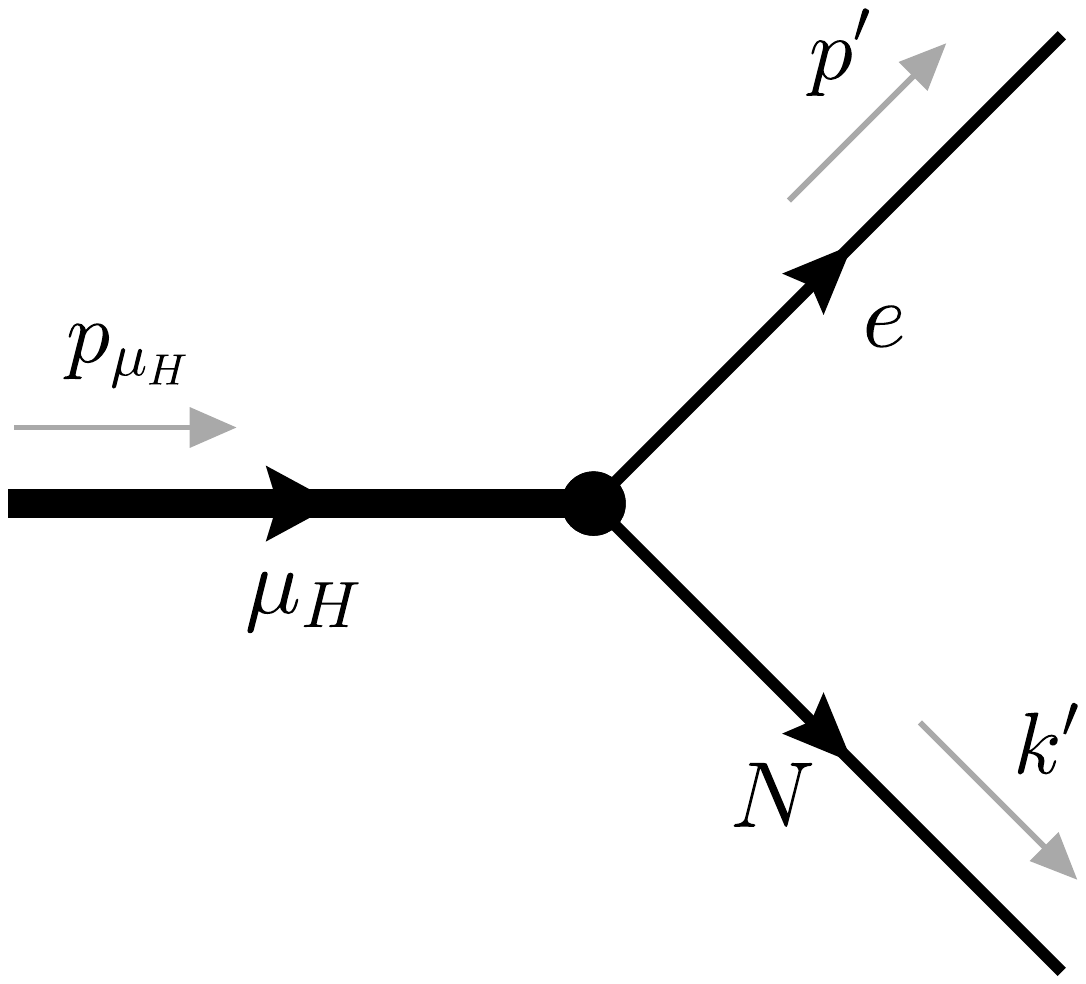}
        \vspace{-18mm}
        \caption{The decay $\mu_{H} \to e N$}
    \end{subfigure}
    \hspace{1mm}
    \begin{subfigure}[t]{0.4\textwidth}
        \centering
        \includegraphics[width=0.70\textwidth]{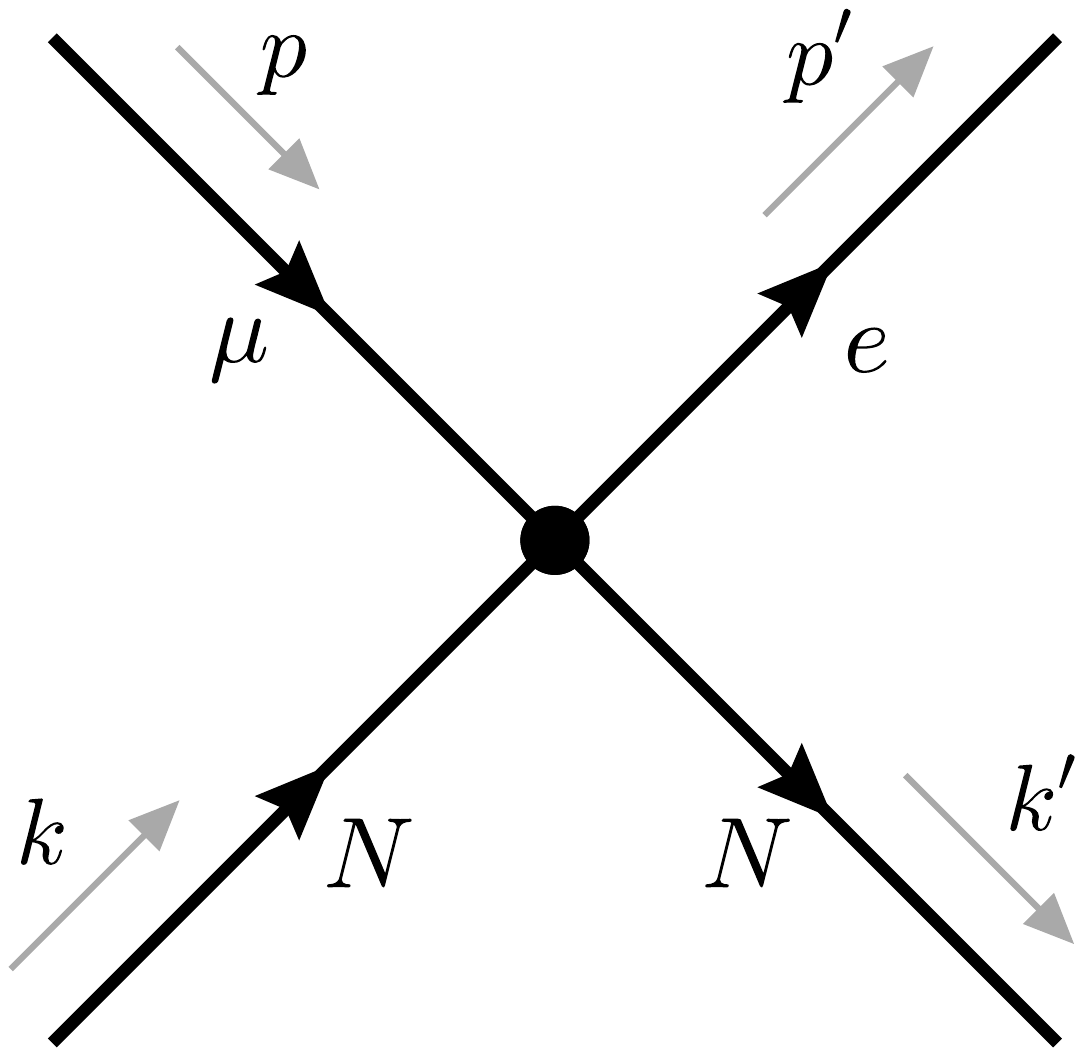}
        \vspace{-18mm}
        \caption{The scattering $\mu N \to e N$}
    \end{subfigure}
\vspace{-12mm}
\caption{Kinematics for direct muon conversion. In diagram (i), the incoming muon and nucleus are bound, forming a muonic hydrogen, $\mu_{H}$, while in diagram (ii) they are free.}
\label{fig:kinematics}
\end{figure}

Figure \ref{fig:kinematics} illustrates the kinematics for direct muon conversion. Diagram (i) presents the bound case, in which the initial state muon and nucleus form a hydrogen atom (represented by a thick line), whereas diagram (ii) shows the scattering between a free muon and a nucleus. In both cases, the final state contains the nucleus and a free electron. We will ultimately be concerned with the bound case --- which is, as will be seen, unavoidable at sufficiently low energies. However, since our analysis starts at the nuclear energies (in which the muon is not yet bound), and since the EFT can be more easily constructed by analyzing diagram (ii), we focus on the kinematics of the free case.
\renewcommand{\thesubfigure}{\roman{subfigure}}

The four-momenta $k$ and $p$ characterize the incoming nucleus and muon, respectively, and the four-momenta $k'$ and $p'$ the outgoing nucleus and electron, respectively. We work in the frame in which the initial nucleus is at rest, and we choose for convenience the electron momentum to lie along the $z$ axis. If we ignored nuclear recoil corrections, binding effects and real radiation, the electron energy would be equal to the muon energy. In that case, and since the muon energy is --- up to corrections suppressed by the muon velocity --- given by the muon mass, we would have $E_e = m_\mu$.
However, both nuclear recoil corrections and binding effects decrease the maximal energy available for the electron. More importantly, because the electron can also lose energy due to real radiation, we consider $E_e$ to be a free parameter, and interpret the quantity $\Delta E$ mentioned in the Introduction as the total energy lost by the electron due to real radiation. Then, ignoring nuclear recoil corrections, we define $\Delta E := m_\mu + E_b - E_e$, where $E_b=\mathcal{O}((Z\alpha)^2 m_\mu)$ is the binding energy. Although $\Delta E\geq 0 $ is, in principle, arbitrary, we choose $\Delta E \sim m_e$ for power counting purposes; this choice agrees with the experimental setup \cite{Mu2e-II:2022blh, Jansen:2023ojv,Bernstein:2019fyh}. Therefore, given $E_b \sim m_e$, we conclude that the electron energy is very close to the muon mass, $E_e \simeq m_{\mu}$.

The explicit parametrization for the four-momenta in diagram (ii) is 
\ali{
\label{eq:momenta}
p &= \Big(\sqrt{m_{\mu}^2 + |\vec{p}|^2}, \vec{p}\Big),
&
p' &= \Big(E_e,0,0,-\sqrt{E_e^2 - m_e^2 }\Big), \nonumber \\
k &= (M_N,\vec{0}),
&
k' &= (\sqrt{M_N^2 + |\vec{k'}|^2},\vec{k'}),
}
where we assume that the muon is non-relativistic and its momentum is of the order of the inverse Bohr radius,
\ali{
\label{eq:kinematics-basic}
|\vec{p}| = \mathcal{O}(m_{\mu} Z \alpha),
}
while the momentum of the outgoing nucleus is determined from momentum conservation,
\ali{
|\vec{k'}| = \mathcal{O}(m_{\mu}) = \mathcal{O}(E_e).
}

Besides the loop expansion --- organized by powers of the fine structure constant $\alpha$ --- we consider two other expansions, which we dub the \textit{recoil} expansion and the \textit{power} expansion. They are respectively organized by powers of the parameters $\lambda_R$ and $\lambda$, obeying the scalings
\ali{
\label{eq:scaling-def}
	\lambda_R \sim \dfrac{m_{\mu}}{M_N},
	\qquad
	\lambda \sim Z \alpha \sim \sqrt{\dfrac{m_e}{m_{\mu}}}.
}
The electron of momentum $p'$ is ultra-relativistic. In processes involving such particles, it is convenient to decompose the momentum along the light-cone. For this purpose, we define the light-like four-vectors $n_+ = (1, 0, 0, 1)$ and $n_- = (1, 0, 0, -1)$, which implies
%
\begin{align}
\label{eq:ids}
&n_-p' = \frac{m_e^2}{2 E_e} + \mathcal{O}(\lambda),&
&n_+p' = 2 E_e + \mathcal{O}(\lambda). 
\end{align}
Throughout the paper, we will interchangeably use three different ways to denote an arbitrary four-vector $l$:
\begin{itemize}
    \item explicitly write all four components, in which case $l$ is a four-component object:  $l = (l_0, l_1, l_2, l_3)$;
    \item resort to the light-cone basis, so that $l$ is represented by a three-component object (two light-cone coordinates and a perp component): $l = (n_+l, l_\perp,$ $n_-l)$;
    \item separate the temporal and the spatial components, in which case $l$ is represented by a two-component object: $l = (l_0, \vec{l})$.
\end{itemize}
Since the main complications in our formalism come from the presence of massive particles, we must define additional two time-like reference  four-vectors,  
\ali{
v_{}=(1,0,0,0),
\qquad
v_{e}=\left(\frac{2 m_\mu}{m_e},0,\frac{m_e}{2 m_\mu}\right),
}
as well as residual momenta $\hat{p}$, $\hat{p}'_{}$, $\hat{k}$ and $\hat{k'}$, which are such that
\ali{
\label{eq:hatted-def}
p_{} &= m_{\mu} v_{} + \hat{p}_{},
&
p'_{} &= m_{e} v_{e} + \hat{p}'_{}, \nonumber  \\
k_{} &= M_N v_{} + \hat{k}_{},
&
k'_{} &=M_N v_{} + \hat{k}'_{}.
}
Given eq.~(\ref{eq:momenta}) and eq.~(\ref{eq:scaling-def}), we find the following scaling for the components of the residual momenta:
\ali{
\label{eq:hatted-guys}
\hat{p} &\sim \left(\dfrac{|\vec{p}|^2}{m_{\mu}},\vec{p}\right) \sim m_{\mu} \left(\lambda^2, \lambda \right),
&
\hat{p}'&\sim m_{\mu} \, \lambda^2 \left(1,\lambda^2, \lambda^4\right),
&
\hat{k}'
&\sim m_{\mu} \left( \lambda_R,1\right).
}
Furthermore, for our choice of reference frame, $\hat{k}=0$; yet, we keep using $\hat{k}$ to enable a description in an arbitrary reference frame. 
Eq.~(\ref{eq:hatted-def}) splits each momentum into two terms: a mass multiplied by a time-like vector and a residual momentum denoted with a hat. In each case, the residual momentum is much smaller than the first term, and describes small fluctuations of the momentum of the respective field around its on-shell value.

\subsection{Regions, modes and scales}
\label{sec:Framework}

We now pave the way for building an EFT framework that correctly describes muon conversion and DIO near its endpoint. Crucial to this construction is an analysis of the momentum regions that contribute to the loop integrals \cite{Beneke:1997zp, Beneke:1999zr, Jantzen:2011nz}. This analysis is described in detail in Appendix~\ref{app:regions}. In this section, we summarize its results.

\begin{figure}[htb!]
    \renewcommand{\thesubfigure}{\alph{subfigure}}
    \centering
    \begin{subfigure}[t]{0.28\textwidth}
        \centering
        \includegraphics[width=1\textwidth]{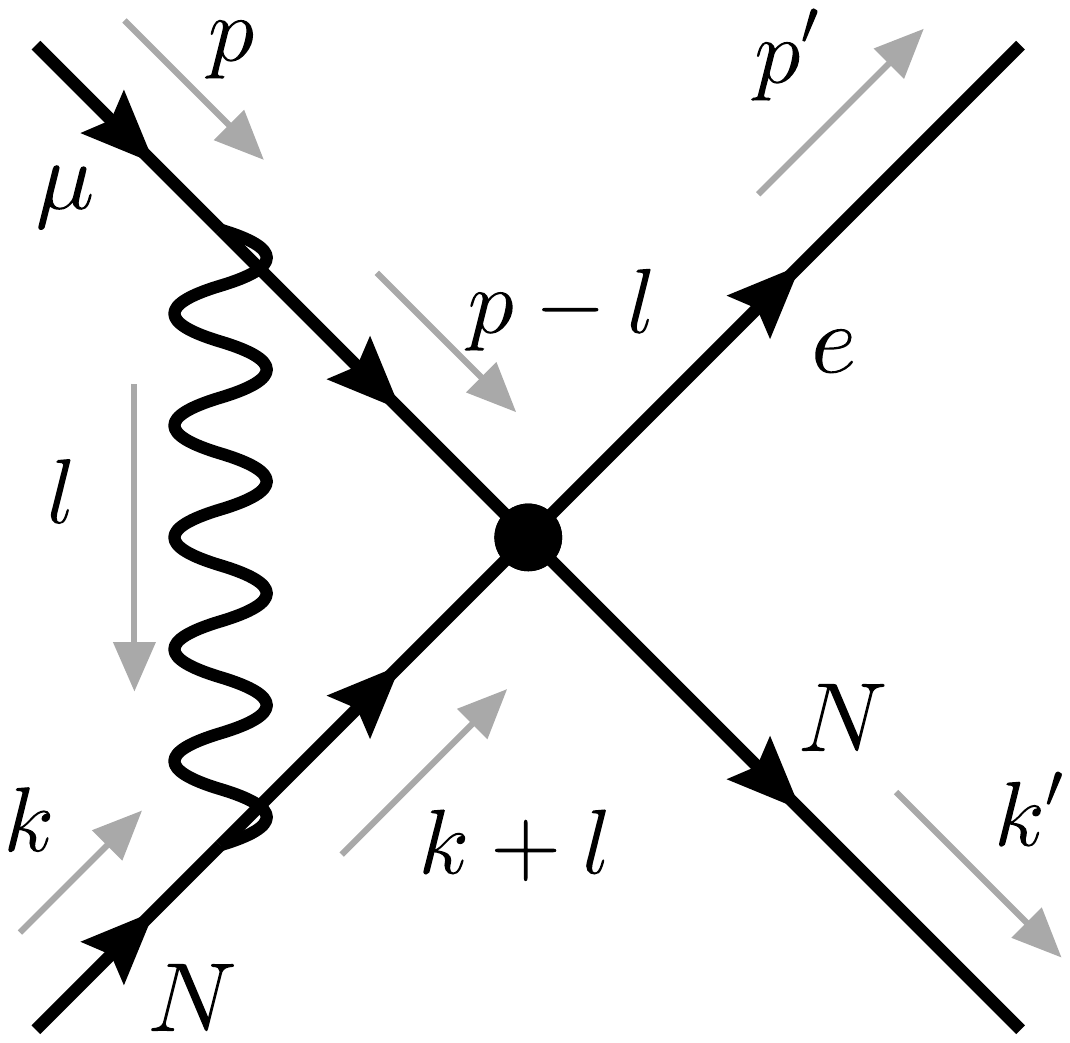}
        \vspace{-22mm}
        \caption{}
    \end{subfigure}
    \hspace{1mm}
    \begin{subfigure}[t]{0.28\textwidth}
        \centering
        \includegraphics[width=1\textwidth]{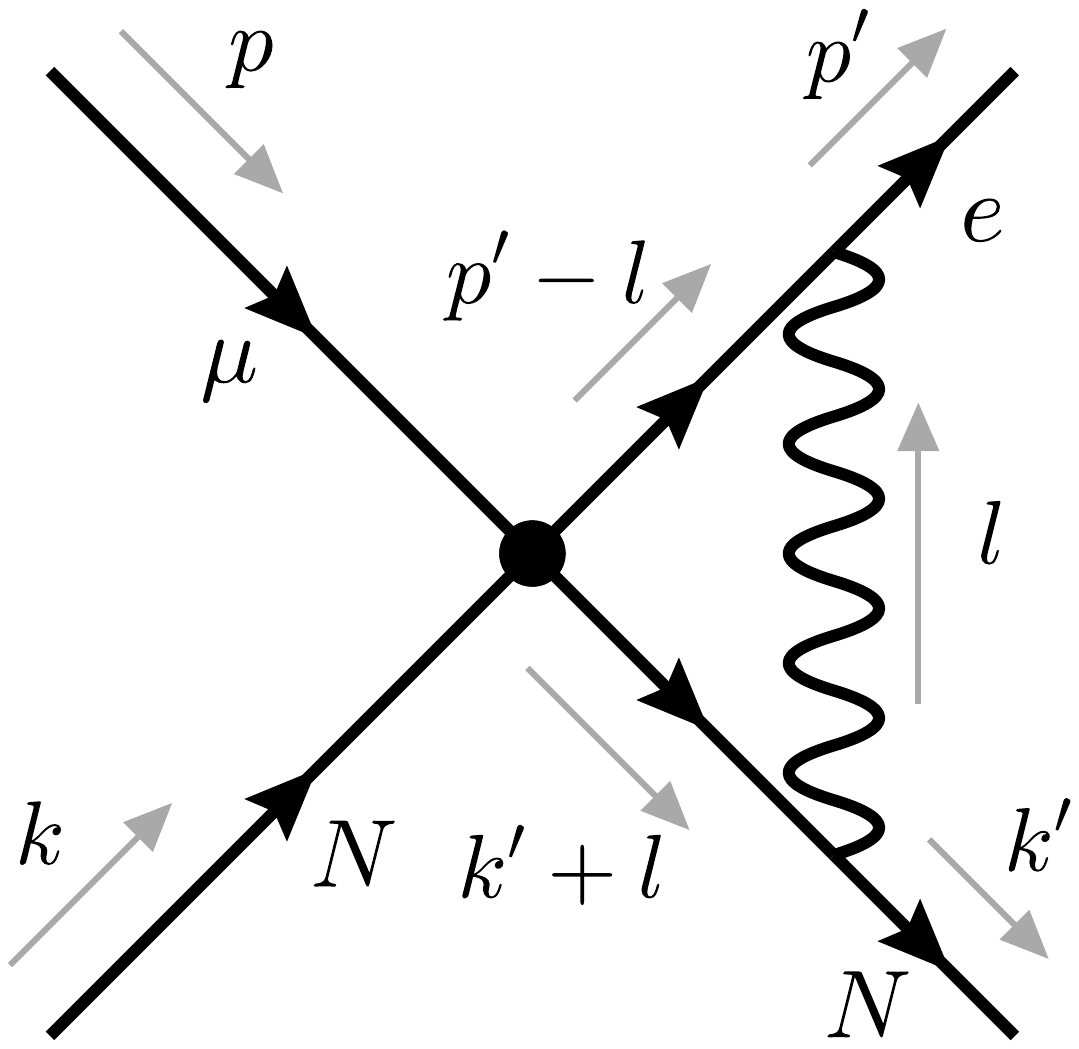}
        \vspace{-22mm}
        \caption{}
    \end{subfigure}
    \hspace{1mm}
    \begin{subfigure}[t]{0.28\textwidth}
        \centering
        \includegraphics[width=1\textwidth]{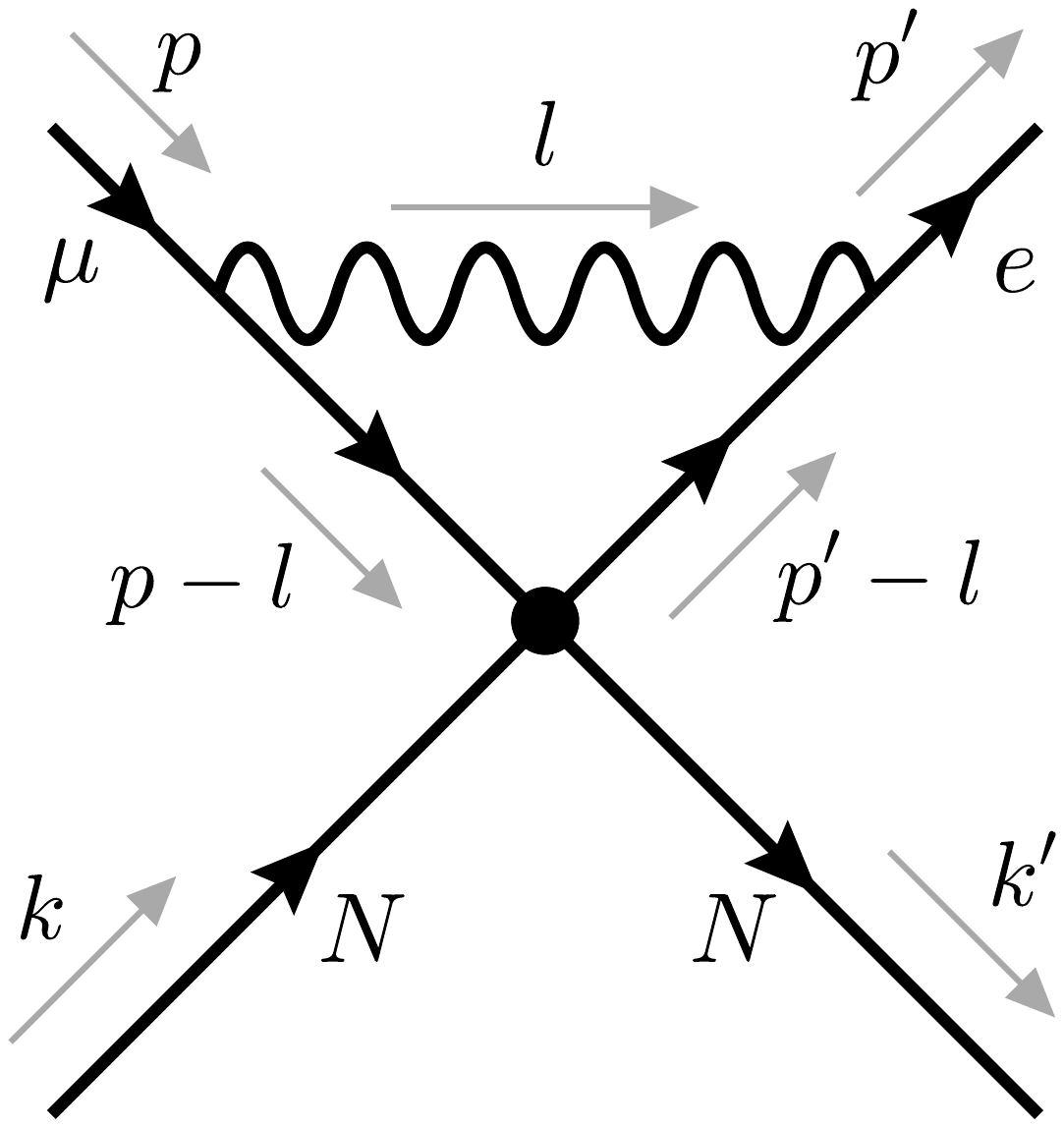}
        \vspace{-22mm}
        \caption{}
    \end{subfigure}
    \\[-7mm]
    \begin{subfigure}[t]{0.28\textwidth}
        \centering
        \includegraphics[width=1\textwidth]{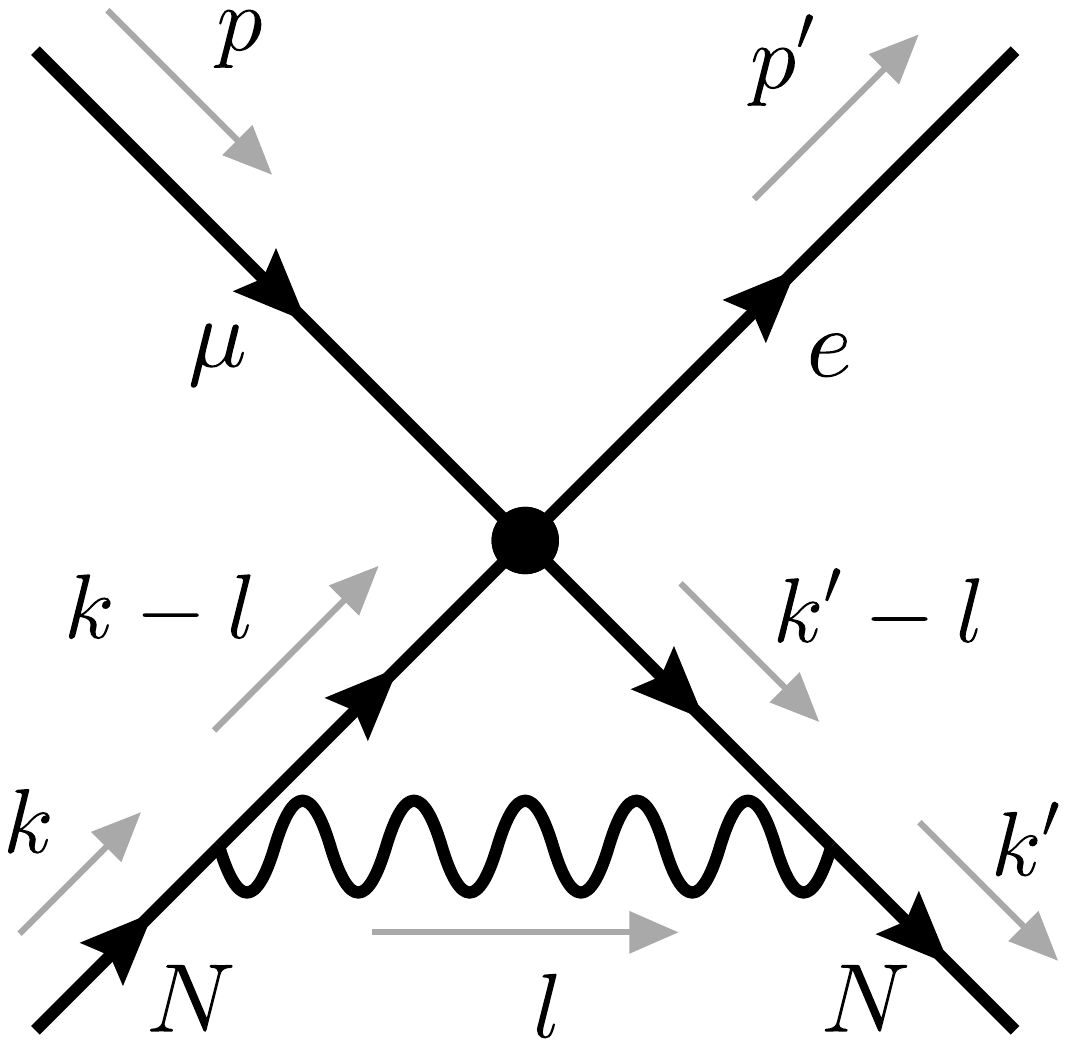}
        \vspace{-22mm}
        \caption{}
    \end{subfigure}
    \hspace{1mm}
    \begin{subfigure}[t]{0.28\textwidth}
        \centering
        \includegraphics[width=1\textwidth]{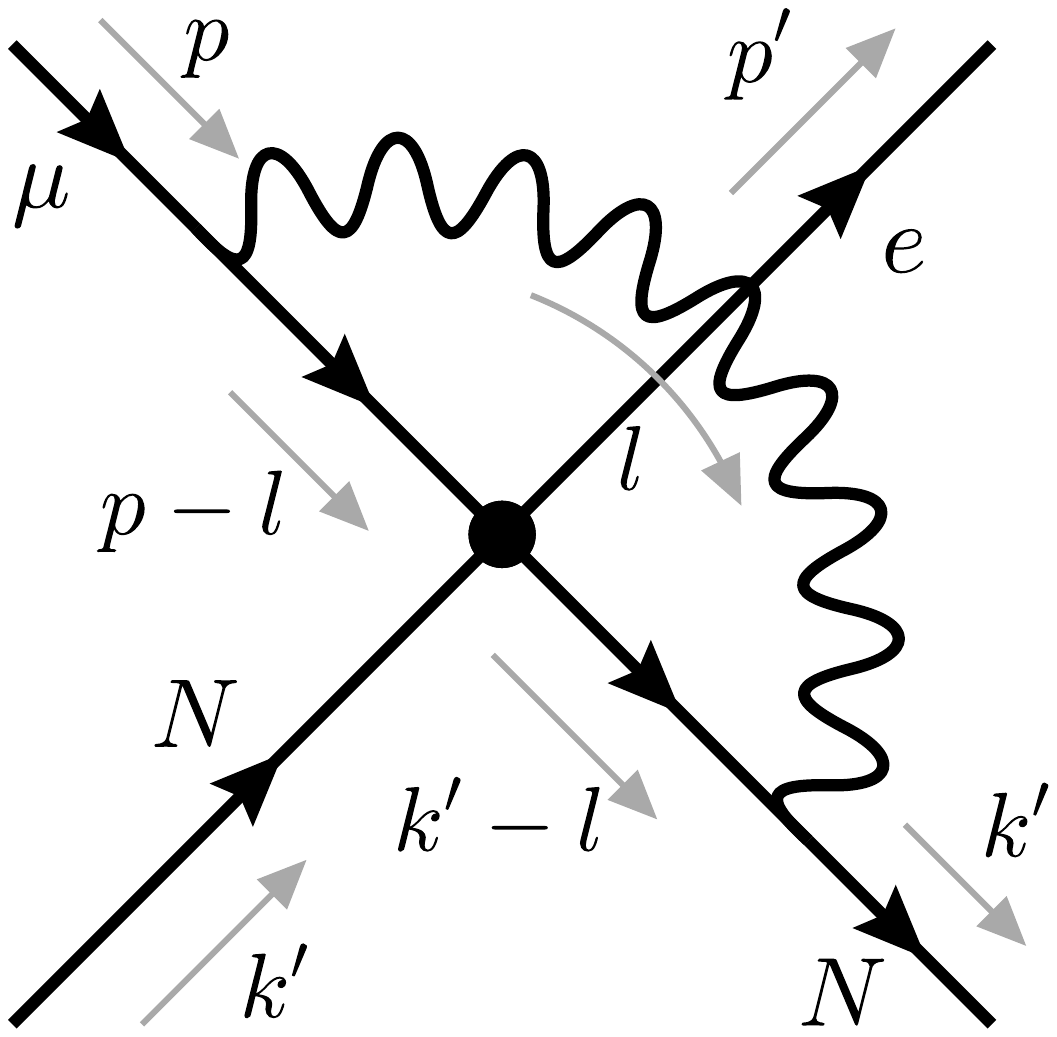}
        \vspace{-22mm}
        \caption{}
    \end{subfigure}
    \hspace{1mm}
    \begin{subfigure}[t]{0.28\textwidth}
        \centering
        \includegraphics[width=1\textwidth]{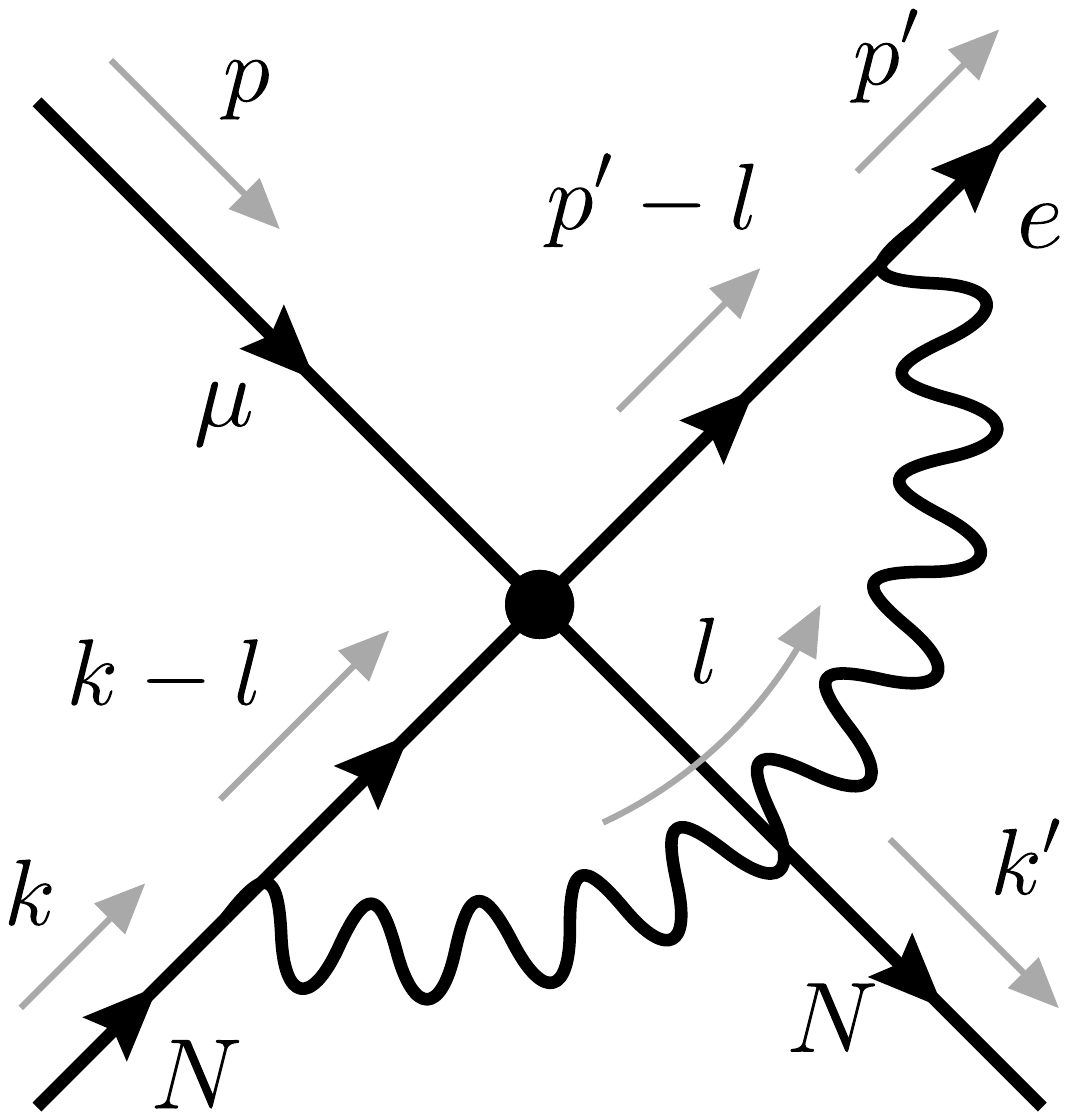}
        \vspace{-22mm}
        \caption{}
    \end{subfigure}
    \vspace{-10mm}
    \caption{Feynman diagrams for the virtual corrections of $\mu N \to e N$ in the full theory.}
    \label{fig:virtual-diags-full}
\end{figure}

\begin{figure}[htb!]
    \centering
    \begin{subfigure}[t]{0.28\textwidth}
    \renewcommand{\thesubfigure}{\alph{subfigure}}
    \setcounter{subfigure}{6}    
        \centering
        \vspace{0.2mm}
        \includegraphics[width=1\textwidth]{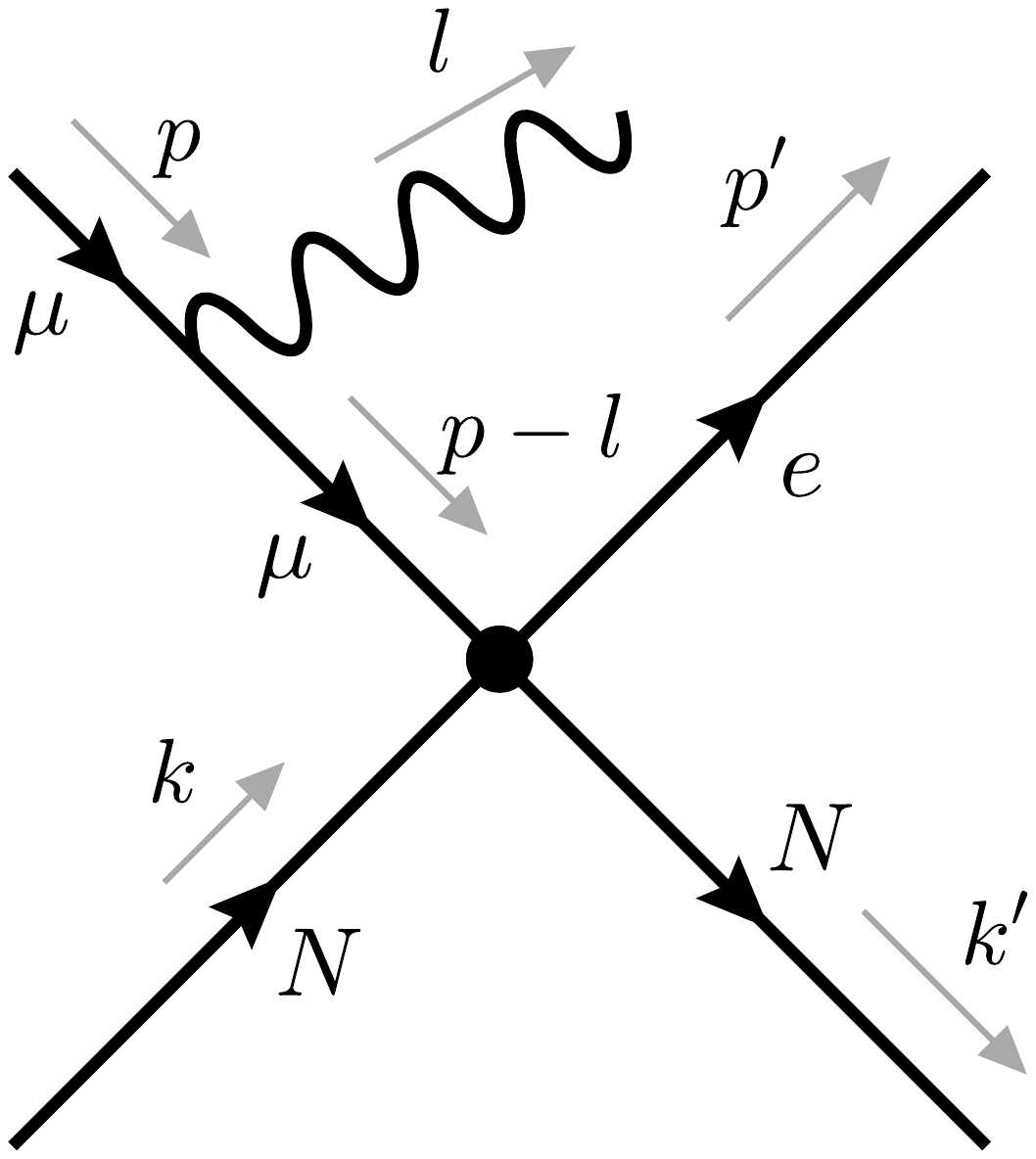}
        \vspace{-18mm}
        \caption{}
    \end{subfigure}
    \hspace{1mm}
    \begin{subfigure}[t]{0.28\textwidth}
    \renewcommand{\thesubfigure}{\alph{subfigure}}
    \setcounter{subfigure}{7}    
        \centering
        \vspace{4.7mm}
        \includegraphics[width=1\textwidth]{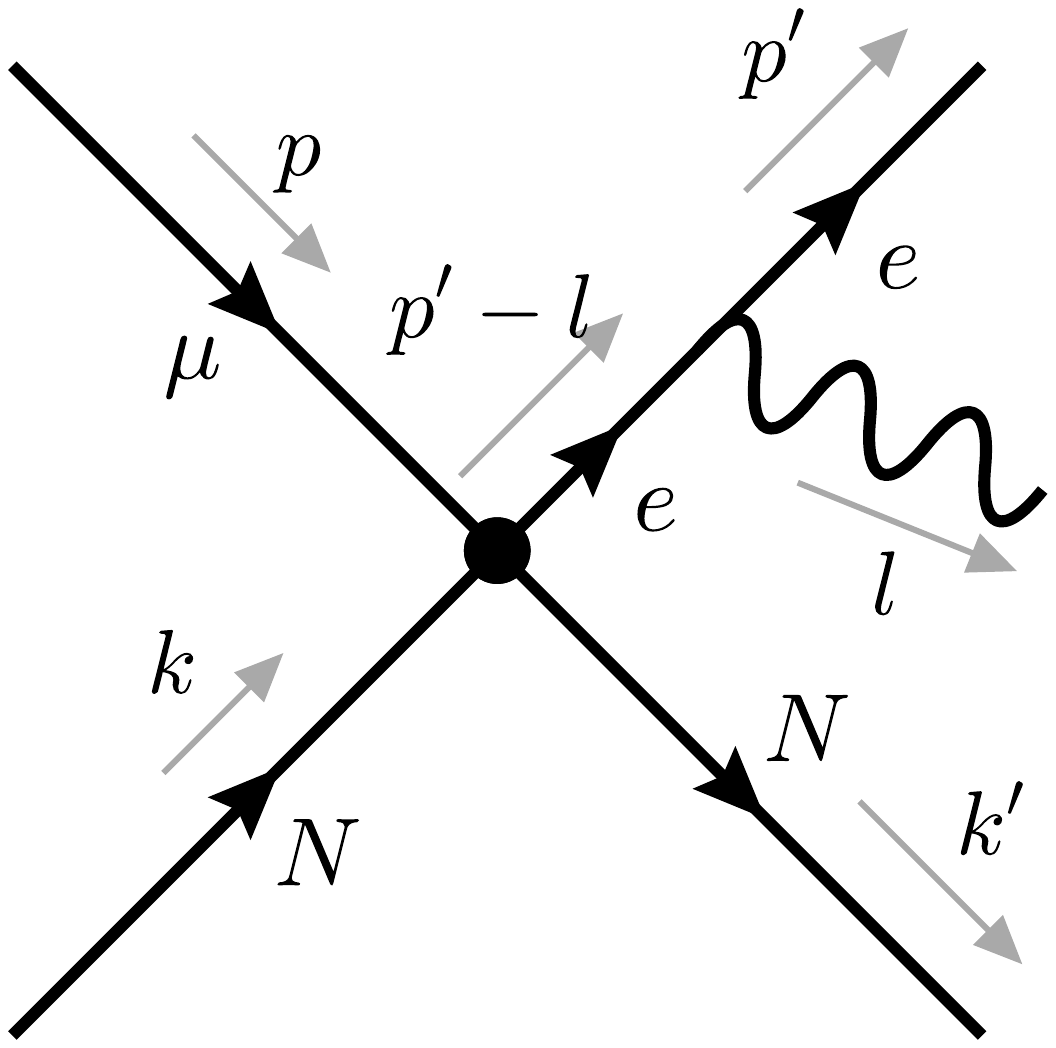}
        \vspace{-22.5mm}
        \caption{}
    \end{subfigure}
    \vspace{-0.1mm}
    \caption{Feynman diagrams for the real corrections of $\mu N \to e N$ in the full theory.}
    \label{fig:real-diags-full}
\end{figure}

The analysis of the regions applies not only the virtual diagrams but also to the phase-space integrals in the real corrections. In figure \ref{fig:virtual-diags-full}, we present the complete set of one-loop virtual electromagnetic corrections to direct muon conversion, while in figure \ref{fig:real-diags-full} we show the relevant real corrections (the photon emissions from the nucleus legs vanish in the Feynman gauge ---  which we employ throughout the paper --- in the limit of infinite nucleus mass).
All these diagrams belong to NEFT, which we identify in what follows as the \textit{full theory}. This is the theory in which all the states of energy above the nuclear level were integrated out. Consequently, the full theory treats disparate scales (such as $M_N$ and $m_e$) at the same level, thus lacking both a homogeneous power counting and a scale separation. It constitutes the starting point of our analysis, from which we will eventually obtain a consistent description.

In table \ref{table:regions}, we show the relevant momentum \textit{regions} that contribute to each of the diagrams of figures \ref{fig:virtual-diags-full} and \ref{fig:real-diags-full} (in the case of the real radiation, we consider the terms obtained after squaring the real emission amplitude); see appendix \ref{app:regions} for details. 
\begin{table}[h!]
\begin{center}
\begin{tabular}{lll}
\hlinewd{1.1pt} \\[-5mm]
Type of correction & Diagram & Regions \\ \midrule
\multirow{6}{*}{Virtual} & (a) & potential, hard \\
 & (b) & hard, collinear \\
 & (c) & hard, collinear \\
 & (d) & hard-nuclear \\
 & (e) & hard \\
 & (f) & hard, collinear \\[1mm]
\cdashline{1-3} \\[-5mm]
\multirow{3}{*}{Real} & (g)$\times$(g) & soft \\
& (h)$\times$(h) & soft-collinear \\
& (g)$\times$(h) & soft, soft-collinear \\[0.3mm]
\hlinewd{1.1pt}
\end{tabular}
\caption{Relevant regions of both virtual and real corrections to muon conversion and DIO. The letters correspond to  diagrams in figures \ref{fig:virtual-diags-full} and \ref{fig:real-diags-full}. The different regions in the third column correspond to modes of the fields, defined in table \ref{table:modes}. Details of the analysis of momentum regions can be found in appendix \ref{app:regions}.}
\label{table:regions}
\end{center}
\vspace{-5mm}
\end{table}
The momentum regions correspond to different possibilities of \textit{scaling} for the loop momentum $l$, i.e., different ways in which the components of $l$ can scale with the power-counting parameter. Though the regions are properties of the loop integrals, they can be set in correspondence with \textit{modes} of the fields. This happens in such a way that the names of the scaling for $l$ will also describe the corresponding modes of the fields. Table~\ref{table:modes} describes the modes corresponding to the regions found in table \ref{table:regions}.
\begin{table}[h!]
\begin{center}
\begin{tabular}{lcll}
\hlinewd{1.1pt} \\[-5mm]
Mode name & Abbreviation & Momentum scaling & Virtuality \\ \midrule
hard-nuclear & $\hn$ & $l^{\hn} \sim (1 \oj, 1 \oj, 1 \oj) \, M_N$ & \hspace{3mm} $M_N^2$\\[0.5mm]
hard & $\h$ & $l^{\h}  \hspace{1.9mm} \sim (1 \oj, 1 \oj, 1 \oj) \, m_{\mu}$ & \hspace{3mm} $m_{\mu}^2$\\[0.5mm]
semi-hard & $\sh$ & $l^{\sh} \hspace{0.5mm} \sim (\lambda \lj, \lambda \lj, \lambda \lj) \, m_{\mu} $ & $\lambda^2 \, m_{\mu}^2$\\[0.5mm]
hard-collinear & $\hc$ & $l^{\hc} \hspace{0.6mm} \sim (1 \oj, \lambda \lj, \lambda^2) \, m_{\mu} $ & $\lambda^2 \, m_{\mu}^2$\\[0.5mm]
potential & $\p$ & $l^{\p} \hspace{2.1mm} \sim (\lambda^2, \lambda) \, m_{\mu} $ & $\lambda^2 \, m_{\mu}^2$\\[0.5mm]
soft & $\s$ & $l^{\s} \hspace{2.3mm} \sim (\lambda^2, \lambda^2, \lambda^2) \, m_{\mu} $ & $\lambda^4 \, m_{\mu}^2$ \\[0.5mm]
collinear & $\c$ & $l^{\c} \hspace{2.3mm} \sim (1 \hspace{1.9mm}, \lambda^2, \lambda^4) \, m_{\mu} $ & $\lambda^4 \, m_{\mu}^2$ \\[0.5mm]
soft-collinear & $\sc$ & $l^{\sc} \hspace{0.9mm} \sim (\lambda^2, \lambda^4, \lambda^6) \, m_{\mu} $ & $\lambda^8 \, m_{\mu}^2$ \\[0.5mm]
\hlinewd{1.1pt}
\end{tabular}
\caption{Relevant modes for muon conversion and DIO.}
\label{table:modes}
\end{center}
\end{table}
For each mode, we show the name, the abbreviation adopted in the rest of the paper, the momentum scaling and the virtuality of a photon with that scaling.%
\fn{The names of the modes and regions adopted in tables \ref{table:regions} and \ref{table:modes} combine the standard terminology of SCET with that of pNRQCD. We adopted the parameter $\lambda$ for the power expansion; the standard pNRQCD power expansion can be obtained by identifying $\lambda$ with the velocity $v$ of the muon.}
The table also describes the semi-hard mode, which does not correspond to any region in table \ref{table:regions}, but which is necessary for the construction of NRQED.

Based on the modes in the table \ref{table:modes}, as well as on the hierarchy~(\ref{eq:hierarchy}), we define several physical \textit{scales} that we will reference throughout the paper:
\ali{
\hspace{25mm} \textrm{hard-nuclear scale:}& \hspace{-2mm} &\mu_{\hnvar} &\sim M_N, \nonumber \\[1mm]
\hspace{25mm} \textrm{hard scale:}& \hspace{-2mm} &\mu_{\hvar} &\sim m_{\mu} \simeq E_e  = \mathcal{O}(\lambda_R \, M_N), \nonumber \\[1mm]
\hspace{25mm} \textrm{semi-hard scale:}& \hspace{-2mm} &\mu_{\shvar} &\sim Z \alpha m_{\mu} = \mathcal{O}(\lambda \, \mu_{\hvar}), \nonumber \\[1mm]
\hspace{25mm} \textrm{soft scale:}& \hspace{-2mm} &\mu_{\svar} &\sim (Z \alpha)^2 m_{\mu} \simeq m_e \simeq \Delta E = \mathcal{O}(\lambda^2 \, \mu_{\hvar}), \nonumber \\[-0.2mm]
\hspace{25mm} \textrm{soft-collinear scale:}& \hspace{-2mm} &\mu_{\scvar} &\sim m_e \frac{\Delta E}{m_{\mu}} = \mathcal{O}(\lambda^4 \, \mu_{\hvar})\nonumber,
}
so that their hierarchy is 
\ali{
\mu_{\hnvar}  \gg \mu_{\hvar} \gg \mu_{\shvar} \gg \mu_{\svar} \gg \mu_{\scvar}.  
}

\section{EFT}
\label{sec:EFT}

In this section, we develop the EFT approach that corresponds to the analysis of regions summarized in the previous section.
Using EFTs instead of a mere analysis of regions brings multiple advantages. Decisively, an EFT allows for a systematic organization of the various contributions --- clearly separating quantities of different scales --- as well as the derivation of all-orders factorization theorems. Furthermore, the EFT contains only the relevant modes, providing Feynman rules that reproduce the full theory expressions order by order in the EFT expansion. This happens in such a way that the information about the physics associated with short-distance modes is codified in matching coefficients and is obtained via matching. As mentioned before, this offers a solution to the problem of large logarithms that would otherwise plague the calculation.
The EFT method is especially convenient for bound-state problems, offering an elegant bridge between quantum field theory and non-relativistic quantum mechanics. In particular, the Schrödinger equation naturally follows from the rules of pNRQED \cite{Beneke:1998jj, Beneke:1999qg, Beneke:2013jia}. In this way, bound-state wave functions are harmoniously included in the quantum field theory.

Another decisive element of the EFT approach is that, for each EFT, both the operator basis and the Lagrangian have a well-defined and systematic expansion in powers of a particular expansion parameter. Whereas in the traditional EFT approach this parameter is usually associated to a mass dimension, this does not have to be the case in modern EFTs. Given the complexity of our problem, we will focus only on the LP terms. The subleading terms correspond to perturbations, and become relevant as we move away from the endpoint of the spectrum. A systematic extension to next-to-leading power (NLP) introduces further conceptual and technical challenges in various EFTs \cite{Beneke:2007pj, Moult:2018jjd,Beneke:2019kgv, Moult:2019mog, Moult:2019uhz, Liu:2019oav, Wang:2019mym, Beneke:2020ibj, Liu:2020tzd,vanBeekveld:2021mxn, Inglis-Whalen:2021bea, Bell:2022ott, Beneke:2022obx, Liu:2022ajh, Cornella:2022ubo, Hurth:2023paz, Schoenleber:2024ihr}. Though such extension is possible with the current state-of-the-art methods within the framework introduced here, we leave it for future work.

\begin{figure}[htb!]
\centering
\includegraphics[width=1\textwidth]{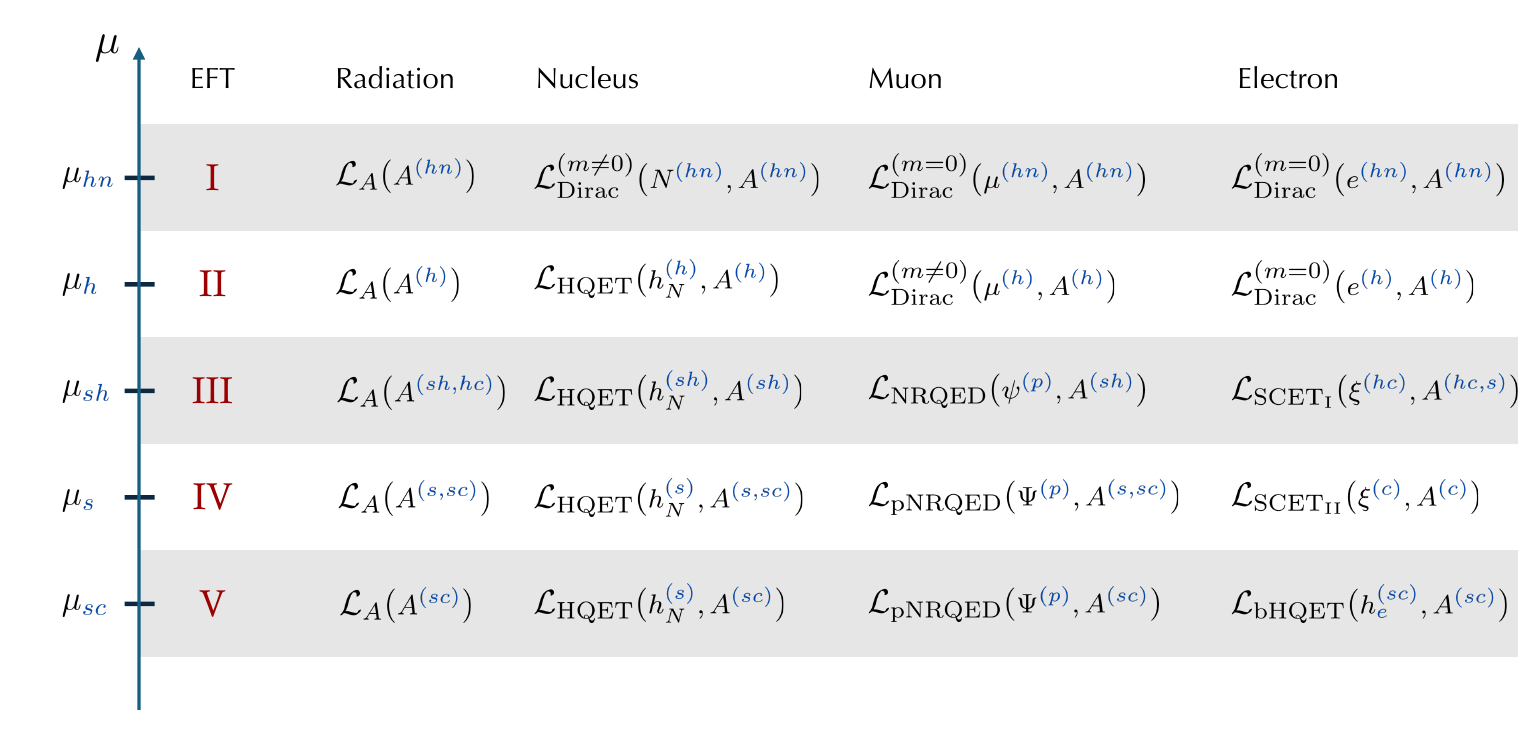}
\vspace{-10mm}
\caption{Chart of the multiple scales relevant for muon conversion and DIO. Each scale is associated with an EFT. For each EFT, we show the Lagrangian governing the relevant fields. See text for details.}
\label{fig:EFT-chart}
\end{figure}

In the EFT approach, the different scales discussed in the analysis of regions are set in correspondence with different EFTs. This is schematically illustrated in figure \ref{fig:EFT-chart}, which shows a range of energies, starting with the hard-nuclear scale, and going down to the soft-collinear one.
A sequence of five EFTs encompasses this range precisely in a one-to-one correspondence with the scales identified above. Each EFT has its own Lagrangian and currents.  The latter refers to interactions of matter fields connecting different modes, obtained after integrating out modes of higher virtuality and, in the case of muon conversion, violating lepton flavor.%
\fn{
This distinction is common, e.g., in the SCET or HQET literature, and enables a systematic organization of the interactions. We note that the potential terms, which correspond to non-local four-fermion operators mediating instantaneous interactions below the semi-hard scale, are part of the Lagrangian $\mathcal{L}_{\rm pNRQED}$.  
}
The last four columns of the figure describe, for each EFT, the Lagrangian that governs the fields directly involved in muon conversion. There, $A$ is the photon field; as for the remaining fields, while some have already been introduced in the previous sections, others will be introduced throughout section \ref{sec:EFT}. Here and in what follows, superscripts refer to the mode at stake.

The rest of the section is devoted to characterizing the subsequent EFTs. For each one, we start by discussing the power counting of the fields, which allows us to determine the LP Lagrangian. Whenever relevant, we identify the proper low-energy modes which describe small oscillations of the fields around the vacuum, and which thus lead to an adequate perturbative description (free from large logarithmic corrections).%
\fn{Conversely, all modes that put particles far off-shell correspond to large fluctuations and must, for that reason, be integrated out.}
After writing the Lagrangian, we present the current, fix the matching conditions, provide the one-loop matching coefficients (for which we assume the $\overline{\rm MS}$ scheme) and discuss RG evolution. 
We complement the equations with several discussions and clarifications.

\subsection{EFT I: \texorpdfstring{$\mu \sim \mu_{\hnvar}$}{mu ≃ muhn}}

EFT I describes physics at and somewhat below the hard-nuclear scale. Accordingly, we consider the loop momentum $l$ of the diagrams of figure \ref{fig:virtual-diags-full} to scale as $l \sim l^{\hn}$ (see table \ref{table:modes}). The propagators of the different fields, before expansion, thus read 
\bs
\label{eq:propagator-to-the-counting-EFT-I}
\ali{
\langle 0|T\{N^{\hn}\ar{0} \bar{N}^{\hn}\ar{x}\}| 0\rangle &= \int \frac{d^4 l^{\hn}}{(2 \pi)^4} \frac{i}{\slashed{l}^{\hn}-M_N+i \varepsilon} e^{i l^{\hn}\cdot x} \sim M_N^3, \\
\langle 0|T\{\mu^{\hn}\ar{0} \bar{\mu}^{\hn}\ar{x}\}| 0\rangle &= \int \frac{d^4 l^{\hn}}{(2 \pi)^4} \frac{i}{\slashed{l}^{\hn}-m_{\mu}+i \varepsilon} e^{i l^{\hn}\cdot x} \sim M_N^3, \\
\langle 0|T\{e^{\hn}\ar{0} \bar{e}^{\hn}\ar{x}\}| 0\rangle &= \int \frac{d^4 l^{\hn}}{(2 \pi)^4} \frac{i}{\slashed{l}^{\hn}-m_e+i \varepsilon} e^{i l^{\hn} \cdot x} \sim M_N^3, \\
\langle 0|T\{A^{\hn}_{\mu}\ar{0} A^{\hn}_{\nu}\ar{x}\}| 0\rangle &= \int \frac{d^4 l^{\hn}}{(2 \pi)^4} \frac{-i}{(l^{\hn})^2+i \varepsilon} g^{\mu \nu}  e^{i l^{\hn} \cdot x} \sim M_N^2,
}
\es
where the photon propagator is written in the Feynman gauge, and where we indicated the order of the leading term only. 
We conclude that $A^{\hn}_{\mu} \sim M_N$ and $N^{\hn} \sim \mu^{\hn} \sim e^{\hn} \sim M_N^{3/2}$. It follows that the derivative terms for the leptons and nucleus are all of $\mathcal{O}(M_N^4)$, that is,  $\bar{N}^{\hn} \slashed{\partial} N^{\hn} \sim \bar{\mu}^{\hn} \slashed{\partial} \mu^{\hn} \sim \bar{e}^{\hn} \slashed{\partial} e^{\hn} \sim M_N^4$. 
The nucleus mass term is also of this order, $\bar{N}^{\hn} N^{\hn} M_N \sim M_N^4$. By contrast, $\bar{\mu}^{\hn} \mu^{\hn} m_{\mu} \sim M_N^3 m_{\mu} \sim M_N^4 \lambda_R \ll M_N^4$, which means that the muon mass term is recoil suppressed when compared to the derivative terms; this is even more true for the electron mass term. Hence, the Lagrangian of EFT I can be written as 
\ali{
\label{eq:Lag-I}
	\mathcal{L}^{\I} = \mathcal{L}^{\I\rm,LR} + \mathcal{O}(\lambda_R),
}
where $\mathcal{L}^{\I\rm,LR}$ is the leading recoil Lagrangian, which is given by
\be
\label{eq:LagI}
\mathcal{L}^{\I\rm,LR} = \mathcal{L}_{A}^{\I\rm,LR} +
\mathcal{L}_{N}^{\I\rm,LR} + \mathcal{L}_{\mu}^{\I\rm,LR} + \mathcal{L}_{e}^{\I\rm,LR} \, ,
\ee
with
\ali{
\label{eq:3.4}
\mathcal{L}_{A}^{\I\rm,LR} &= -\dfrac{1}{4} F_{\mu\nu}^{\hn} F^{\mu\nu\hn},
&
\mathcal{L}_{N}^{\I\rm,LR} &= \bar{N}^{\hn} (i \slashed{D}^{\hn} - M_{N}) N^{\hn}, \nonumber \\
\mathcal{L}_{\mu}^{\I\rm,LR} &= \bar{\mu}^{\hn} i \slashed{D}^{\hn} \mu^{\hn},
&
\mathcal{L}_{e}^{\I\rm,LR} &= \bar{e}^{\hn} i \slashed{D}^{\hn} e^{\hn}.
}
Here, $F_{\mu\nu}^{\hn} = \partial_{\mu} A_{\mu}^{\hn} - \partial_{\nu} A_{\mu}^{\hn}$ and $\slashed{D}^{\hn} = \gamma^{\mu} D^{\hn}_{\mu}$, where the covariant derivative is 
\ali{
\label{eq:cov-der-definition}
D^{\hn}_{\mu} = \partial_{\mu} + i Q e A_{\mu}^{\hn},
}
with $e= \sqrt{4 \pi \alpha}$ being the electromagnetic coupling and $Q$ the charge of the particle involved ($Q=Z$ for the nucleus and $Q=-1$ for the leptons). All fields describe hard-nuclear modes. Recoil corrections are $\mathcal{O}(\lambda_R) = \mathcal{O}\left(10^{-3}\right)$ and we neglect them in what follows. The leading recoil corrections can be easily restored \cite{Czarnecki:2011mx} and, since EFT I obeys a simple dimensional counting, a systematic extension beyond leading recoil is also feasible. 

Before we proceed, an important comment is in order, concerning the separation of modes in the EFT framework that we are discussing. At the hard-nuclear scale, the EFT does not exhibit homogeneous power counting. The reason is that the separation of the modes is not complete. This incompleteness arises because, in our approach, modes of higher virtuality contain lower virtuality modes as submodes.

Let us illustrate this aspect by taking the muon mass term as an example. In the action, that term reads
\begin{align}
   \int d^4 x_{\hn} \, m_\mu {\bar{\mu}}^{\hn} \mu^{\hn}\sim \frac{m_\mu} {M_N}. 
\end{align}
This term is power suppressed only when we assume a strict separation of modes, that is, when we assume that the muon field contains hard-nuclear modes only.
By contrast, if the muon field is assumed to contain not only the hard-nuclear mode but also submodes such as the hard one, we have
\begin{align}
\label{eq:submodes-case}
   \int d^4 x_{\hn} \, m_\mu {\bar{\mu}}^{\hn} \mu^{\hn}  \supset \int d^4 \, x_{\h} m_\mu \bar{\mu}^{\h} \mu^{\h} \sim 1 \,,
\end{align}
in which case the mass term is not suppressed. Now, if we were to adopt the first approach --- with a strict separation of modes --- 
we would have to write a different field for each possible mode explicitly: a hard-nuclear muon $\mu^{\hn}$, a hard muon $\mu^{\h}$, etc. This is clearly not practical. Instead, we adopt the approach of implicitly including the submodes, while ensuring that they are explicitly written only when considering modes with the corresponding virtuality. So, for example, in EFT I (aimed at the hard-nuclear scale), we omit the hard submodes, even though they contribute at LP.
While this distinction may seem pedantic at this point, we adopt it to make the notation more concise; it will become crucial at lower scales, where several modes with the same virtuality appear. The strategy of the submodes also simplifies matching; for example, the hard-nuclear region directly corresponds to Lagrangian~(\ref{eq:LagI}), without the need to expand the loop integrals in the small parameters.  

Among the terms of eq.~(\ref{eq:total-current}), only the scalar and the vector operators are relevant at LP. We thus write the current in EFT I as 
\be
\label{eq:current-I}
\mathcal{J}^{\I}\ar{0}
= -\frac{4 G_F}{\sqrt{2}} \bigg\{ C_{SX}^{\I} \mathcal{O}_{SX}^{\I}\ar{0}
+ C_{VX}^{\I} \mathcal{O}_{VX}^{\I}\ar{0} \bigg\} + \rm h.c. ,
\ee
with
\ali{
\label{eq:current-Ia}
\mathcal{O}_{SX}^{\I} \equiv \bar{N}^{\hn} N^{\hn} \, \bar{e}^{\hn} P_X \mu^{\hn},
\qquad
\mathcal{O}_{VX}^{\I} \equiv \bar{N}^{\hn} \gamma_{\alpha} N^{\hn} \, \bar{e}^{\hn} \gamma_{\alpha} P_X \mu^{\hn}.
}
Here and in what follows, we assume Einstein's summation convention for the chiralities, and omit position-space arguments whenever all the fields are evaluated at the same space-time point. The current is evaluated at position zero for later convenience.

As discussed in section \ref{sec:nucleus}, non-perturbative nuclear physics is beyond the scope of this work. Consequently, we take EFT I, with eqs.~(\ref{eq:Lag-I}) and~(\ref{eq:current-I}), as the starting point of our analysis. The power counting of EFT I and electromagnetic gauge invariance ensure that any operators with multiple photons with hard-nuclear virtuality will be power suppressed. However, to account for subleading nuclear effects \cite{Bartolotta:2017mff} and combine them with the higher-order QED effects discussed here, such operators must be included.
\subsection{EFT II: \texorpdfstring{$\mu \sim \mu_{\hvar}$}{mu ≃ muh}}
\label{sec:EFT-II}

EFT II takes EFT I and integrates out modes with virtualities $M_N^2$ and $M_N m_{\mu}$. The focus of EFT II is thus the hard scale, where the typical momentum scales as $l \sim l^{\h}$. With this scaling, the Dirac propagator for $N$ needs to be expanded. As a result, the nucleus field is now HQET-like, and is described by a field $h_N^{\h}$ such that
\ali{
\label{eq:propagator-to-the-counting-EFT-II}
\langle 0|T\{h_N^{\h}\ar{0} \bar{h}^{\h}_N\ar{x}\}| 0\rangle &= \int \frac{d^4 l^{\h}}{(2 \pi)^4} \frac{i}{v \cdot l^{\h} +i \varepsilon} e^{i l^{\h} \cdot x} \sim m_{\mu}^3,
}
which implies $h_N^{\h} \sim m_{\mu}^{3/2}$.
Following a reasoning analogous to that of eq.~(\ref{eq:propagator-to-the-counting-EFT-I}), we also find $\mu^{\h} \sim e^{\h} \sim m_{\mu}^{3/2}$. Then, the muon mass term is of leading power, $\int d^4x \, \bar{\mu}^{\h} \mu^{\h} m_{\mu} \sim \mathcal{O}(m_{\mu}^0)$, whereas the electron mass term is power suppressed, $\int d^4x \, \bar{e}^{\h} e^{\h} m_{e} \sim \mathcal{O} (m_e/m_\mu)$ $ \sim \lambda^2 $.
The Lagrangian for EFT II can then be written as 
\ali{
\label{eq:Lag-II}
	\mathcal{L}^{\II} = \mathcal{L}^{\II\rm,LP} + \mathcal{O}(\lambda),
}
where $\mathcal{L}^{\II\rm,LP}$ is the LP Lagrangian, given by
\be
\mathcal{L}^{\II\rm,LP} = 
\mathcal{L}_{A}^{\II\rm,LP} + 
\mathcal{L}_{h_N}^{\II\rm,LP} + \mathcal{L}_{\mu}^{\II\rm,LP} + \mathcal{L}_{e}^{\II\rm,LP} \, ,
\ee
with
\ali{
\label{eq:sub-Lags-EFT-II}
\mathcal{L}_{A}^{\II\rm,LP} &= -\dfrac{1}{4} F_{\mu\nu}^{\h} F^{\mu\nu\h},
&
\mathcal{L}_{h_N}^{\II\rm,LP} &= \bar{h}_N^{\h} i v \cdot D^{\h} h_N^{\h}, \nonumber \\
\mathcal{L}_{\mu}^{\II\rm,LP} &= \bar{\mu}^{\h} (i \slashed{D}^{\h} - m_{\mu}) \mu^{\h},
&
\mathcal{L}_{e}^{\II\rm,LP} &= \bar{e}^{\h} i \slashed{D}^{\h} e^{\h}.
}
The covariant derivative $D^{\h}$ is equivalent to eq.~(\ref{eq:cov-der-definition}), but for hard photons.
In fact, all the fields in eq.~(\ref{eq:sub-Lags-EFT-II}) describe hard fluctuations. We note that even though $\mathcal{L}^{\II}$ must be matched to $\mathcal{L}^{\I}$, the LP terms discussed here are protected by symmetries and do not receive non-trivial matching coefficients.  
We also note that the nucleus described by the $\mathcal{L}_{h_N}^{\II\rm,LP}$ does not propagate in space, a direct consequence of treating it as a static (i.e., infinitely heavy) source of electric potential.
The NLP corrections to $\mathcal{L}_{h_N}^{\II}$ would allow us to reproduce the recoil corrections. 

At leading recoil, only two operators in EFT II contribute to coherent muon conversion --- specifically, the scalar left- and right-handed ones, given by
\be
\label{eq:OSX-II}
\mathcal{O}_{X}^{\II} \equiv \bar{h}_N^{\h} h_N^{\h} \, \bar{e}^{\h} P_X \mu^{\h}.
\ee
The matching condition between EFTs I and II reads %
\ali{
\label{eq:matching-I-II}
C_{SX}^{\I} \mathcal{O}_{SX}^{\I}\ar{0} + C_{VX}^{\I} \mathcal{O}_{VX}^{\I}\ar{0} = C_{X}^{\II} \, \mathcal{O}_{X}^{\II}\ar{0},
}
where all the quantities are understood to be evaluated at the matching scale $\mu= \mu_{\hnvar}$. 
Introducing $\bar{X}$ to denote the alternative chirality to that of $X$ (i.e., $\bar{L} = R$ and $\bar{R} = L$), the matching coefficients at one-loop accuracy are%
\fn{To simplify the reading, we omit the scale at which $\alpha$ is evaluated (which is the scale of the matching --- in this case, $\mu_{\hnvar}$).}
\ali{
\label{eq:matching-I-II-result}
C_{X}^{\II}(\mu_{\hnvar}) &= C_{SX}^{\I}(\mu_{\hnvar}) + C_{V\bar{X}}^{\I}(\mu_{\hnvar}) + \dfrac{Z \alpha}{2 \pi} \Bigg[2 \, C_{SX}^{\I}(\mu_{\hnvar}) \ln \dfrac{M_N^2}{\mu_{\hnvar}^2} - 5 \, C_{SX}^{\I}(\mu_{\hnvar}) \hspace{15mm} \nonumber \\
& \hspace{25mm} + 7 \, C_{V\bar{X}}^{\I}(\mu_{\hnvar}) \Bigg]  - \dfrac{Z^2 \alpha}{4 \pi} C_{SX}^{\I}(\mu_{\hnvar}) \left( 3 \ln \dfrac{M_N^2}{\mu_{\hnvar}^2} + 2\right).
}
It is clear that only corrections proportional to $Z$ exist. On the other hand, these corrections cannot be calculated perturbatively. More generally, hard-nuclear loops involving the nucleus field cannot be computed perturbatively in an EFT with a point-like nucleus at the nuclear-scale $\mu_{\hn} \sim M_N$. Instead, they should be replaced by a suitable generalization of EFT I, with a proper non-perturbative matching (including QED effects, see refs. \cite{Beneke:2017vpq,Beneke:2019slt,Cornella:2022ubo,Boer:2023vsg} for a corresponding discussion in the case of heavy mesons).

Such loops belong to what we identified before as nuclear effects (i.e., terms with non-zero powers of $Z$), and we will discuss them in future work. For the scales which we are here interested in --- much lower than $\mu_{\hn}$ (namely, $\mu_{\h}$ and below) --- and at LP, it is sufficient to include a nuclear form-factor to describe the leading nuclear effects.
Moreover, we adopt the point of view according to which those effects are perturbative in QED, but need to be calculated non-perturbatively in QCD. 
They depend on non-perturbative scales (such as $\Lambda_{\rm QCD}$) and hadron masses, and we assume that they are included in the matching coefficients of EFT II, defined at the hard scale.

All of this implies that we effectively take the hard scale as the actual starting point for a \emph{perturbative analysis} of QED effects in muon conversion and DIO. That is to say, we take the matching coefficients at the muon mass scale, $C_{X}^{\II}(\mu_{\h})$, as input parameters of our analysis, and assume that they are determined either experimentally or using lattice field theory.
This approach is consistent with the circumstance that the RG running of $C_{X}^{\II}(\mu_{\h})$ is not affected by nuclear effects. Indeed, the RGE for the matching coefficient reads 
\ali{
\label{eq:RGE-II}
\mu \dfrac{d}{d \mu} C_{X}^{\II}\bigg(\mu, \alpha(\mu), C^{\I}(\mu)\bigg) =  -\dfrac{3 \alpha(\mu)}{2\pi} C_{X}^{\II}\bigg(\mu, \alpha(\mu), C^{\I}(\mu)\bigg) + \mathcal{O}(\alpha^2).
}
We adopt a notation that explicitly reflects the fact that the matching coefficient depends on the scale $\mu$ not only directly, but also indirectly through both $\alpha$ and the coefficients of the prior EFT. As a result, the RGE~(\ref{eq:RGE-II}) contains information not only about the explicit $\mu$ dependence of the matching equation, but also about the running of both $\alpha$ and the coefficients $C_{SX}^{\I}$ and $C_{VX}^{\I}$.
And whereas the latter does not run (as is well known), $C_{SX}^{\I}$ does --- this being the origin of the result of the right-hand side of eq.~(\ref{eq:RGE-II}), which is thus inherited from EFT I and derives from the photon exchange between the two leptons.
This exchange gives no contribution to the matching, eq.~(\ref{eq:matching-I-II}), since the amplitude for it (the photon exchange) is scaleless in EFT I.
It is also the running of $C_{SX}^{\I}$ that explains the absence of terms of $\mathcal{O}(Z^1\alpha)$ and $\mathcal{O}(Z^2\alpha)$ in eq.~(\ref{eq:RGE-II}), since that running cancels exactly the ones coming from the explicit $\ln(\mu_{\hnvar}^2)$ dependence in eq.~(\ref{eq:matching-I-II-result}).

\subsection{EFT III: \texorpdfstring{$\mu \sim \mu_{\shvar}$}{mu ≃ mush}}

EFT III is obtained after integrating out the hard modes.%
\fn{In order not to complicate the  formalism any further, we have refrained from introducing another intermediate EFT at scale $\mu^2 \sim \lambda m_\mu^2$, which does not generate any non-trivial contributions at LP.} 
The modes characterizing the nucleus and pure radiation are now the semi-hard. For the nucleus, lowering the scale below $\mu \simeq \mu_{\hvar}$ does not lead to a different expansion of the propagator. Here and in what follows, then, the nucleus will still be described by an HQET field $h_N$. 
In the case of the electron (i.e., the electron of momentum $p'$, whose energies are of the order of the muon mass), the semi-hard modes put the field far off-shell, so that they are also integrated out. The on-shell modes are the hard-collinear ones, implying that the electron is described by SCET. More specifically, since $m_e^2 \ll \mu_{\shvar}^2$, the electron is effectively massless, described by a $\rm SCET_{I}$ field coupled to soft photons. We describe this hard-collinear electron by the field $\xi$. 
The theory also contains (non-energetic) semi-hard electrons; they are described by a massless Dirac Lagrangian, and we denote their field by $e$. Finally, in the muon case, the on-shell mode has the potential scaling; it is described by NRQED, and we identify its field by $\psi$. Since NRQED does not have homogeneous power counting, the potential muon can interact with a semi-hard photon, creating a (off-shell) semi-hard muon. The Lagrangian in the muon sector can be constructed using standard techniques \cite{Beneke:2013jia, Beneke:2024sfa}.

In sum, the fields for the nucleus, muon and the energetic electron are now those of HQET, NRQED and $\rm SCET_{I}$, respectively, and the field for a non-energetic electron is that of massless QED. As before, we obtain the counting of the fields by considering their propagators. For the nucleus, we have $l \sim l^{\sh}$, so that an equation equivalent to~(\ref{eq:propagator-to-the-counting-EFT-II}) leads to $h_N^{\sh} \sim (\lambda m_{\mu})^{3/2}$. As for the muon and the electron of momentum $p'$, we have
\bs
\label{eq:propagator-to-the-counting-EFT-III}
\ali{
\langle 0|T\{\psi^{\p}\ar{0} \bar{\psi}^{\p}\ar{x}\}| 0\rangle &= \int \frac{d^4 l^{\p}}{(2 \pi)^4} \frac{2 m_\mu i}{2 m_\mu v \cdot l^{\p}-(\vec{l}^{\p})^2 +i \varepsilon} e^{i l^{\p} \cdot x} \sim (\lambda m_{\mu})^3, \\
\label{eq:propagator-to-the-counting-EFT-III-b}
\langle 0|T\{\xi^{\hc}\ar{0} \bar{\xi}^{\hc}\ar{x}\}| 0\rangle &= \int \frac{d^4 l^{\hc}}{(2 \pi)^4} \frac{\slashed{n}_-}{2}\frac{i \, n_+ l^{\hc}}{(l^{\hc})^2 + i \varepsilon} e^{i l^{\hc} \cdot x} \sim \lambda^2 m_{\mu}^3,
}
\es
so that $\psi^{\p} \sim (\lambda m_{\mu})^{3/2}$ and 
$\xi^{\hc} \sim \lambda m_{\mu}^{3/2}$.%
\fn{An electron mass term is power suppressed for hard-collinear modes.  For details, see ref. \cite{Leibovich:2003jd}.} 
Then, the Lagrangian can be written as 
\ali{
\label{eq:Lag-III}
	\mathcal{L}^{\III} = \mathcal{L}^{\III\rm,LP} + \mathcal{O}(\lambda),
}
where $\mathcal{L}^{\III\rm, LP}$ represents the LP Lagrangian of EFT III and is given by
\be
\label{eq:L-parts-III}
\mathcal{L}^{\III\rm,LP} = \mathcal{L}_{A}^{\III\rm,LP} +\mathcal{L}_{h_N}^{\III\rm,LP} + \mathcal{L}_{\psi}^{\III\rm,LP} + 
\mathcal{L}_{e}^{\III\rm,LP} + \mathcal{L}_{\xi}^{\III\rm,LP}
\, ,
\ee
with 
\bs
\ali{
\mathcal{L}_{A}^{\III\rm,LP} &= -\dfrac{1}{4} F_{\mu\nu}^{\sh} F^{\mu\nu\sh} -\dfrac{1}{4} F_{\mu\nu}^{\hc} F^{\mu\nu\hc},
\qquad
\mathcal{L}_{h_N}^{\III\rm,LP} = \bar{h}_N^{\sh} i v \cdot D^{\sh} h_N^{\sh}, \\
\mathcal{L}_{\psi}^{\III\rm,LP} &= 
\bar{\psi}^{\p} \left(i v \cdot D^{\sh} + \frac{(\vec{D}^{\sh})^2}{2m_\mu} \right)\psi^{\p},
\qquad
\mathcal{L}_{e}^{\III\rm,LP} = \bar{e}^{\sh} i \slashed{D}^{\sh} e^{\sh},
\\
\mathcal{L}_{\xi}^{\III\rm,LP} &=
\bar{\xi}^{\hc} \frac{\slashed{n}_{+}}{2}\bigg[ i n_{-}D^{\textcolor{myblue}{(hc+s)}} +  i \slashed{D}_{\perp}^{\hc} \frac{1}{i n_{+}D^{\hc}} i\slashed{D}_{\perp}^{\hc} \bigg]\xi^{\hc},
}
\es
where $D^{\sh}$ and $D^{\hc}$ are equivalent to eq.~(\ref{eq:cov-der-definition}), but for semi-hard and hard-collinear photons, respectively, and
\ali{
\label{eq:cov-deriv-SCET-I}
i n_{-}D^{\textcolor{myblue}{(hc+s)}} = i n_{-}D^{\textcolor{myblue}{(hc+s)}}\ar{x} = i n_{-} \partial - e \left[n_{-} A^{\hc}\ar{x} + n_{-} A^{\s}\ar{x_-}\right],
}
with $(x_-)^{\mu} \equiv (n_+x) \frac{(n_-)^{\mu}}{2}$.
The hard-collinear electrons do not interact with semi-hard photons.%
\footnote{Our semi-hard photons correspond to soft photons in the standard NRQED/NRQCD literature.}
Instead, we find soft photons in $\mathcal{L}_{\xi}^{\III\rm,LP}$, since $n_-A^{\s} \sim n_-A^{\hc}$ ($A^{\s}$ has argument $x_-$ as a consequence of multipole expansion \cite{Beneke:2002ni}). The soft photons are indeed necessary to reproduce corrections to Coulomb potential, which are well understood in the QED case.

The LP matching condition between EFTs II and III is
\ali{
\label{eq:matching-II-III}
C_{X}^{\II} \mathcal{O}_{X}^{\II}\ar{0} = \int dt \, C_{X}^{\III}\ar{t} \, \mathcal{O}_{X}^{\III}\ar{t},
}
with
\be
\label{eq:OIII}
\mathcal{O}_{X}^{\III}\ar{t} \equiv \bar{h}_N^{\sh}\ar{0} h_N^{\sh}\ar{0} \, \left[\bar{\xi}^{\hc} W^{\hc}\right]\ar{t n_+} P_X Y^{\sh \dagger}_{n_-}\ar{0} \psi^{\p}\ar{0}.
\ee
This expression contains two Wilson lines;
$W^{\hc}$ is a hard-collinear Wilson line, which we define as 
\ali{
\label{eq:hc-Wilson-line}
W^{\hc}\ar{x} \equiv \exp \left[i e Q \int_{-\infty}^0 ds \, n_{+} \, A^{\hc}\ar{x+s \, n_{+}}\right],
}
and $Y^{\sh\dagger}_{n_-}$ is a semi-hard Wilson line, which we define as 
\ali{
Y^{\sh\dagger}_{n_-}\ar{x} &\equiv \textrm{exp} \left[ i Q e \int_0^{\infty} ds \, n_-  \, A^{\sh}\ar{x + s \,n_-} \, e^{-\varepsilon s} \right].
}
Eq.~(\ref{eq:OIII}) is manifestly gauge invariant under all relevant gauge transformations: semi-hard, hard-collinear and soft. Regarding invariance under semi-hard gauge transformations, this is trivial for the nucleus fields (which appear in pairs), but not for the leptons, because only the muon field couples to semi-hard photons. In this case, the semi-hard Wilson line ensures gauge invariance. As for the hard-collinear gauge transformations, they only affect the electron field, and the invariance of eq.~(\ref{eq:OIII}) under them results from the inclusion of the hard-collinear Wilson line. Note that the block $\left[\bar{\xi}^{\hc} W^{\hc}\right]$ is non-local along the light-cone (i.e., it has argument $t n_+$) to account for an arbitrary number of unsuppressed $n_+\partial$ derivatives. Finally, soft gauge transformations affect both the (potential) muon and the (hard-collinear) electron fields. To ensure manifest gauge invariance under soft gauge transformations, we choose $x=0$ (which corresponds to the location of the hard interaction) for the whole operator \cite{Beneke:2017ztn, Beneke:2017mmf}. 

Having seen this, we can finally consider the expression for the matching coefficient. Defining its Fourier transform as 
\begin{align}
\int dt e^{i t n_+p'} C_{X}^{\III}\ar{t} = C_{X}^{\III}(n_+p', m_\mu),
\end{align}
we write the matching expression as 
\ali{
\label{eq:CIII}
C_{X}^{\III}(2 E_e, m_\mu;\mu_{\hvar}) &= C_{X}^{\II}(\mu_{\hvar}) \, \mathcal{H}(2 E_e, m_\mu;\mu_{\hvar}),
}
where we define $\mathcal{H}$ to be the \textit{hard function}, given by
\ali{
\label{eq:hard-function}
\mathcal{H}(2 E_e, m_\mu;\mu_{\hvar}) &=
1 -\frac{\alpha}{4 \pi }  \Bigg\{ \frac{m_{\mu}
   \ln \left(\frac{16 E_e^4 \mu_{\hvar}^2}{m_{\mu}^6}\right)}{4 E_e - 2 m_{\mu}} - \ln\left(\frac{\mu_{\hvar}}{m_{\mu}}\right) \ln \left(\frac{\mu_{\hvar}^7}{4 E_e^2 m_{\mu}^5}\right) \nonumber \\ 
   & \hspace{-15mm} - 2
   \operatorname{Li}_2\left(1-\frac{m_{\mu}}{2
   E_e}\right) + \ln^2 \left(\frac{2
   E_e m_{\mu}^2}{\mu_{\hvar}^3}\right) + \frac{2
   E_e \ln \left(\frac{m_{\mu}}{\mu_{\hvar}}\right)}{2
   E_e-m_{\mu}}+\frac{\pi^2}{12} \Bigg\} +\mathcal{O}(\alpha^2).
}
The one-loop matching coefficient was obtained before, in ref.~\cite{Bauer:2000yr}, in the context of heavy-to-light factorization in SCET. The NLP matching has also been considered in refs.~\cite{Hill:2004if, Beneke:2005gs}.
Both the renormalization constant and the anomalous dimension of the operator~(\ref{eq:OIII}) can be obtained using standard techniques (see refs.~\cite{Beneke:2017ztn, Beneke:2018rbh, Beneke:2019kgv} for details on the renormalization of the SCET currents, computational methods and SCET Feynman rules).  
The RGE has the typical structure encountered in the heavy-to-light transitions, involving the cusp anomalous dimension \cite{Korchemsky:1987wg}. In our case, it reads 
\ali{
\mu \dfrac{d}{d \mu} C_{X}^{\III}(2E_e, m_\mu;\mu) 
 = \Bigg[ \Gamma_{\rm cusp}^{\III} \ln \left(\dfrac{2E_e}{\mu}\right) + \gamma^{\III} \Bigg]
C_{X}^{\III}(2E_e, m_\mu;\mu),
}
with the cusp and non-cusp parts respectively given by \cite{Frenkel:1984pz,Korchemsky:1987wg,Henn:2019swt}
\ali{
\label{eq:cusp-and-non-cusp}
\Gamma_{\rm cusp}^{\III} = \dfrac{\alpha}{\pi} - \left(\dfrac{\alpha}{\pi}\right)^2 \dfrac{5}{9}+ \mathcal{O}(\alpha^3),
\qquad
\gamma^{\III} = - \dfrac{5 \alpha}{4 \pi} +\mathcal{O}(\alpha^2).
}
The solution is 
\ali{
\label{eq:CIII-running}
C_{X}^{\III}(\mu) = U_{\hvar}(\mu_{\hvar},\mu) \, C_{X}^{\III}(\mu_{\hvar}),
}
with
\ali{
U_{\hvar}(\mu_{\hvar},\mu) &= \operatorname{exp} \Bigg\{-\frac{\alpha(\mu)  (5 \alpha(\mu_{\hvar})+9
   \pi ) \ln \left(\frac{\alpha(\mu) }{\alpha(\mu_{\hvar})}\right)-(5 \alpha(\mu) +9 \pi ) (\alpha(\mu) -\alpha(\mu_{\hvar}))}{4 \alpha(\mu)  \alpha(\mu_{\hvar})} \Bigg\} \nonumber \\
& \hspace{15mm} \times \left(\frac{\alpha(\mu_{\hvar}) }{\alpha(\mu)}\right)^{15/8} \left(\frac{(2 E_e)^2}{\mu_{\hvar}^2}\right)^{\frac{9 \pi  \ln
   \left(\frac{\alpha(\mu) }{\alpha(\mu_{\hvar})}\right)-5
   \alpha(\mu) +5 \alpha(\mu_{\hvar})}{12 \pi}}.
}
Considering the limit in which the running of the electromagnetic coupling $\alpha$ is neglected (i.e., setting $\beta(\alpha) =0$), the evolution factor simplifies to
\begin{align}
 U_{\hvar}(\mu_{\hvar},\mu) = \operatorname{exp}\Bigg\{
\dfrac{\alpha}{36 \pi } \ln\left(\frac{\mu_{\hvar}}{\mu }\right) \left[45-2 \left(9-\frac{5 \alpha }{\pi }\right) \left(\ln \frac{2 E_e}{\mu } + 
   \ln \frac{2 E_e}{\mu_{\hvar}}\right)\right]
\Bigg\} \, .
\end{align}

We conclude this section by noting that the soft photons present in $\mathcal{L}_{\xi}^{\III\rm, LP}$ can be decoupled via the field redefinition \cite{Bauer:2001yt}
\ali{
\label{soft-decoupling-xi}
\bar{\xi}^{\hc}\ar{x} &= \bar{\xi}^{\hc}_{\zz}\ar{x} Y_{n_-}^{\s\dagger}\ar{x_-},
}
where $\xi^{\hc}_{\zz}$ is the soft-decoupled hard-collinear outgoing electron field and $Y_{n_-}^{\s\dagger}$ is a soft Wilson line. For convenience, given an arbitrary charge $Q$ and an arbitrary four-vector $u$, we define soft Wilson lines for outgoing and incoming particles respectively as 
\bs
\label{eq:Ys-definition}
\ali{
Y^{\s\dagger}_{u}\ar{x} &\equiv \textrm{exp} \left( i Q e \int_0^{\infty} ds \, u \cdot \, A^{\s}\ar{x + s \,u} \, e^{-\varepsilon s} \right), \\
\overline{Y}_{u}^{\s}\ar{x} &\equiv \textrm{exp} \left( i Q \, e \int_{-\infty}^{0} ds \, u \cdot A^{\s}\ar{x + s \, u} \, e^{\varepsilon s} \right).
}
\es
As a result of eq.~(\ref{soft-decoupling-xi}), the information about the interactions between hard-collinear electrons and soft photons is moved from the Lagrangian to the current. $\mathcal{L}_{\xi}^{\III\rm,LP}$ thus contains no LP interactions between the soft and the hard-collinear modes,
\ali{
\mathcal{L}_{\xi}^{\III\rm,LP} &=
\bar{\xi}_{\zz}^{\hc} \frac{\slashed{n}_{+}}{2}\bigg[ i n_{-}D^{\hc} +  i \slashed{D}_{\perp}^{\hc} \frac{1}{i n_{+}D^{\hc}} i\slashed{D}_{\perp}^{\hc} \bigg]\xi_{\zz}^{\hc}.
}
On the other hand, the current after decoupling is such that
\ali{
\label{eq:O-III-after-decoupling}
\mathcal{O}_{X}^{\III}\ar{t}  =  \, \bar{h}_N^{\sh}\ar{0} \, h_N^{\sh}\ar{0}  \left[\bar{\xi}_{\zz}^{\hc} W^{\hc}\right]\ar{t n_+}  P_X Y_{n_-}^{\s\dagger}\ar{0} Y^{\sh \dagger}_{n_-}\ar{0} \psi^{\p}\ar{0} \,.
}

In the next step, we will integrate out semi-hard photons. Technically, the potential muon field should continue to be dressed by the semi-hard Wilson line $Y^{\sh \dagger}_{n_-}$. However, since we are interested in the endpoint region of muon conversion and DIO --- where the real radiation has a much lower energy than the semi-hard scale --- that Wilson line will not contribute to the decay rate. Therefore, we will omit it from now on.

\subsection{EFT IV: \texorpdfstring{$\mu \sim \mu_{\svar}$}{mu ≃ mus}}
To construct EFT IV, we start with EFT III and integrate different modes for the various fields.
Concerning the muon, we integrate out the (off-shell) semi-hard modes while keeping the potential ones. 
In this case, the Coulomb interactions between the muon and the nucleus can no longer be treated perturbatively, and thus become part of the LP Lagrangian \cite{Beneke:1999zr}. This leads to muon-nucleus bound states appearing in the spectrum, as well as to Coulomb enhancements near the threshold. The adequate EFT to describe this spectrum region is pNRQED, with the muon described by a potential field $\Psi$.
As for the nucleus, EFT IV integrates out both the semi-hard and the potential modes of EFT III. The relevant mode is then the soft one, leading to $h_N^{\s} \sim \lambda^3 m_{\mu}^{3/2}$ and $\bar{h}_N^{\s} v \cdot D^{\s} h_N^{\s} \sim \lambda^8 m_{\mu}^{4}$.
Regarding the electron of momentum $p'$, both the hard-collinear and the soft modes are integrated out, so the collinear modes are now relevant. Following the reasoning similar to eq.~(\ref{eq:propagator-to-the-counting-EFT-III-b}), we conclude that $\xi^{\c} \sim \lambda^2 m_{\mu}^3$, and the electron mass term is no longer power suppressed. Finally, the theory also contains (non-energetic) soft electrons described by a Dirac Lagrangian, which are now massive.
Taking all of this into account, the Lagrangian of EFT IV can be written as 
\ali{
\label{eq:Lag-IV}
	\mathcal{L}^{\IV} = \mathcal{L}^{\IV\rm,LP} + \mathcal{O}(\lambda),
}
where $\mathcal{L}^{\IV\rm,LP}$ is the LP Lagrangian and is given by
\be
\mathcal{L}^{\IV\rm,LP} = 
\mathcal{L}_{A}^{\IV\rm,LP} + \mathcal{L}_{h_N}^{\IV\rm,LP} + \mathcal{L}_{\Psi}^{\IV\rm,LP} + 
\mathcal{L}_{e}^{\IV\rm,LP} +
\mathcal{L}_{\xi}^{\IV\rm,LP} \, ,
\ee
with
\bs
\ali{
\mathcal{L}_{A}^{\IV\rm,LP} &= -\dfrac{1}{4} F_{\mu\nu}^{\s} F^{\mu\nu\s}- \dfrac{1}{4} F_{\mu\nu}^{\sc} F^{\mu\nu\sc}, 
\hspace{10mm}
\mathcal{L}_{h_N}^{\IV\rm,LP} =
\bar{h}_N^{\s} i v \cdot D^{\textcolor{myblue}{(s+sc)}} h_N^{\s},
\\
\label{eq:Lag-muon-IV}
\mathcal{L}_{\Psi}^{\IV\rm,LP} &= \bar{\Psi}^{\p}\ar{x} \left( i v \cdot D^{\textcolor{myblue}{(s+sc)}}\ar{x_0} + \dfrac{\vec{\nabla}^2}{2 m_{\mu}}
\right) \Psi^{\p}\ar{x} \nonumber \\
& \hspace{25mm} + \int d^3 r \, \, \, \bar{h}_N^{\s}\ar{x} h_N^{\s}\ar{x} V\ar{\vec{r}} \bar{\Psi}^{\p}\ar{x+\vec{r}} \Psi^{\p}\ar{x+\vec{r}}, \\
\label{eq:Lag-soft-e}
\mathcal{L}_{e}^{\IV\rm,LP} &= \bar{e}^{\s} (i \slashed{D}^{\textcolor{myblue}{(s+sc)}} - m_{e}) e^{\s}, \\
\label{eq:Lag-SCET-II}
\mathcal{L}_{\xi}^{\IV\rm,LP} &=
\bar{\xi}^{\c} \frac{\slashed{n}_{+}}{2}\bigg[ i n_{-}D^{\c} 
+ \Big( i \slashed{D}_{\perp}^{\c} - m_e \Big)
\frac{1}{i n_{+}D^{\c}}\Big( i\slashed{D}_{\perp}^{\c} + m_e \Big)  \bigg]\xi^{\c},
}
\es
where $D^{\c}$ is equivalent to eq.~(\ref{eq:cov-der-definition}), but with collinear photons, and
\ali{
D_{\mu}^{\textcolor{myblue}{(s+sc)}}\ar{x} &= \partial_{\mu} + i Q e \bigg[A_{\mu}^{\s}\ar{x} + n_+ \cdot A^{\sc}\ar{x_+} \dfrac{{n_-}_{\mu}}{2} \bigg].
}
The second line in eq.~(\ref{eq:Lag-muon-IV}) is the so-called potential term. Its leading order contribution is the Coulomb potential, $V\ar{\vec{r}} = -Z \alpha/r$, with $r=|\vec{r}|$. Since we are consistently including the first QED corrections in the different steps of our framework, we must also consider the one-loop $\mathcal{O}(\alpha)$ corrections to $V\ar{\vec{r}}$. These are due to the Uehling term \cite{Uehling:1935uj}, and are such that, with on-shell subtraction and including power-suppressed terms,%
\footnote{The one-loop QED correction to Coulomb potential comes from the vacuum polarization diagram which is composed of a closed loop of massive electrons, and which corrects a soft photon that propagates between the muon and the nucleus. (In the usual pNRQCD nomenclature, the soft regions is dubbed ultra-soft, and in QCD it leads to ultra-soft corrections appearing at the third order \cite{Beneke:2007pj}.)  
This leads to eq.~(\ref{eq:inhomogeneous-potential}). However, this equation does not have homogeneous power-counting. Indeed, $r$ scales as the inverse of the muon momentum, so that $m_e r \sim \lambda \ll 1$. If we were to derive the potential strictly using EFT, we would have obtained the potential $V\ar{\vec{r}}= -\frac{Z\alpha}{r}\left(1- \frac{2}{3}\frac{\alpha}{\pi}(\ln \mu r + \gamma_E + \frac{5}{6})\right)$ and the soft matrix element correction
$\delta_s\ar{\vec{r}}= \frac{Z\alpha^2}{r} \frac{2}{3\pi}\ln \frac{m_e}{ \mu} $. 
This treatment would imply a much more complicated structure for the potential terms, as well as soft matrix elements with closed fermion loops.   
To avoid these complications, we choose here to retain power-suppressed terms in the Uehling potential, not distinguishing the soft from the potential contributions.
(This simplification finds support in the fact that, at LP, a proper EFT treatment would only affect the normalization of the muon conversion spectrum --- and not the shape, which is our focus here.)
Similar effects of the long-distance contributions induced by closed fermion loops were previously discussed in the context of the NLO corrections for the potential in the full SM at the electroweak scale \cite{Beneke:2019qaa, Beneke:2020vff, Urban:2021cdu}.}
the corrected potential reads 
\begin{align}
\label{eq:inhomogeneous-potential}
    V\ar{\vec{r}} &= 
-\frac{Z \alpha}{r}\left(1+\frac{2\alpha}{3\pi}\int_{1}^{\infty} d
xe^{-2m_{e}rx}\frac{2x^{2}+1}{2x^{4}}\sqrt{x^{2}-1}\right)+ \mathcal{O}(Z\alpha^3) \, .
\end{align}
At higher orders in $\alpha$, there are also long-distance corrections to the potential due to the electron-induced light-by-light scattering \cite{Korzinin:2018tnx}.

The soft mode is not present in $\mathcal{L}_{\xi}^{\IV\rm,LP}$, in contrast with $\mathcal{L}_{A}^{\IV\rm,LP}$, $\mathcal{L}_{h_N}^{\IV\rm,LP}$ and $\mathcal{L}_{\Psi}^{\IV\rm,LP}$.
In $\mathcal{L}_{h_N}^{\IV\rm, LP}$ and $\mathcal{L}_{\Psi}^{\IV\rm, LP}$, besides soft modes for the photons, we write explicitly soft-collinear modes as well. This is done for convenience, and will become clear when we integrate out the soft physics (note that, in temporal-spatial notation, the soft-collinear mode obeys $l^{\sc} \sim (\lambda^2, \lambda^2) \, m_{\mu}$, just like the soft mode). Finally, in $\mathcal{L}_{\Psi}^{\IV\rm,LP}$, the photons fields are evaluated at position argument $x_0$, due to multipole expansion of pNRQED.

The LP matching condition between EFTs III and IV is 
\ali{
\label{eq:matching-III-IV}
\int dt \, C_{X}^{\III}\ar{t} \, \mathcal{O}_{X}^{\III}\ar{t} = \int dt \, C_{X}^{\IV}\ar{t} \, \mathcal{O}_{X}^{\IV}\ar{t},
}
with 
\be
\mathcal{O}_{X}^{\IV} \ar{t}\equiv \bar{h}_N^{\s}\ar{0} h_N^{\s}\ar{0} \, \left[\bar{\xi}^{\c} W^{\c}\right]\ar{t n_+} P_X Y_{n_-}^{\s\dagger} \Psi^{\p}\ar{0},
\ee
where $W^{\c}$ is a collinear Wilson line (equivalent to eq.~(\ref{eq:hc-Wilson-line}), but for collinear photons).
Due to the decoupling transformation, the LP matching for the current is trivial. In fact, the soft modes cannot induce a hard-collinear scale inside the loops at LP. This situation is similar to Drell-Yan threshold factorization \cite{Beneke:2018gvs, Beneke:2019oqx}, where non-trivial matching coefficients (commonly referred to as radiative jet functions \cite{Beneke:2019oqx,Moult:2019mog,Liu:2021mac,Liu:2020ydl}) start contributing only beyond LP. Consequently, at LP we have
\begin{align}
  \left[ \bar{\xi}_{\zz}^{\hc} W^{\hc}\right]\ar{x} =   \left[\bar{\xi}^{\c} W^{\c}\right]\ar{x},
\end{align}
This has two implications. The first one is a trivial relation between the Fourier transform of the matching coefficients,
\ali{
\label{eq:CIV}
C_{X}^{\IV}( 2E_e, m_\mu;\mu_{\shvar}) = C_{X}^{\III}( 2E_e, m_\mu;\mu_{\shvar}).
}
The second one is that the running between $\mu_{\shvar}$ and $\mu_{\svar}$ is still dictated by $U_{\hvar}$, so that
\ali{
\label{eq:CIV-running}
C_{X}^{\IV}( 2E_e, m_\mu;\mu_{\svar}) = U_{\hvar}(\mu_{\hvar}, \mu_{\svar}) \, C_{X}^{\III}( 2E_e, m_\mu;\mu_{\hvar}).
}

At last, the remaining soft modes of $\mathcal{L}^{\IV\rm,LP}$ are decoupled via the field redefinitions
\bs
\label{eq:soft-decoupling}
\ali{
\bar{h}_{N}^{\s} &= \bar{h}^{\s}_{N\zz} Y_{v}^{\s\dagger} , 
&
\bar{\Psi}^{\p} &= \Psi^{\p}_{\zz} Y_{v}^{\s\dagger},
&
\bar{e}^{\s} &= \bar{e}^{\s}_{\zz} Y_{v}^{\s\dagger}, \\
h_{N}^{\s} &= \overline{Y}_{v}^{\s} h_{N\zz}^{\s},
&
\Psi^{\p} &= \overline{Y}_{v}^{\s} \Psi^{\p}_{\zz},
&
e^{\s} &= \overline{Y}_{v}^{\s} e_{\zz}^{\s},
}
\es
where the fields with index zero are the soft-decoupled versions of the corresponding fields without index, and where the soft Wilson lines follow the definition~(\ref{eq:Ys-definition}). As a result of these redefinitions, and continuing what was done at the end of the previous section, the remaining information about the interactions between soft fermions and soft photons is moved from the Lagrangian to the current. The decoupled versions of $\mathcal{L}_{h_N}^{\IV\rm,LP}$, $\mathcal{L}_{e}^{\IV\rm,LP}$ and $\mathcal{L}_{\Psi}^{\IV\rm,LP}$ read 
\bs
\ali{
\mathcal{L}_{h_N}^{\IV\rm,LP} &=
\bar{h}_{N\zz}^{\s} i v \cdot D^{\sc} \, h_{N\zz}^{\s},
\hspace{15mm}
\mathcal{L}_{e}^{\IV\rm,LP} =
\bar{e}_{\zz}^{\s} (i \slashed{D}^{\textcolor{myblue}{(sc)}} - m_{e}) e^{\s}_{\zz}, \\
\mathcal{L}_{\Psi}^{\IV\rm,LP} &= \bar{\Psi}_{\zz}^{\p}\ar{x} \left( i v \cdot D^{\sc}\ar{x_0} + \dfrac{\vec{\nabla}^2}{2 m_{\mu}} \right) \Psi_{\zz}^{\p}\ar{x} \nonumber \\
& \hspace{25mm} + \int d^3  r\, \bar{h}_{N\zz}^{\s}\ar{x} h_{N\zz}^{\s}\ar{x} V\ar{\vec{r}} \bar{\Psi}_{\zz}^{\p}\ar{x+\vec{r}} \Psi_{\zz}^{\p}\ar{x+\vec{r}},
}
\es
with $D^{\sc}$ equivalent to eq.~(\ref{eq:cov-der-definition}), but for soft-collinear photons.  We note that the soft Wilson lines are absent in the potential term.
Finally, the current is such that
\ali{
\label{eq:O-IV-after-decoupling}
\mathcal{O}_{X}^{\IV}\ar{t}
&= 
\left[\bar{h}_{N\zz}^{\s} Y_{v}^{\s\dagger}\right]\ar{0} \left[\overline{Y}_{v}^{\s} h_{N\zz}^{\s}\right]\ar{0} \left[\bar{\xi}^{\c} W^{\c}\right]\ar{t n_+} P_X Y_{n_-}^{\s\dagger} \left[\overline{Y}_{v}^{\s} \Psi_{\zz}^{\p}\right]\ar{0}.
}
For later convenience, we rewrite this operator as 
\ali{
\label{eq:to-analogy}
\mathcal{O}_{X}^{\IV}\ar{t} &= \mathcal{O}_{\svar}\ar{0} \, \mathcal{O}_{X\zz}^{\IV}\ar{t},
}
where we define the soft operator $\mathcal{O}_{\svar}$ as the product of all soft Wilson lines,
\ali{
\label{eq:soft-operator}
\mathcal{O}_{\svar}\ar{x} = \left[Y_{v}^{\s\dagger} \overline{Y}_{v}^{\s} \, Y_{n_-}^{\s\dagger} \overline{Y}_{v}^{\s} \right]\ar{x},
}
and the operator with all fields soft-decoupled as 
\ali{
\mathcal{O}_{X\zz}^{\IV}\ar{t} = \bar{h}_{N\zz}^{\s}\ar{0} \, h_{N\zz}^{\s}\ar{0} \, \left[\bar{\xi}^{\c} W^{\c}\right]\ar{t n_+} P_X \Psi_{\zz}^{\p}\ar{0}.
}
At this point, we start to see factorization, as the soft matrix element can be taken independently from the rest. To complete the factorization, we need to disentangle the soft-collinear interaction from all the other sectors of the theory. This will be done in the final step of our EFT construction, to which we now turn.

\subsection{EFT V: \texorpdfstring{$\mu \sim \mu_{\scvar}$}{mu ≃ musc}}
\label{sec:EFT-V}

The final EFT is adequate for the soft-collinear scale. 
In the case of the muon --- which is still described by potential modes --- the only LP interaction is with soft-collinear photons. This happens through the covariant derivative, and is possible due to the large component of the soft-collinear modes. Similarly, the (soft) nucleus interacts solely with soft-collinear photons. 
Regarding the electron, the collinear modes of EFT IV are integrated out, such that the soft-collinear ones are now the relevant modes. Since the scale is lower than the electron mass, the electron cannot have any large collinear fluctuations, and its large momentum component is fixed up to small soft-collinear fluctuations. The appropriate formalism for such a situation has been developed in the context of heavy quarks and is known as boosted HQET \cite{Fleming:2007qr, Fleming:2007xt, Dai:2021mxb,Beneke:2023nmj}. Finally, the non-energetic electron that has been present in the previous EFTs is now wholly integrated out.

In sum, the Lagrangian is 
\ali{
\label{eq:Lag-V}
	\mathcal{L}^{\V} = \mathcal{L}^{\V\rm,LP} + \mathcal{O}(\lambda),
}
where $\mathcal{L}^{\V\rm,LP}$ is the LP Lagrangian, given by
\be
\mathcal{L}^{\V\rm,LP} = \mathcal{L}_{A}^{\V\rm,LP} + \mathcal{L}_{h_N}^{\V\rm,LP} + \mathcal{L}_{\Psi}^{\V\rm,LP} + \mathcal{L}_{h_e}^{\V\rm,LP} \, ,
\ee
with
\bs
\ali{
\mathcal{L}_{A}^{\V\rm,LP} &= -\dfrac{1}{4} F_{\mu\nu}^{\sc} F^{\mu\nu\sc}, 
\hspace{30mm}
\mathcal{L}_{h_N}^{\V\rm,LP} =
\bar{h}_{N\zz}^{\s}  i v\cdot   D^{\textcolor{myblue}{(sc)}}\ar{x} h_{N\zz}^{\s},
\\
\mathcal{L}_{\Psi}^{\V\rm,LP} &= \bar{\Psi}_{\zz}^{\p}\ar{x} \left( i v \cdot D^{\textcolor{myblue}{(sc)}}\ar{x_0} + \dfrac{\vec{\nabla}^2}{2 m_{\mu}}
\right) \Psi^{\p}_{\zz}\ar{x} \nonumber \\
& \hspace{25mm} + \int d^3 r \, \, \, \bar{h}_N^{\s}\ar{x} h_N^{\s}\ar{x} V\ar{\vec{r}} \bar{\Psi}_{\zz}^{\p}\ar{x+\vec{r}} \Psi_{\zz}^{\p}\ar{x+\vec{r}}, \\
\mathcal{L}_{h_e}^{\V\rm,LP} &= \bar{h}_e^{\sc} \, i v_e \cdot D_e^{\sc} \, \frac{\slashed n_+}{2} h_e^{\sc}.
}
\es
The soft-collinear electron field is related to the collinear field as 
\ali{
\xi^{\c}\ar{x} = \sqrt{\frac{n_+ \cdot v_e}{2}}e^{- i m_e v_e \cdot x} h^{\sc}_e\ar{x}.
}
In $\mathcal{L}_{{h_e}}^{\V\rm,LP}$, the covariant derivative is as in eq.~(\ref{eq:cov-der-definition}), but for soft-collinear photons. On the other hand, the soft-collinear derivative acting on the soft or potential fermions contains only the large component of the photon field; this is because $\partial_\mu$ acting on the soft and potential fields scales as $p_\mu^{\s}$ and $p_\mu^{\p}$, respectively, so that only $n_+A$ is $\mathcal{O}(\lambda^2)$. Hence,
\begin{align}
i  D^{\sc}_\mu\ar{x}   = i \partial_\mu +  Q e \frac{(n_-)_{\mu}}{2}n_+\cdot A_{\sc}\ar{x_+} \;,
\end{align}
where the soft-collinear field is evaluated at the light-cone position $x_+$.
The matching between EFTs IV and V must be performed in the presence of the soft operator $\mathcal{O}_{\svar}$, since the soft modes couple to the soft-collinear ones.%
\fn{In QED, this effect is only relevant at the two-loop level due to the soft-collinear emission from a soft electron loop.}
The matching equation thus reads 
\ali{
\label{eq:matching-IV-V}
\mathcal{O}_{\svar}\ar{0} \int dt \, C_{X}^{\IV}\ar{t} \mathcal{O}_{X\zz}^{\IV}\ar{t} = \mathcal{O}_{\svar}\ar{0} \, \int dt C_{X}^{\IV}\ar{t}e^{i m_e  n_+ \cdot v_e t } C_m(m_e) \,\mathcal{O}_{X}^{\V}\ar{0},
}
where $\mathcal{O}_{X}^{\V}$ is the same as $\mathcal{O}_{X\zz}^{\IV}$, but with the collinear electron field replaced by the soft-collinear field; that is,
\be
\mathcal{O}_{X}^{\V} \equiv \bar{h}_{N\zz}^{\s} h_{N\zz}^{\s} \, \bar{h}_e^{\sc} P_X \Psi^{\p}_{\zz}.
\ee
We note that this operator is local: the large momentum component of the electron is no longer dynamic and has been relegated to the role of the label. 
The matching coefficient $C_m$, occasionally referred to in the literature as a (radiative) jet function, can be obtained by taking the on-shell matrix element of the collinear electron field.  Up to one-loop accuracy, it reads 
\be
\label{eq:Cm}
C_m(m_e;\mu_{\svar}) = 1 + \frac{\alpha}{4 \pi} \Bigg\{ 2 \ln^2\left(\frac{m_e}{\mu_{\svar}} \right) - \ln \left(\frac{m_e}{\mu_{\svar}} \right) + \frac{\pi^2}{12} + 2 \Bigg\}.
\ee
For later convenience, we define $C_{X}^{\V}$ as 
\begin{align}
\label{eq:CV}
   C_{X}^{\V}(\mu_{\svar}) =   C_{X}^{\IV}(2 m_{\mu},m_\mu;\mu_s) C_m(m_e; \mu_{\svar}).
\end{align}
The large component of the electron momentum is now fixed, and no longer corresponds to twice the electron energy. 
Beyond next-to-leading logarithmic (NLL) accuracy (see discussion in section \ref{sec:we-love-latin} below), the consistent derivation of the RG evolution between the soft and soft-collinear scales requires a soft rearrangement analogous to the one introduced in refs.~\cite{Beneke:2019slt, Beneke:2024cpq}. In that case, the resummation of logarithms induced by the insertion of the vacuum polarization with electron loops requires a rapidity RG. Since we work up to NLL accuracy (see again section \ref{sec:we-love-latin}), we leave to future work the resummation of the rapidity logarithms for $C_{X}^{\IV}(\mu_{\svar})$, and we rely on abelian exponentiation below the electron mass scale. This would still leave us the task of calculating the RG evolution between the soft and soft-collinear scales at NLL accuracy. Yet, we circumvent this by choosing $\mu=m_e$. That is, we choose the soft scale as the common scale to which all the different functions will be evolved. In that case, we do not need to calculate such RG evolution.

Similarly to the soft modes in the previous EFT, now the soft-collinear modes are decoupled from the Lagrangian via the definitions
\bs
\label{eq:sc-decoupling}
\ali{
\bar{h}_{N\zz}^{\s} &= \bar{h}^{\s}_{N\zzzz} Y_{n_+}^{\sc\dagger}, 
&
\bar{\Psi}_{\zz}^{\p} &= \bar{\Psi}^{\p}_{\zzzz} Y_{n_+}^{\sc\dagger}, 
&
\bar{h}_{e}^{\sc} &= \bar{h}^{\s}_{e\zz} Y_{v_e}^{\sc\dagger}, \\
h_{N\zz}^{\s} &= \overline{Y}_{n_+}^{\sc} h_{N\zzzz}^{\s},
&
\Psi_{\zz}^{\p} &= \overline{Y}_{n_+}^{\sc} \Psi^{\p}_{\zzzz},
&
h_{e}^{\sc} &= \overline{Y}_{v_e}^{\sc} h_{e\zz}^{\sc},
}
\es
where we are using a compact notation: soft and potential fields with one index zero are the soft-decoupled fields, soft and potential fields with two indices zero represent soft-decoupled and soft-collinear-decoupled fields, and soft-collinear fields with one index zero denote soft-collinear-decoupled fields.
The various $Y$ operators in eq.~(\ref{eq:sc-decoupling}) are soft-collinear Wilson lines, whose definition is equivalent to that of eq.~(\ref{eq:Ys-definition}), but for soft-collinear photons:
\bs
\label{eq:Ysc-definition}
\ali{
Y^{\sc\dagger}_{u} &\equiv \textrm{exp} \left( i Q e \int_0^{\infty} ds \, u \cdot \, A^{\sc}\ar{x + s \,u} \, e^{-\varepsilon s} \right), \\
\overline{Y}_{u}^{\sc} &\equiv \textrm{exp} \left( i Q \, e \int_{-\infty}^{0} ds \, u \cdot A^{\sc}\ar{x + s \, u} \, e^{\varepsilon s} \right).
}
\es
The decoupled versions of 
$\mathcal{L}_{h_N}^{\V\rm,LP}$, $\mathcal{L}_{\Psi}^{\V\rm,LP}$ and $\mathcal{L}_{{h_e}}^{\V\rm,LP}$ read 
\bs
\label{eq:Lags-EFTV-after-sc-decoupling}
\ali{
\mathcal{L}_{h_N}^{\V\rm,LP} &=
\bar{h}_{N\zzzz}^{\s} i v \cdot \partial h_{N\zzzz}^{\s},
\\
\mathcal{L}_{\Psi}^{\V\rm,LP} &= \bar{\Psi}_{\zzzz}^{\p}\ar{x} \left( i v \cdot \partial + \dfrac{\vec{\nabla}^2}{2 m_{\mu}}
\right) \Psi^{\p}_{\zzzz}\ar{x} \nonumber \\
& \hspace{25mm} + \int d^3 r \, \, \, \bar{h}_{N\zzzz}^{\s}\ar{x} h_{N\zzzz}^{\s}\ar{x} V\ar{\vec{r}} \bar{\Psi}_{\zzzz}^{\p}\ar{x+\vec{r}} \Psi_{\zzzz}^{\p}\ar{x+\vec{r}}, 
\label{eq:LpNRQED}\\
\mathcal{L}_{h_e}^{\V\rm,LP} &= \bar{h}_{e\zz}^{\sc} \, i v_e \cdot \partial \, h_{e\zz}^{\sc}.
}
\es
After the soft-collinear decoupling, the different sectors in eqs.~(\ref{eq:Lags-EFTV-after-sc-decoupling}) do not interact with each other at LP. This means that we achieved factorization for bare quantities: all possible states in the theory are formally defined with respect to the LP Lagrangian. Hence, the latter is now a sum of independent terms to all orders in perturbation theory. This also justifies our distinction between currents and Lagrangian. Indeed, since currents are always power suppressed, the amplitude for muon conversion at any order in perturbation theory contains a single insertion of the current, but allows for an unlimited number of insertions of the LP Lagrangian.

As with soft decoupling, soft-collinear decoupling also transfers the information about interactions between the fermions and soft-collinear photons from the Lagrangian to the current. In analogy with eq.~(\ref{eq:to-analogy}), we write
\ali{
\label{eq:O-V-after-decoupling}
\mathcal{O}_{X}^{\V} &= \mathcal{O}_{\scvar}\ar{0} \, \mathcal{O}_{X\zz}^{\V},
}
with the soft-collinear operator
\ali{
\mathcal{O}_{\scvar}\ar{x} = 
\left[Y^{\sc\dagger}_{n_+} \overline{Y}_{n_+}^{\sc} Y^{\sc\dagger}_{v_e} \overline{Y}_{n_+}^{\sc}\right]\ar{x}
}
and 
\ali{
\mathcal{O}_{X\zz}^{\V}
=
\bar{h}_{N\zzzz}^{\s} h_{N\zzzz}^{\s} \, \bar{h}_{e\zz}^{\sc} P_X \Psi_{\zzzz}^{\p}.	
}
It is worth emphasizing that problems with massive external fermions induce an infinite cascade of modes with increasingly lower virtuality. This potentially spoils all-order factorization of the amplitudes. However, such a cascade of modes is tamed by the measurement function. The reason is that this function constrains the real radiation, in the sense that the latter cannot resolve wavelengths longer than a certain amount. As a consequence, the Wilson lines associated with long-wavelength modes on both sides of the cut are evaluated at the same position, and thus cancel exactly after squaring the amplitude. Such a mechanism ensures that only a finite number of modes contributes to a physical observable, thus restoring the factorization for physical observables.

\section{Factorization}
\label{sec:Factorization}

\subsection{Factorization theorem}

\label{subsec:Factorization-theorem}

The EFT discussed in section \ref{sec:EFT-V} is the final step to establish a proper formalism for muon conversion and DIO. After the soft-collinear-decoupling transformation in eq.~(\ref{eq:sc-decoupling}), indeed, all the relevant modes for the two processes are completely separated at the level of the LP Lagrangian, $\mathcal{L}^{\V\rm,LP}$. As we suggested, this implies that the eigenstates of the LP Hamiltonian factorize into a product of states, each one belonging to a subspace defined by the LP Hamiltonian of a given mode.

We are, therefore, ready to take the matrix element of the effective current, which will be the essential ingredient to compute the decay rate. 
As discussed in section \ref{sec:basics}, we focus here on direct muon conversion. To calculate the matrix element, we need to specify the initial and final states, as well as the current.
The initial state contains the muonic atom in the ground state; this is a composite state involving the muon and the nucleus, $\left | \mu_H \right\rangle =\left | \mu^{\p} N^{\s} \right\rangle $.
The final state consists of a heavy nucleus $\left|N^{\s} \right \rangle$ and a single soft-collinear electron  $\left|e^{\sc} \right \rangle$, as well as arbitrary radiation $\left|\mathcal{X} \right \rangle$ consistent with the experimental cut. The radiation is composed of both soft radiation and soft-collinear radiation; the former includes soft photons and soft electrons, and we denote it by $\left|\mathcal{X}^{\s} \right \rangle$, whereas the latter is purely photonic and we denote it by $\left|\mathcal{X}^{\sc} \right \rangle$. Accordingly, we write $\left|\mathcal{X} \right\rangle  = \left|\mathcal{X}^{\s} \right\rangle \otimes \left|\mathcal{X}^{\sc} \right\rangle$.
Finally, concerning the current, we combine eq.~(\ref{eq:current-I}) with the multiple matching equations given throughout section \ref{sec:EFT} and get
\ali{
\mathcal{J}\ar{0} = -\frac{4 G_F}{\sqrt{2}}
C_{X}^{\V} \mathcal{O}_{\svar}\ar{0} \mathcal{O}_{\scvar}\ar{0} \mathcal{O}_{X\zz}^{\V}\ar{0}.
}
At last, then, the matrix element for direct muon conversion reads 
\ali{
& i \mathcal{M}_{\mu_{H} \to e N \mathcal{X}} = \Big\langle e^{\sc} N^{\s} \mathcal{X}^{\s}  \mathcal{X}^{\sc} \Big| i \mathcal{J}\ar{0} \Big| \mu_H \Big\rangle \nonumber \\
=&  -i \frac{4 G_F}{\sqrt{2}} C_{X}^{\V}
\,\,
\Big\langle \mathcal{X}^{\sc} \Big| \mathcal{O}_{\scvar}\ar{0} \Big| 0 \Big\rangle
\,\,
\Big\langle e^{\sc} \Big| \left[\bar{h}_{e\zz}^{\sc}\ar{0} \right]_\alpha \Big| 0 \Big\rangle \, \, \left\langle \mathcal{X}^{\s} \right| \mathcal{O}_{\svar}\ar{0} \Big| 0 \Big\rangle
\hspace{20mm} \nonumber \\
&  \hspace{20mm} \times \Big\langle N^{\s}\Big|\bar{h}_{N\zzzz}^{\s}\ar{0} h_{N\zzzz}^{\s}\ar{0} \, \left[ P_X \Psi_{\zzzz}^{\p}\ar{0}\right]_{\alpha}
\Big| \mu_H \Big\rangle,
}
where $\alpha$ is an index in the Dirac space.
The soft-collinear-decoupled electron is non-interacting, which implies that, to all orders,
\ali{
    \Big\langle e^{\sc} \Big| \Big[ \bar{h}_{e\zz}^{\sc}\ar{0}\Big]_{\alpha} \Big| 0 \Big\rangle &= \left[\bar{u}_{h_e} \right]_\alpha.
}
The soft nucleus fields, together with the potential muon field, yield a muon wave-function at the origin,
\ali{
\label{eq:muH-matrix-element}
\Big\langle N^{\s}\Big|\bar{h}_{N\zzzz}^{\s}\ar{0} h_{N\zzzz}^{\s}\ar{0} \, \left[ P_X \Psi_{\zzzz}^{\p}\ar{0}\right]_{\alpha}
\Big| \mu_H \Big\rangle &= \dfrac{1}{\sqrt{2 m_{\mu}}} \bar{u}_{h_N} u_{h_N} \, \left[P_X u_{\Psi}\right]_\alpha \psi_{\rm Schr.}\ar{0},
}
where $\psi_{\rm Schr.}\ar{x}$ is the $1S$ state position-space wave-function of the muon in the hydrogen-like ion, computed with the potential~(\ref{eq:inhomogeneous-potential}). The derivation of this formula is possible with the help of the spectral representation of the Coulomb Green's function, in a similar fashion to the approach used for the Sommerfeld enhancement factor \cite{Beneke:2014gja}.
Combining the three previous equations, we write the amplitude as 
\ali{
\label{eq:matrix-element-trick}
i \mathcal{M}_{\mu_{H} \to e N \mathcal{X}} &= \mathcal{N} \, \, i \mathcal{M}_{\mu N \to e N},
}
where 
\ali{
\mathcal{N} = \dfrac{\psi_{\rm Schr.}\ar{0}}{\sqrt{2 m_{\mu}}}  \, \langle \mathcal{X}^{\sc}|\mathcal{O}_{\scvar}\ar{0} \left| 0 \right\rangle \langle \mathcal{X}^{\s}|\mathcal{O}_{\svar}\ar{0} \left| 0 \right\rangle,
}
and where $i \mathcal{M}_{\mu N \to e N}$ is the LP scattering amplitude of $\mu N \to e N$, given by 
\ali{
i \mathcal{M}_{\mu N \to e N} = -i \frac{4 G_F}{\sqrt{2}} \, C_{X}^{\V} \, \bar{u}_{h_N} u_{h_N} \bar{u}_{h_e} P_X u_{\Psi}.
}
For convenience (and a more straightforward comparison with the existing literature), we keep a relativistic normalization of the states; we also use the full theory spinors, but with the momentum corresponding to the LP kinematics. 
Extending the kinematics of figure \ref{fig:kinematics} (i) to account for the final state $\mathcal{X}^{\s}  \otimes \mathcal{X}^{\sc} = \sum_i \mathcal{X}_i^{\s} \otimes \sum_j \mathcal{X}_j^{\sc}$ (such that $\mathcal{X}_i^{\s}$ has momentum $p_{\mathcal{X}_i^{\s}}$ and energy $E_{\mathcal{X}_i^{\s}}$, and equivalently for $\mathcal{X}_j^{\sc}$), the decay rate for muon conversion reads
\ali{
\label{eq:decay-width}
\Gamma_{\mu_{H} \to e N \mathcal{X}}
&= \dfrac{1}{2 M_{\mu_{H}}} \int (2 \pi)^4 \delta^{(4)}\left(p_{\mu_{H}}-p'-k'-\sum_{i}p_{\mathcal{X}i}\right) \dfrac{d^3 k'}{(2 \pi)^3 2 M_N} \dfrac{d^3 p'}{(2 \pi)^3 2 E_e} \nonumber \\
& \hspace{40mm} \times d\mathcal{P}^{\s} \, d\mathcal{P}^{\sc} \,\, |\overline{\mathcal{M}}_{\mu_{H} \to e N \mathcal{X}}|^2,
}
with
$d\mathcal{P}^{\s} = \prod_i d^{d-1} p_{\mathcal{X}_i^{\s}}\Big/\Big[(2 \pi)^{d-1} 2 E_{\mathcal{X}_i^{\s}}\Big]$
and
$d\mathcal{P}^{\sc} = \prod_j d^{d-1} p_{\mathcal{X}_j^{\sc}}\Big/\Big[(2 \pi)^{d-1} 2 E_{\mathcal{X}_j^{\sc}}\Big]$
being the phase space factors of the emitted radiation.%
\fn{
They need to be calculated in $d$ dimensions, since the real emission diagrams generate infrared (IR) divergences. These precisely cancel those coming from the virtual corrections, thus leading to a final IR-finite result. Afterwards, the remaining integrals of eq.~(\ref{eq:decay-width}) can be calculated in $d=4$ dimensions.
}
$M_{\mu_{H}}$ is the mass of the muonic atom $\mu_H$, which can be written as $M_{\mu_{H}} = M_N + m_\mu +E_b$.
Note that since we are using the relativistic state normalization for all the spinors, then a field $f$ of mass $m_f$ obeys
$
\sum_{\rm spins} \bar{u}_{f} u_{f} = 2 m_f,
$
which implies
\ali{
|\overline{\mathcal{M}}_{\mu_{H} \to e N \mathcal{X}}|^2 &= 
\dfrac{|\mathcal{N}|^2}{4} \sum_{\rm spins} |\mathcal{M}_{\mu N \to e N}|^2
=|\mathcal{N}|^2 \, 32 G_F^2 M_N^2 m_{\mu}^2 \left(\left|C_{L}^{\V}\right|^2 + \left|C_{R}^{\V}\right|^2 \right),
}
with
\ali{
|\mathcal{N}|^2 = \dfrac{|\psi_{\rm Schr.}\ar{0}|^2}{2 m_{\mu}}  \sum_{\mathcal{X}^{\sc}} \langle 0 |\mathcal{O}^\dagger_{\scvar}\ar{0} | \mathcal{X}^{\sc} \rangle  \langle \mathcal{X}^{\sc}|\mathcal{O}_{\scvar}\ar{0} \left| 0 \right\rangle \, \sum_{\mathcal{X}^{\s}} \langle 0 |\mathcal{O}^\dagger_{\svar}\ar{0} | \mathcal{X}^{\s} \rangle  \langle \mathcal{X}^{\s}|\mathcal{O}_{\svar}\ar{0} \left| 0 \right\rangle.
}
Now, using for compactness the total energy of the soft and soft collinear radiation $E_{\mathcal{X}^{\s}} \equiv \sum_i E_{\mathcal{X}_i^{\s}}$ and $E_{\mathcal{X}^{\sc}} \equiv \sum_j E_{\mathcal{X}_j^{\sc}}$, we write
\ali{
1= \int dE_{\svar} \delta(E_{\svar} - E_{\mathcal{X}^{\s}}) \int dE_{\scvar} \delta(E_{\scvar} - E_{\mathcal{X}^{\sc}})
}
to define the \textit{soft function} and the \textit{soft-collinear function} respectively as 
\bs
\label{eq:S-and-SC-functions}
\begin{align}
\label{eq:S-function}
\mathcal{S}(E_{\svar}) &\equiv \sum_{\mathcal{X}^{\s}}\int \prod_{i} \frac{d^{d-1} p_{\mathcal{X}_i^{\s}}}{(2\pi)^{d-1}2E_{\mathcal{X}_i^{\s}}} 
\delta(E_{\svar} - E_{\mathcal{X}^{\s}}) \langle 0 |\mathcal{O}^\dagger_{\svar}\ar{0} | \mathcal{X}^{\s} \rangle  \langle \mathcal{X}^{\s}|\mathcal{O}_{\svar}\ar{0} \left| 0 \right\rangle \nonumber \\
&= \int \frac{dt}{2\pi} e^{i t E_{\svar}}\langle 0 |\mathcal{O}_{\svar}\ar{t}  \mathcal{O}_{\svar}\ar{0} \left| 0 \right\rangle, \\
\label{eq:SC-function}
\mathcal{SC}(E_{\scvar}) &\equiv \sum_{\mathcal{X}^{\sc}}\int \prod_{j} \frac{d^{d-1} p_{\mathcal{X}_j^{\sc}}}{(2\pi)^{d-1}2E_{\mathcal{X}_j^{\sc}}}  \delta(E_{\scvar} - E_{\mathcal{X}^{\sc}}) \langle 0 |\mathcal{O}^\dagger_{\scvar}\ar{0} | \mathcal{X}^{\sc} \rangle  \langle \mathcal{X}^{\sc}|\mathcal{O}_{\scvar}\ar{0} \left| 0 \right\rangle \nonumber \\
&= \int \frac{dt}{2\pi} e^{i t E_{\scvar}}\langle 0 |\mathcal{O}_{\scvar}\ar{t}  \mathcal{O}_{\scvar}\ar{0} \left| 0 \right\rangle.
\end{align}
\es
These definitions allow us to write the normalized all-order differential rate for muon conversion as  
\ali{
\label{eq:NLO-rate}
\frac{1}{\Gamma_{\rm{LO}}}
\frac{d \Gamma_{\mu_{H} \to e N \mathcal{X}}}{dE_e} &= |\psi_{\rm corr}|^2 |C_{\rm corr}|^2 \int  dE_{\scvar} \, dE_{\svar} \, \delta(\Delta E - E_{\scvar} -E_{\svar})
\mathcal{S}(E_{\svar})  \mathcal{SC}(E_{\scvar}).
}
Here, $\Gamma_{\rm{LO}}$ is the LO conversion rate, given by
\begin{align}
    \Gamma_{\rm{LO}} = \dfrac{2 G_F^2 m_{\mu}^2}{\pi} |\psi_{\rm Schr.}\ar{0}|_{\rm LO}^2 \left(\left|C_{L,{\rm LO}}^{\V}\right|^2 + \left|C_{R,{\rm LO}}^{\V}\right|^2 \right),
\end{align}
where the quantities with index LO are calculated at LO in perturbation theory.
Moreover, the corrected quantities in eq.~(\ref{eq:NLO-rate}) (with index `corr') are defined with reference to the LO ones:
\begin{align}
\label{eq:extra-definitions}
|\psi_{\rm corr}|^2 \equiv \frac{|\psi_{\rm Schr.}\ar{0}|^2}{|\psi_{\rm Schr.}\ar{0}|_{\rm LO}^2},
\qquad
|C_{\rm corr}|^2 \equiv \frac{\left(\left|C_{L}^{\V}\right|^2 + \left|C_{R}^{\V}\right|^2 \right)}{\left(\left|C_{L,{\rm LO}}^{\V}\right|^2 + \left|C_{R,{\rm LO}}^{\V}\right|^2 \right)} .
\end{align}

Let us now discuss in detail eq.~(\ref{eq:NLO-rate}). 
The first factor in the right-hand side is obtained after solving the Schrödinger equation with the NLO potential~(\ref{eq:inhomogeneous-potential}).
Since we chose to calculate that potential --- and subsequently solve the Schrödinger equation --- in the on-shell renormalization scheme, we do not include RGE running effects (this has no practical consequence, as the numerical impact of those effects is negligible). We find
\begin{align}
\label{eq:psi-corr}
|\psi_{\rm corr}|^2 = 1 + \frac{\alpha}{\pi} \delta_{\rm pot},
\end{align}
such that, for aluminum, $\delta_{\rm pot} = 6.4$ \cite{Szafron:2015kja}.

The second factor in the right-hand side of eq.~(\ref{eq:NLO-rate}) is $|C_{\rm corr}|^2$, defined in eq.~(\ref{eq:extra-definitions}). It captures the virtual corrections which are codified in the subsequent matching coefficients, and which are resummed via the RGEs for those coefficients. As discussed in section \ref{sec:EFT-II}, we take $C_{X}^{\II}(\mu_{\hvar})$ as the input parameters of our analysis.
Then, combining eqs.~(\ref{eq:CV}), (\ref{eq:CIV-running}) and (\ref{eq:CIII}), we find
\ali{
	C_{X}^{\V}(\mu_{\svar}) &= C_{X}^{\II}(\mu_{\hvar}) \, \mathcal{H}(2 m_{\mu}, m_\mu;\mu_{\hvar}) \, U_{\hvar}(\mu_{\hvar},\mu_{\svar}) \, C_m(m_e,\mu_{\svar}).
}
Without $\mathcal{O}(\alpha)$ matching corrections, this equation implies $C_{X,{\rm LO}}^{\V} = C_{X,{\rm LO}}^{\II}$. Hence,
\ali{
|C_{\rm corr}|^2  = 
|\mathcal{H}(2 m_{\mu}, m_\mu;\mu_{\hvar})|^2 
\, \,
|U_{\hvar}(\mu_{\hvar},\mu_{\svar})|^2
\, \,
\left| C_m(m_e,\mu_{\svar}) \right|^2.
}
This formula condensates in an elegant way three contributions: the matching at the hard scale, the running between the hard and the soft scales, and the matching at the soft scale.

The remaining factors in the right-hand side of eq.~(\ref{eq:NLO-rate}) originate at the soft and soft-collinear scales, and are respectively encoded in the functions $\mathcal{S}$ and $\mathcal{SC}$.  
We start by recalling section \ref{sec:Kinematics} to realize that, without real photon emission, the maximum energy allowed to the electron is $E_e^{\rm max} := m_\mu + E_b$. On the other hand, a non-zero $\Delta E$ implies $E_e = E_e^{\rm max} - \Delta E$; conversely, $\Delta E = E_e^{\rm max} - E_e$, which means that a logarithm whose argument involves $\Delta E$ will introduce a non-trivial dependence on $E_e$, and thus modify the shape of the spectrum.
We assume $\Delta E \sim m_e$ for scaling purposes, but we treat $\Delta E$ as an independent parameter. We then consider the normalized cumulant distribution,
\ali{
\label{eq:cumulant-definition}
	\frac{\Gamma_{\rm{cumul}}}{\Gamma_{\rm{LO}}}
	&= \int_{E_e^{\rm max} - \Delta E}^{E_e^{\rm max}}  dE'_e\frac{1}{\Gamma_{\rm{LO}}}
\frac{d \Gamma_{\mu_{H} \to e N \mathcal{X}}}{dE'_e} \nonumber \\
&= |\psi_{\rm corr}|^2 |C_{\rm corr}|^2  \int dE_{\scvar} \, dE_{\svar} \, \theta(\Delta E - E_{\scvar} -E_{\svar})
\mathcal{S}(E_{\svar})  \mathcal{SC}(E_{\scvar}) \nonumber \\
&= |\psi_{\rm corr}|^2 |C_{\rm corr}|^2 \, \mathsf{\Sigma}.
}
We introduced $\mathsf{\Sigma}$ as the radiator factor, which we define as 
\ali{
\mathsf{\Sigma} = \mathsf{\Sigma}(\Delta E, \mu) =  \int_0^{\Delta E} d \mathcal{E} \, 
\mathcal{S}(\mathcal{E}) \, \widehat{\mathcal{SC}}(\Delta E-\mathcal{E}),
}
where the cumulant soft-collinear function is given by
\label{eq:SC-functions}
\begin{align}
\label{eq:S-function-C}
\widehat{\mathcal{SC}}(E_{\scvar}) &\equiv \sum_{\mathcal{X}^{\sc}} \int \prod_{j} \frac{d^{d-1} p_{\mathcal{X}_j^{\sc}}}{(2\pi)^{d-1}2E_{\mathcal{X}_j^{\sc}}} \theta(E_{\scvar} - E_{\mathcal{X}^{\sc}}) \langle 0 |\mathcal{O}^\dagger_{\scvar}\ar{0} | \mathcal{X}^{\sc} \rangle  \langle \mathcal{X}^{\sc}|\mathcal{O}_{\scvar}\ar{0} \left| 0 \right\rangle.
\end{align}
Now, as mentioned in section \ref{sec:EFT-V}, we choose the soft scale $\mu_{\svar}$ as the common scale, to which all objects are evolved. Then, and since the logarithmic corrections to $\mathcal{S}$ at one-loop accuracy are small at the scale $\mu_s$, there is no need to calculate the RGE for $\mathcal{S}$.%
\fn{The RGE can be obtained using standard methods, recalling the soft rearrangement necessary to separately define soft and collinear anomalous dimensions \cite{Beneke:2019slt,Beneke:2024cpq} and endpoint subtraction \cite{Beneke:2020ibj,Beneke:2022obx}.}
Using dimensional regularization with $d=4-2\epsilon$ dimensions, the bare soft function reads 
\ali{
    \label{eq:S-bare}
    \mathcal{S}_0(E_{\svar}) &= \delta(E_{\svar}) + 
\dfrac{\alpha}{E_{\svar}} \left(\dfrac{\mu}{2 E_{\svar}}\right)^{2 \epsilon} \dfrac{e^{\gamma_E  \epsilon } (\epsilon -1)
   \csc (\pi  \epsilon)}{\Gamma (2-2 \epsilon ) \Gamma (\epsilon +1)} + \mathcal{O}(\alpha^2),
}
where $\gamma_E$ is Euler's constant. Concerning $\widehat{\mathcal{SC}}(E_{\scvar})$, and as suggested in section \ref{sec:EFT-V}, we can obtain an all-order result with the help of abelian exponentiation \cite{Yennie:1961ad}. The reason is that the soft-collinear function is defined below the electron mass scale, so that no electrons can be radiatively generated. In other words, the theory at the soft-collinear scale contains only photons, implying that $\widehat{\mathcal{SC}}(E_{\scvar})$ can be directly exponentiated.%
\fn{By contrast, the soft function does not exponentiate directly, as electrons can still be radiatively generated at the soft scale.
This is reflected at the level of the Lagrangians: whereas EFT IV contains a term for a radiatively generated (non-energetic) electron, eq.~(\ref{eq:Lag-soft-e}), such a term is absent in EFT V.}
We find
\begin{align}
\label{eq:resummed-SC}
  \widehat{\mathcal{SC}}(E_{\scvar}) = \exp\Bigg[ -\frac{\alpha}{\pi} \Bigg\{ \ln^2 \Bigg(\frac{\Delta E m_e}{m_{\mu} \, \mu} \Bigg) + \ln\left( \frac{\Delta E \, m_e}{m_{\mu} \, \mu}\right) + \frac{\pi^2}{24}  \Bigg\} \Bigg].  
\end{align}
Finally, the resummed radiator factor evaluated at the scale $\mu_{\svar}$ reads  
\begin{align}
     \mathsf{\Sigma}(\Delta E, \mu_{\svar}) &= 
\exp
\Bigg\{ -\frac{\alpha}{\pi} \Bigg[ \ln^2 \Bigg(\frac{\Delta E m_e}{m_{\mu} \, \mu_{\svar}} \Bigg) + \ln\left( \frac{\Delta E \, m_e}{m_{\mu} \, \mu_{\svar}}\right) + \frac{\pi^2}{24} \Bigg]  \Bigg\}\nonumber\\
& \hspace{15mm} \times \Bigg\{1+ \frac{\alpha}{ \pi } \left[\ln^2 \left(\dfrac{2 \Delta E}{\mu_{\svar}}\right) -  \ln \left(\dfrac{2 \Delta E}{\mu_{\svar}}\right) + 1 - \dfrac{\pi^2}{8} \right] + \mathcal{O}(\alpha^2)\Bigg\} .
\end{align}
For completeness, we also provide the bare soft-collinear function,
\ali{
\mathcal{SC}_0(E_{\scvar}) &= \delta(E_{\scvar})
- \dfrac{\alpha}{E_{\scvar}} \left(\dfrac{m_{\mu} \, \mu}{E_{\scvar} m_e}\right)^{2 \epsilon} \dfrac{e^{\gamma_E \epsilon} (\epsilon -1) \csc (\pi \epsilon)}{ \Gamma (1-\epsilon)}  + \mathcal{O}(\alpha^2).
}

\subsection{Formul\ae \, for the cumulant distribution}
\label{sec:we-love-latin}

We can now derive compact formul\ae \, for the cumulant distribution.
Since we are normalizing the cumulant distribution to the LO rate, the LO result of eq.~(\ref{eq:cumulant-definition}) is trivial,
\ali{
\label{eq:trivial}
\frac{\Gamma_{\rm{cumul}}}{\Gamma_{\rm{LO}}} \bigg|_{\mathrm{LO}} = 1.
}
Let us then discuss the NLO rate. 
We start by considering the result which takes a fixed-order (FO) in the coupling constant $\alpha$. This result refrains from using RGE running, and sets all the different scales $\mu_{\hvar}$, $\mu_{\scvar}$, etc. to be equal to a generic $\mu$. We find%
\fn{All the four functions of eq.~(\ref{eq:FO-part-2}) are renormalized. 
We checked the IR-finiteness of the result; i.e., 
the IR divergences of the bare version of $|\mathcal{H}(2 m_{\mu}, m_\mu;\mu)|^2_{\rm{NLO}} \times \left|C_m(m_e,\mu)\right|^2_{\rm{NLO}}$ cancel with those of the bare version of $\left[  \int_0^{\Delta E} \mathcal{S}(\mathcal{E}) \right]_{\rm{NLO}} \times \left[ \,  \widehat{\mathcal{SC}}(\Delta E)\right]_{\rm{NLO}}$.}
\ali{
\label{eq:FO-part-1}
\frac{\Gamma_{\rm{cumul}}}{\Gamma_{\rm{LO}}} \bigg|_{\mathrm{FO}} &=
|\psi_{\rm corr}|^2_{\rm{NLO}}
\, \times \,
|\mathcal{H}(2 m_{\mu}, m_\mu;\mu)|^2_{\rm{NLO}} 
\, \times \,
\left| C_m(m_e,\mu) \right|^2_{\rm{NLO}} \nonumber \\
& \hspace{15mm} \times \, \left[ \int_0^{\Delta E} d\mathcal{E} \, \mathcal{S}(\mathcal{E}) \right]_{\rm{NLO}} \times \left[ \,  \widehat{\mathcal{SC}}(\Delta E)\right]_{\rm{NLO}},
}
where we consistently drop terms beyond the NLO accuracy when multiplying out the terms. In this equation, 
$|\psi_{\rm corr}|^2_{\rm{NLO}}$ is given in eq.~(\ref{eq:psi-corr}), and the different functions read 
\bs
\label{eq:FO-part-2}
\ali{
|\mathcal{H}(2 m_{\mu}, m_\mu;\mu)|^2_{\rm{NLO}} &=
1 -\dfrac{\alpha}{2 \pi }  \Bigg\{\ln^2\left(\frac{m_{\mu}}{\mu}\right) + \ln^2\left(\frac{4m_{\mu}}{\mu}\right) + \ln \left( \frac{4m_{\mu}}{\mu} \right) \nonumber \\
& \hspace{50mm} - \frac{\pi^2}{12} - 2 \ln^2(2)\Bigg\} + \mathcal{O}(\alpha^2), \\[2mm]
\left|C_m(m_e,\mu)\right|^2_{\rm{NLO}}
&= 1 + \frac{\alpha}{2 \pi} \Bigg\{ 2 \ln^2\left(\frac{m_e}{\mu} \right) - \ln \left(\frac{m_e}{\mu} \right) + \frac{\pi^2}{12} + 2 \Bigg\} + \mathcal{O}(\alpha^2), \\[2mm]
\left[ \int_0^{\Delta E} \hspace{-2mm} d\mathcal{E}  \,\mathcal{S}(\mathcal{E}) \right]_{\rm{NLO}} &= 1+\frac{\alpha}{ \pi } \Bigg\{  \ln^2 \left(\dfrac{2 \Delta E}{\mu}\right) -  \ln \left(\dfrac{2 \Delta E}{\mu}\right) + 1 - \dfrac{\pi^2}{8}\Bigg\} + \mathcal{O}(\alpha^2), \\[2mm]
\left[ \,  \widehat{\mathcal{SC}}(\Delta E)\right]_{\rm{NLO}} &= 1 -\frac{\alpha}{\pi} \Bigg\{ \ln^2 \Bigg(\frac{\Delta E m_e}{m_{\mu} \, \mu} \Bigg) + \ln\left( \frac{\Delta E \, m_e}{m_{\mu} \, \mu}\right) + \frac{\pi^2}{24}  \Bigg\} + \mathcal{O}(\alpha^2).
}
\es
By not exploiting the advantages of the EFT approach --- namely, scale separation and RG running --- the FO result of eqs.~(\ref{eq:FO-part-1}) and (\ref{eq:FO-part-2}) is simply the full theory result. As expected, it is plagued by large logarithms of different types, all of which spoil the convergence of perturbation theory. Finally, it is clear that the last two functions, which depend on $\Delta E$, will be those responsible for a non-trivial shape of the rate.

Instead of the FO result, we should consider the resummed one. This is characterized by two main features: the use of canonical scales and the inclusion of RG evolutions. The first feature sets $\mu_{\hvar}=2 m_{\mu}, \mu_{\svar}=m_e$ and $\mu_{\scvar}=\Delta E \, m_e/m_{\mu}$, which implies the disappearance of large logarithms in the different functions of eq.~(\ref{eq:FO-part-2}). The second feature resums those logarithms to all orders in perturbation theory. As discussed before, we choose $\mu_{\svar}=m_e$ as the common scale (to which all others evolve), so that only the running of the hard and soft-collinear functions is necessary. Neglecting the running of $\alpha$ between $\mu_{\hvar}$ and $\mu_{\svar}$, we find
\ali{
\label{eq:resummed}
\hspace{-0.8mm} \frac{\Gamma_{\rm{cumul}}}{\Gamma_{\rm{LO}}} \bigg|_{\mathrm{Resu.}} &= \Bigg\{1 + \frac{\alpha}{\pi} \left[\delta_{\rm pot} + \ln^2\left(\frac{m_e}{2
\Delta E}\right) + \ln \left(\frac{m_e}{2\Delta E}\right) - \frac{\ln(2)}{2} - \frac{\pi^2}{12} + 2\right]\Bigg\} \, \times  \nonumber \\
& \hspace{-20mm} {\mathrm{exp}}\Bigg\{\frac{\alpha}{2\pi}
\left[
\left( \dfrac{10 \alpha}{9 \pi} - 2\right)\ln ^2\left(\frac{2 m_{\mu}}{m_e}\right) +5 \ln \left(\frac{2 m_{\mu}}{m_e}\right) -2\ln^2\left(\frac{m_{\mu}}{\Delta E}\right)+2
   \ln \left(\frac{m_{\mu}}{\Delta E}\right)
\right]
\Bigg\}.
}
On the right-hand side, the first line describes the remains of the FO result after the use of the canonical scales, whereas the second line shows the all-order RG evolutions. Note that the large logarithms are only contained in the second line, and are thus resummed to all orders in $\alpha$ at NLL accuracy. The logarithms in the first line are small (recall $\Delta E \sim m_e$), so that perturbation theory is safe. 
The (leading) double logarithmic terms in the second line can be also computed using a collinear factorization approach \cite{Szafron:2016cbv, Szafron:2017guu}; however, the latter is not valid in general, as it fails to systematically account for bound state physics in the initial state and Coulomb interactions in the final state.

Given that eq.~(\ref{eq:resummed}) involves all orders in $\alpha$, it is convenient to investigate different approximations. To that end, we need to determine how large the logarithms of the second line are when compared to the (small) coupling constant $\alpha$. This discussion on the logarithmic counting might differ in the QED and QCD literature, given the difference of magnitude between the coupling constants $\alpha$ and $\alpha_s$. To understand this aspect, let $L$ designate a generic large logarithm (in the second line of eq.~(\ref{eq:resummed}), $L \sim \ln(m_{\mu}/\Delta E) \sim \ln(m_{\mu}/m_e)$). Then, in QCD the usual assumption is that $\alpha_s \times L \sim 1$. In QED, it is sometimes argued that it suffices to take $\alpha \times L^2 \sim 1$ \cite{Frixione:2022ofv}. This approach enables considering only the leading (double) logarithmic terms in the exponent, and thus treating the remaining terms in a perturbative fashion.
Here, since we are dealing with an observable which intrinsically involves double logarithms, we take the usual QCD convention and apply it to the QED coupling constant $\alpha$. 
We therefore assume the following counting:
\begin{align}
\label{eq:log-counting}
\begin{array}{
c@{\hskip 1em}
c@{\hskip 1em}
c@{\hskip 1em}
c@{\hskip 1em}
c@{\hskip 1em}
c@{\hskip 1em}
c@{\hskip 1em}
c}
d\Gamma = \sum\limits_{n} &
\alpha^n L^{2n} &+ & \alpha^n L^{2n-1} &+ &\alpha^n L^{2n-2} &+& \dotsb \\[-6mm]
& \underbrace{\hspace{1.1cm}}_{\text{LL/DL}} & & \underbrace{\hspace{1.5cm}}_{\text{NLL/NDL}} & & \underbrace{\hspace{1.5cm}}_{\text{NNLL/NNDL}} & & 
\end{array}
\end{align}
In this expression, LL refers to the leading-logarithmic approximation, NLL to the next-to-leading-logarithmic approximation, and so on. On the other hand, DL refers to double-logarithmic approximation, NDL to the next-to-double-logarithmic approximation, and so on. The difference between the LL series (LL, NLL, NNLL, etc.) and the DL series (DL, NDL, NNDL, etc.) is that the former includes the running of the coupling constant, while the latter is obtained for a fixed value of $\alpha$. 

According to the counting of eq.~(\ref{eq:log-counting}), the first approximation (LL/DL) involves only the one-loop cusp anomalous dimension of the different RG evolutions. NLL/NDL accuracy requires, in addition, the two-loop cusp and one-loop non-cusp anomalous dimensions. In this paper, we have all the elements to consider the latter accuracy: on the one hand, and as has been discussed, only the running of the hard and soft-collinear functions is needed; on the other hand, the RG evolution of the hard function was given at the two-loop cusp and one-loop non-cusp anomalous dimensions in eq.~(\ref{eq:cusp-and-non-cusp}), whereas the evolution of the soft-collinear function is given in eq.~(\ref{eq:resummed-SC}) to all orders.
As discussed in section \ref{sec:EFT-V}, a complete analysis beyond NLL requires rapidity RG, which we do not consider here. Instead, and as usual in the literature, we consider results at NLL' accuracy; in addition to the NLL results, this include the remains of the FO results after the use of the canonical scales.

Before we present our numerical results, let us briefly comment on how to incorporate finite-size corrections to the nucleus.

\subsection{Finite nucleus size effects}
\label{subsec:FSZ}

Corrections to the point-like assumption for the nucleus are crucial in both the muon conversion and the muon DIO decay: not only for correctly understanding the phenomenology, but also for a proper calculation of the rates \cite{Cirigliano:2009bz,Heeck:2022wer, Heeck:2021adh, Borrel:2024ylg}.
In the EFT language, a nucleus with finite size implies that, in the current $\mathcal{J}$, the nucleus fields become spatially non-local. This approach would require assumptions on how the nucleus radius scales with $\lambda$. We leave this approach for future work. 

Rather than a rigorous EFT construction, therefore, we adopt here a more phenomenological treatment.
Accordingly, and as mentioned in section \ref{sec:nucleus}, we describe finite-size corrections with form factors. At leading recoil, it is sufficient to introduce one form factor $\rho\ar{\vec{r}}$, describing the nucleus charge density. This leads to two modifications. The first one is that the Coulomb potential, instead of eq.~(\ref{eq:inhomogeneous-potential}), now reads 
\begin{align}
\label{eq:corrected-V}
V\ar{\vec{r}} &= -\int d^3 r' \rho \ar{\vec{r}'}
\frac{Z \alpha}{|\vec{r}-\vec{r}'|}\left(1+\frac{2\alpha}{3\pi}\int_{1}^{\infty} d
xe^{-2m_{e}|\vec{r}-\vec{r}'|x}\frac{2x^{2}+1}{2x^{4}}\sqrt{x^{2}-1}\right)\,,
\end{align}
and it can be interpreted as a convolution of the point-like potential with the charge density. 
The second modification concerns the potential matrix element, which is now such that
\begin{align}
\label{eq:LO}
    \langle N^{\s} | \bar{h}_{N\zzzz}^{\s} h_{N\zzzz}^{\s} \,  P_X \Psi_{\zzzz}^{\p} | \mu_H \rangle &= \dfrac{1}{\sqrt{2 m_{\mu}}} \left[\bar{u}_{h_N} u_{h_N}\right] \left[P_X u_{\Psi}\right]_\alpha \int d^3 x e^{i m_e \vec{v}_e\vec{x}} \rho\ar{\vec{x}}\psi_{\rm Schr.}\ar{\vec{x}} \,. 
\end{align}
In the point-like limit, $\rho\ar{\vec{x}} = \delta^{(3)}\ar{x}$, we recover eq.~(\ref{eq:muH-matrix-element}). Eq.~(\ref{eq:LO}) agrees with the LO expression for the muon conversion \cite{Feinberg:1959ui}. It is clear that, as mentioned in the Introduction, the finite nuclear size effects do not modify the shape of the spectrum, but only affect the total normalization.
Beyond LP, the dipole soft interactions become relevant  and will introduce corrections that depend on the non-trivial nuclear structure. A detailed EFT analysis of the interplay between the effects of a finite nuclear size and QED corrections is left for future work.

\section{Numerical results}
\label{sec:Numerical-results}

We now turn to the numerical results. A proper phenomenological analysis involves a study of scale dependencies, effects of a finite nuclear size and the impact of the background (given by muon DIO near its endpoint). This is beyond the scope of our work and is left for the future. Here, and as suggested before, we restrict ourselves to a preliminary investigation, focused on the impact of the NLO QED corrections to the signal of muon conversion.
More specifically, we are interested in exploring the different formul\ae \,discussed in section \ref{sec:we-love-latin}. They depend on $\Delta E$ --- which we treat, as referred in section \ref{subsec:Factorization-theorem}, as an independent parameter. On the other hand, the EFT framework developed here assumes $\Delta E \sim m_e$. In the present section, then, we investigate the numerical values of the formul\ae \, of section \ref{sec:we-love-latin} for values of $\Delta E$ close to $m_e$.

We start by considering the NLO FO result. In this case, we assume a renormalization scale equal to the muon mass, $\mu = m_{\mu}$. We show in table \ref{table:numbers} results for different values of $\Delta E$. We decompose the total result according to its different contributions, as described in eq.~(\ref{eq:FO-part-1}).
\begin{table}[h!]
\begin{center}
\begin{tabular}{lrrrr}
\hlinewd{1.1pt} \\[-5mm]
$\Delta E \, ({\rm MeV})$ & \multicolumn{1}{c}{0.25} & \multicolumn{1}{c}{0.5} & \multicolumn{1}{c}{1} & \multicolumn{1}{c}{2} \\ \midrule \midrule
$|\psi_{\rm corr}|^2_{\rm{NLO}} - 1$ &1.49\% &1.49\% &1.49\% &1.49\% \\[1mm]
$|\mathcal{H}(2 m_{\mu}, m_\mu;\mu_{\hvar})|^2_{\rm{NLO}} - 1$ &$-0.18\%$ &$-0.18\%$ &$-0.18\%$ &$-0.18\%$ \\[1mm]
%
$\left|C_m(m_e,\mu_{\hvar})\right|^2_{\rm{NLO}} - 1$ &7.54\% &7.54\% &7.54\% &7.54\% \\[1mm]
$\left[ \int_0^{\Delta E} d\mathcal{E} \, \mathcal{S}(\mathcal{E}) \right]_{\rm{NLO}} - 1$ & $7.83\%$ & $6.06\%$ & $4.51\%$ & $3.19\%$ \\[1.5mm]
$\widehat{\mathcal{SC}}(\Delta E)_{\rm{NLO}} - 1$ & $-27.47\%$ & $-24.08\%$ & $-20.92\%$ & $-17.97\%$ \\[1mm] \midrule
%
$\left(\Gamma_{\rm{cumul}}/\Gamma_{\rm{LO}}\right) \big|_{\mathrm{FO}} - 1$ & $-10.79\%$ & $-9.18\%$ & $-7.56\%$ & $-5.94\%$ \\[2mm]
\hlinewd{1.1pt}
\end{tabular}
\caption{Numerical results for the NLO FO approach, assuming $\mu = m_{\mu}$. For different values of $\Delta E$ (line 1), lines 2 to 5 describe the different functions that compose the result (according to eq.~(\ref{eq:FO-part-1})), while line 6 shows the total result.}
\label{table:numbers}
\end{center}
\end{table}
We see that, for the values considered, the total NLO FO result corresponds to negative corrections, which means that the decay rate is decreased from its LO value. For the specific case of $\Delta E = m_e$, we find a correction of around $-9\%$. While the three first functions of the right-hand side of eq.~(\ref{eq:FO-part-1}) obviously do not depend on $\Delta E$ (recall eq.~(\ref{eq:FO-part-2})), the last two depend significantly on it. This happens in such a way that total FO result eventually becomes singular in the limit $\Delta E \to 0$. 

\begin{figure}[h!]
\centering
\includegraphics[width=0.68\textwidth]{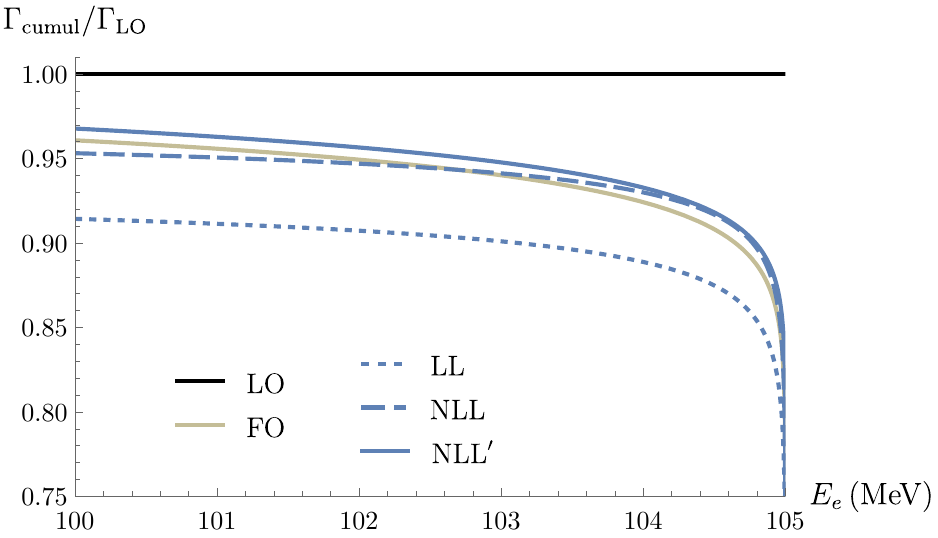}
\caption{Cumulant distribution (normalized to the LO result) against $E_e$, for different approaches. See text for details.}
\label{fig:LO-vs-NLO}
\end{figure}

Let us now compare the different approaches discussed in \ref{sec:we-love-latin}. Figure \ref{fig:LO-vs-NLO} shows the normalized cumulant distribution for different results: the LO, the total NLO FO, and three resummed results --- LL, NLL and NLL'.
As seen in eq.~(\ref{eq:trivial}), the LO result is just a horizontal line.
According to table \ref{table:numbers}, this trivial shape would also be found in the FO and resummed approaches if we ignored the real photon radiation. In other words, the QED virtual corrections only modify the normalization of the rate.  By contrast, the real radiation introduces the non-trivial dependence on the electron energy in both the FO and resummed approaches observed in figure \ref{fig:LO-vs-NLO}.
Concerning the FO result, and recalling $E_e = E_e^{\rm max} - \Delta E$, we recognize the results shown in table~\ref{table:numbers}. As for the resummed results, there is a significant change between LL and NLL, but only a minor one between NLL and NLL' --- as expected from a perturbative LL series. 

Given the small difference between the NLL' and the FO curves, one might wonder if the EFT approach is really necessary. After all, how can we justify the derivation of the intricate sequence of EFTs described in section \ref{sec:EFT}, if its phenomenological impact seems so small? The answer is multifold.
The first and main aspect is that the EFT derived in this paper is a systematically improvable framework; even though the one-loop result might be similar to the fixed order result, only through such framework is a consistent and systematic description of higher orders possible. 
Secondly, the EFT approach allows the introduction of objects like the Schrödinger wave-function in a proper Quantum Field Theory language.
More than this, the EFT is the framework that validates the FO LP result; for example, given that muon conversion involves bound states, it is not self-evident that one can calculate low-energy real radiation by simply resorting to the expressions for non-bound states \cite{Denner:2019vbn}.
The deviations from the simple generalization of the formulae derived for the scattering states will be apparent for subleading power corrections (e.g. in $Z\alpha$ suppressed terms) and in different kinematical regions of the spectrum.  
It is also the EFT framework that allows a proper and general resummation of large logarithms, as discussed after eq.~(\ref{eq:resummed}).
Finally, as mentioned before, figure \ref{fig:LO-vs-NLO} results from a preliminary investigation; an adequate investigation of scale dependencies is in order, and is left for future work.

\begin{figure}[h!]
\centering
\includegraphics[width=0.68\textwidth]{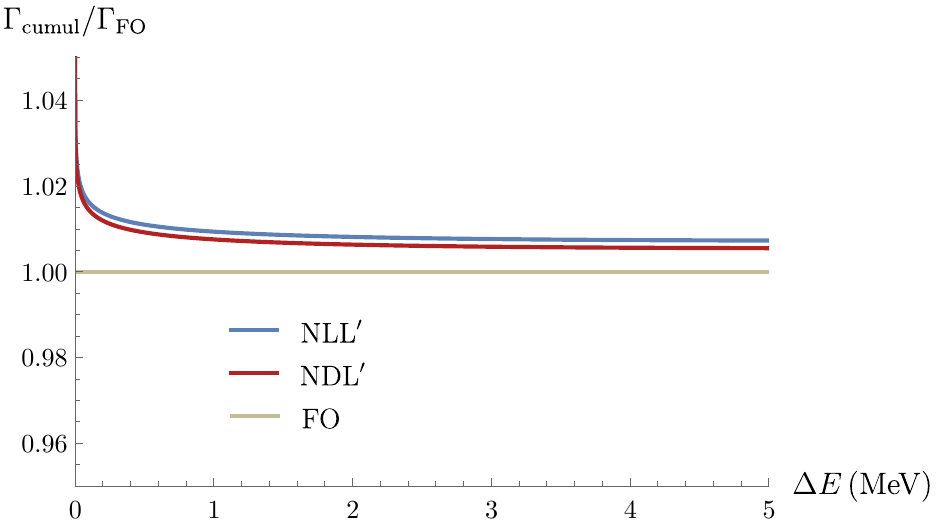}
\caption{Cumulant distribution (normalized to the FO result) against $\Delta E$, for different approaches. See the text for details.}
\label{fig:LO-vs-NLO-DeltaE}
\end{figure}

Here, we briefly ascertain the impact of the running of $\alpha$. We note that since the running of $\mathcal{S}$ and $C_m$ is not needed, and since $\alpha$ does not run below the soft scale, a non-zero $\beta(\alpha)$ only affects the evolution factor $U_{\hvar}$. Moreover, since this factor is evolved in not so large a range (between the hard and the soft scales), we expect a small impact of the running of $\alpha$. This is indeed what we find, as can be seen in figure \ref{fig:LO-vs-NLO-DeltaE}. Here, the vertical axis still contains the cumulant distribution, but now normalized in regards to the NLO FO result (so that the latter is now a horizontal line). The horizontal axis shows $\Delta E$ instead of $E_e$. For $\Delta E=m_e$, the running of $\alpha$ corresponds to a $0.18\%$ effect, such that the NLL' result is $1.09\%$ above the FO one.

\section{Conclusions}
\label{sec:Conclusions}

Muon conversion, the process of a muon decaying within an atomic nucleus into an energetic electron, is one of the most important ways to search for charged lepton flavor violation. The limit on the conversion rate is expected to be improved by four orders of magnitude in upcoming experiments. It is therefore urgent to have precise theoretical predictions. As it turns out, the process represents a fascinating theoretical challenge. While the problem is, in principle, perturbative --- allowing calculations from first principles --- the simultaneous presence of bound states and energetic particles introduces multiple scales that spoil perturbation theory. The presence of masses regularizing collinear physics further complicates the problem, distinguishing it from traditional applications of SCET to QCD. The calculation of QED corrections to muon conversion thus requires a sophisticated theoretical framework.

In this paper, we provided such a framework. Our approach combined various modern EFTs to achieve a systematic power expansion. We derived a factorization theorem that effectively separates the different scales, allowing for a resummation of large logarithms. This has led to a significant improvement in the accuracy of the theoretical predictions for the muon conversion rate. Our numerical results demonstrate the importance of including higher-order QED corrections, whose absolute value can reach 9\%. 
We suggest extending the phenomenological line of investigation, by considering complete BSM models and matching them to our EFT. In this way, one may understand precisely how QED corrections not only alter the interpretation of BSM physics, but also affect the limits on short-distance parameters obtained in the previous and upcoming muon conversion searches. 

Our work provides a solid foundation for further theoretical investigations. The same framework can be applied to the endpoint region of the spectrum of muon decay-in-orbit, which constitutes the only irreducible background in muon conversion searches. A unified treatment of these processes will enable an improved control over QED effects, thus expanding the discovery potential for the experiments.  
The EFT computations presented here can also be extended to include both NLP corrections and other relevant effects (such as nuclear structure and hadronic contributions). This will pave the way for a more precise comparison between theory and experiment in the upcoming searches for charged lepton flavor violation.

This work also provides insights into the development of the EFT framework itself. Problems with relativistic massive particles are notoriously difficult to handle. We plan to improve the logarithmic accuracy of our resummation, which requires not only an extension of the RGE to include rapidity renormalization group \cite{Chiu:2012ir}, but also the computation of certain two-loop diagrams.

\acknowledgments

All Feynman diagrams were drawn with Feyngame \cite{Harlander:2020cyh, Harlander:2024qbn}.
We thank Marvin Schnubel for helpful discussions and comments, and Matthias Neubert for pointing out a typo in our definition of the soft-collinear function. 
D.F. thanks Miko{\l}aj Misiak for discussions, as well as the Mainz Institute for Theoretical Physics (MITP) of the Cluster of Excellence PRISMA+ (Project ID 39083149) for its hospitality and support. RS thanks Martin Beneke for helpful discussions and the Munich Institute for Astro-, Particle and BioPhysics (MIAPbP) which is funded by the Deutsche Forschungsgemeinschaft (DFG, German Research Foundation) under Germany´s Excellence Strategy – EXC-2094 – 390783311, for its hospitality. 
This work was supported by the U.S. Department of Energy under Grant Contract No. DE-SC0012704.

\appendix

\addtocontents{toc}{\protect\setcounter{tocdepth}{2}}

\section{Details on the analysis of the momentum regions}
\label{app:regions}

Section \ref{sec:Framework} above presents the results of an analysis of the relevant momentum regions for electromagnetic corrections to  muon conversion. In this appendix, we present several details of that analysis. We start by discussing the virtual corrections (whose Feynman diagrams are depicted in figure \ref{fig:virtual-diags-full}) and, after that, the real ones (whose Feynman diagrams are depicted in figure \ref{fig:real-diags-full}).

The analysis by regions starts from the presupposition that, if one takes a particular loop integral and expands its result according to some expansion parameter, the expanded result can be decomposed into different contributions. Each contribution is obtained by imposing a certain scaling for the loop momentum before performing the integral. That scaling is called a momentum region, and the analysis by region aims to identify the non-vanishing regions. In other words, the analysis by regions seeks to pinpoint the different possibilities of scaling of the loop momentum that yield a non-vanishing contribution to the expanded result of the integral. 

In what follows, then, we consider two steps for the calculation of each diagram. In the first one, we calculate the expanded amplitude in the full theory with Package-X \cite{Patel:2015tea, Patel:2016fam}. 
More specifically, we use this software to calculate the integral analytically, and \textit{after that} we perform the $\lambda$ expansion --- or the $\lambda_R$ one, in the case of diagram (d) --- defined in section \ref{sec:Kinematics}. In the second step, we assume a particular scaling for the loop momentum \textit{before} performing the integration (see table \ref{table:modes}), and look for the scalings that lead to non-zero results. We resort to HypExp \cite{Huber:2005yg,Huber:2007dx} to expand hypergeometric functions.

We simplify the analysis in several ways. First, we consider only the term of the amplitudes with no powers of loop momentum in the numerator. Accordingly, for a generic diagram $j$, we separate the amplitude into the terms with 0, 1, and 2 powers of loop momentum in the numerator,
\be
\label{eq:generic}
i \mathcal{M}_{(j)} = i \mathcal{M}_{(j,l^0)} + i \mathcal{M}_{(j,l^1)} + i \mathcal{M}_{(j,l^2)},
\ee
and consider only $i \mathcal{M}_{(j,l^0)}$ (which suffices to identify the regions). Second, we assume all particles are free and we describe the amplitudes using spinors for the external particles. Finally, we restrict ourselves to the scalar operator. None of these simplifications affect the identification of the regions. We also define
\ali{
\Gamma_{S} \equiv C_{SX} \, \bar{u}_e(p') P_X u(p) \, \bar{u}_N(k') u_N(k).
}
We use dimensional regularization in our calculations, with $d = 4 - 2 \epsilon$, and always omit terms of $\mathcal{O}(\epsilon)$. The electromagnetic coupling $e= \sqrt{4 \pi \alpha}$ is rendered dimensionless via the usual replacement $e \to e \mu^{\epsilon}$, with $\mu$ being the renormalization scale. We use modified minimal subtraction ($\rm \overline{MS}$) as a subtraction scheme. Finally, to solve the integrals, we resort to the light-cone decomposition
\bs
\ali{
\label{eq:integral-tricks-3}
\int d^d l &= \frac{1}{2} \int d n_+l \int d^{d-2} l_{\perp} \int dn_-l, \\
\label{eq:integral-tricks-4}
\int d^{d-2} l_{\perp} &= \frac{2 \pi^{\frac{d-2}{2}}}{\Gamma(\frac{d-2}{2})} \int_0^{\infty} l_{T}^{d-3} \, \,  dl_{T},
}
\es
where $l_T$ is the magnitude of the transverse spatial momentum, such that
\ali{
\label{eq:pert-to-transv}
l_T^2 = -l_{\perp}^2.
}

\subsection{Virtual corrections}

We give special attention to diagrams (a) and (b) of figure \ref{fig:virtual-diags-full} to illustrate several technical details. The remaining diagrams are discussed in a more abbreviated fashion. In all the diagrams, we use the approximation $E_e = m_{\mu} + \mathcal{O}(\lambda)$, which also does not affect the identification of the regions. 

\subsubsection{Diagram (a)}

The amplitude is 
\be
\label{eq:diag-a-basic}
i \mathcal{M}_{(a,l^0)} = \mu^{2 \epsilon} \int \frac{d^dl}{(2 \pi)^d} (-i e) (i Z e) \, i V_{(a,l^0)} \, \frac{i}{(p-l)^2-m_{\mu}^2+i 0} \frac{i}{(k+l)^2-M_{N}^2+i 0} \frac{i}{l^2+i 0},
\ee
with
\be
V_{(a,l^0)} = - C_{SX} \bar{u}_e(p') P_X (\slashed{p} + m_{\mu}) \gamma_{\nu} u_{\mu}(p) \, \bar{u}_N(k') (\slashed{k} + M_N) \gamma_{\nu} u_N(k).
\ee
After solving the integral analytically, we use $k \cdot p = M_N \sqrt{m_{\mu}^2 + |\vec{p}|^2}$ and expand in $\lambda \sim |\vec{p}|/m_{\mu} \sim Z \alpha$. At the LP, the result is 
\be
i \mathcal{M}_{(a,l^0)}^{\rm{(LP)}} = \frac{Z \alpha}{2} \, \frac{m_{\mu}}{|\vec{p}|} \, \Gamma_{S} \left[ \frac{1}{\epsilon} + \ln\left(\frac{\mu^2}{4 |\vec{p}|^2}\right) + i \pi\right].
\label{eq:diag-a-expanded}
\ee
Note that this result is not suppressed --- i.e., it is of $\mathcal{O}(\lambda^0)$ --- since $|\vec{p}| \simeq m_{\mu} Z \alpha$.
For this reason, we consider also the next-to-LP (NLP) result, $\mathcal{O}(\lambda^1)$. We find
\ali{
\label{eq:MaL0}
i \mathcal{M}_{(a,l^0)}^{\rm{(NLP)}} =
i \frac{Z \alpha}{2 \pi} \, \, \Gamma_{S} \bigg[ \frac{1}{\epsilon} + \ln \left(\frac{\mu^2}{m_{\mu}^2}\right) -2 \bigg].
}
We now perform an analysis of the regions. We consider the $\mathcal{O}(\lambda^0)$ and the $\mathcal{O}(\lambda^1)$ terms separately. It turns out that the potential region gives the former, while the latter is given by the hard region. In what follows, we prove that this is the case.

\smallsection{LP: potential region}
Before solving the integral, we assume $l \sim l^{\p}$. For convenience, we separate the integral into temporal and spatial parts. The amplitude then reads 
\be
i \mathcal{M}_{(a,l^0)}^{({\rm LP},p)} = - Z e^2 \mu^{2 \epsilon} V_{(a,l^0)} \int \frac{d^{d-1}\vec{l}}{(2 \pi)^{d-1}} \frac{1}{\vec{l}^2-i 0} \int \frac{d l_0}{(2 \pi)} \frac{1}{-2 p_0 l_0 + 2 \vec{p}\cdot\vec{l} - \vec{l}^2 + i0} \frac{1}{2 M l_0 + i 0}.
\ee
The temporal part can be solved using residue theorem, leading to
\be
\label{eq:diag-a-pot-part}
i \mathcal{M}_{(a,l^0)}^{({\rm LP},p)} = 2 i Z e^2 m_{\mu} \Gamma_{S} I_l,
\quad
\text{with}
\quad
I_l = \mu^{2 \epsilon} \int \frac{d^{d-1} \vec{l}}{(2 \pi)^{d-1}} \frac{1}{\vec{l}^2 - i0} \frac{1}{(\vec{l}-\vec{p})^2 - \vec{p}^2 - i0}.
\ee
Introducing a Feynman parameter, we find
\be
I_l= - \frac{i}{16 \pi} \frac{1}{|\vec{p}|} \left( \frac{1}{\epsilon} + \ln \frac{\mu^2}{4|\vec{p}|^2} + i \pi\right),
\ee
so that
\be
i \mathcal{M}_{(a,l^0)}^{({\rm LP},p)} = i \mathcal{M}_{(a,l^0)}^{\rm{(LP)}}.
\label{eq:diag-a-pot-final}
\ee

\smallsection{NLP: hard region}
Here, we assume $l \sim l^{\h}$. We resort to light-cone coordinates and expand each of the denominators of the propagators of eq.~(\ref{eq:diag-a-basic}) to first order in $\lambda$.
Therefore, using eq.~(\ref{eq:integral-tricks-3}), we find
%
\bea
i \mathcal{M}_{(a,l^0)}^{({\rm NLP},h)} &=& \frac{Z e^2 V_{(a,l^0)}}{2 (2 \pi)^d} \mu^{2 \epsilon} \int d n_+l\int d^{d-2} l_{\perp} \int dn_-l\frac{1}{l_{\perp}^2 + n_-l\, n_+l+ i 0} \nonumber \\
&& \hspace{-25mm} \times \frac{1}{- n_-p \, n_+l- n_+p \, n_-l+ l_{\perp}^2 + n_-l\, n_+l+ i 0} \,  \frac{1}{ n_+k \, n_-l + n_-k \, n_+l + i 0} \, .
\eea
We start by solving the $d n_-l$ integral using residues. There are three poles,
\ali{
&P_{(a,l^0),1}^{\h} = - \frac{- n_+l n_-p + l_{\perp}^2 + i0}{n_+l - n_+p},&
&P_{(a,l^0),2}^{\h} = - \frac{n_-k \, n_+l + i0}{n_+k},&
&P_{(a,l^0),3}^{\h} = - \frac{l_{\perp}^2 + i0}{n_+l}.&
}
The poles will all be on the same side of the complex plane (so that the integral vanishes by the residue theorem) unless $n_+l<0$
or $0 < n_+l < n_+p$. It turns out that the former leads to a vanishing result. In the latter, only $P_{(a,l^0),1}^{\h}$ is above the real axis. Using the residue theorem, we close the contour in $+i \infty$ and find
\ali{
\label{eq:diaga-comparanda}
i \mathcal{M}_{(a,l^0)}^{({\rm NLP},h)} = -4 m_{\mu} M_N \Gamma_{S} \frac{Z e^2 }{2 (2 \pi)^d} \mu^{2 \epsilon} \int d n_+l \int d^{d-2} l_{\perp} (2 \pi i) \frac{n_+l - m_{\mu}}{M_N m_{\mu} (l_{\perp}^2 - n_+l^2)^2} .
}
Using eqs.~(\ref{eq:integral-tricks-4}) and (\ref{eq:pert-to-transv}), integrating $n_+L$ between $0$ and $n_+p$ and expanding in $\epsilon$, we find
\be
i \mathcal{M}_{(a,l^0)}^{({\rm NLP},h)} = i \mathcal{M}_{(a,l^0)}^{\rm{(NLP)}}.
\ee

\subsubsection{Diagram (b)}

In this case, the amplitude is 
\be
i \mathcal{M}_{(b,l^0)} = \mu^{2 \epsilon} \int \frac{d^dl}{(2 \pi)^d} (-i e) (i Z e) (i V_{(b,l^0)}) \frac{i}{(p'-l)^2-m_e^2+i 0} \frac{i}{(k'+l)^2-M_{N}^2+i 0} \frac{i}{l^2+i 0},
\ee
with
\be
V_{(b,l^0)} = - C_{SX} \, \bar{u}_e(p') \gamma_{\nu} (\slashed{p}' + m_{e}) P_X  u_{\mu}(p) \, \bar{u}_N(k') \gamma_{\nu}  (\slashed{k}' + M_N) u_N(k).
\ee
After solving the integral, we use $k \cdot p' = M_N E_e + \mathcal{O}(\lambda^2)$ and expand in $\lambda$. There is no $\mathcal{O}(\lambda^0)$ contribution. At NLP, we find
\ali{
\label{eq:MbL0}
i \mathcal{M}_{(b,l^0)} &= - i \frac{Z \alpha}{4 \pi} \Gamma_{S}
\Bigg\{
\frac{2}{\epsilon } \left[\ln \left(\frac{m_{e}}{2
m_{\mu}}\right)+i \pi \right]-2 \ln \left(\frac{2
m_{\mu}}{\mu}\right) \left[\ln \left(\frac{\mu}{2
m_{\mu}}\right)+2 i \pi \right]\nonumber \\
& \hspace{35mm} -2 \ln^2\left(\frac{m_{e}}{\mu}\right)-\frac{5 \pi
^2}{3} \Bigg\}.
}
We now show that $i \mathcal{M}_{(b,l^0)}$ is the sum of two regions: the hard one and the collinear one.

\smallsection{Hard region}
Following a procedure similar to the one used to describe the hard region of diagram (a), we find
\bea
\label{eq:diag-b-hard-first}
i \mathcal{M}_{(b,l^0)}^{\h} &=& \dfrac{Z e^2 V_{(b,l^0)}}{2 (2 \pi)^d} \mu^{2 \epsilon} \int d n_+l \int d^{d-2} l_{\perp} \int dn_-l \dfrac{1}{l_{\perp}^2 + n_-l \, n_+l + i 0} \nonumber \\
&& \hspace{1mm} \times \dfrac{1}{l_{\perp}^2 + n_-l \, n_+l - n_+p' \, n_-l + i 0} \,  \dfrac{1}{ n_+k' \, n_-l + n_-k' \, n_+l + i 0} \, .
\eea
As before, we solve the $d n_-l$ integral using residues. There are three poles,
\ali{
P_{(b,l^0),1}^{\h} = - \dfrac{l_{\perp}^2 + i0}{n_+l - n_+p'},
\quad
P_{(b,l^0),2}^{\h} = - \dfrac{n_-k' \, n_+l + i0}{n_+k'},
\quad
P_{(b,l^0),3}^{\h} = - \dfrac{l_{\perp}^2 + i0}{n_+l}.
}
Two disconnected branches lead to a non-zero integral:
\ali{
\textrm{branch 1:} \ \ n_+l < 0,
\qquad \qquad
\textrm{branch 2:} \ \ 0 < n_+l < n_+p'.
}
\noindent \underline{Branch 1}: \\
Here, $P_{(b),1}^{\h}$ and $P_{(b),3}^{\h}$ are located above the real axis. We thus close the contour in $+i \infty$, and find
\bea
i \mathcal{M}_{(b,l^0)}^{(\hvar,l^0,1)} &=& -4 m_{\mu} M_N \Gamma_{S} \dfrac{Z e^2 \mu^{2 \epsilon}}{2 (2 \pi)^d} \int d n_+l \int d^{d-2} l_{\perp} (-2 \pi i) \, n_+k \nonumber \\
&& \hspace{-16mm} \times \dfrac{1}{l_{\perp}^2 n_+k' - n_-k' \, n_+l^2 + i0} \, \dfrac{1}{l_{\perp}^2 \, n_+k' - n_-k' \, n_+l^2 + n_-k' \, n_+l \, n_+p' + i0} \, .
\eea
Using eqs.~(\ref{eq:integral-tricks-4}) and (\ref{eq:pert-to-transv}), integrating $n_+l$ between $-\infty$ and $0$ and expanding in $\epsilon$, we obtain
\be
i \mathcal{M}_{(b,l^0)}^{(\hvar,1)} = - \frac{i Z \alpha}{4 \pi} \, \Gamma_{S} \, 
\Bigg\{-\frac{1}{\epsilon^2} + 
\frac{2 \ln \left(\frac{2 m_{\mu}}{\mu}\right)}{\epsilon}-2 \ln^2\left(\frac{2
m_{\mu}}{\mu}\right)-\frac{5 \pi^2}{12}
\Bigg\}.
\ee

\noindent \underline{Branch 2}:
\\
Here, only $P_{(b),1}^{\h}$ is above the real axis. We close again the contour in $+i \infty$, and find
\bea
i \mathcal{M}_{(b,l^0)}^{(\hvar,2)} &=& -4 m_{\mu} M_N \Gamma_{S} \dfrac{Z e^2 \mu^{2 \epsilon}}{2 (2 \pi)^d} \int d n_+l \int d^{d-2} l_{\perp} (2 \pi i) \, \dfrac{n_+l - n_+p'}{n_+p'} \nonumber \\
&& \hspace{-16mm} \times \dfrac{1}{l_{\perp}^2 + i0} \, \dfrac{1}{l_{\perp}^2 \, n_+k' - n_-k' \, n_+l^2 + n_-k' \, n_+l \, n_+p' + i0} \, .
\eea
Proceeding similarly to that in branch 1 (but integrating $n_+l$ between $0$ and $n_+p'$), we find
\bea
i \mathcal{M}_{(b,l^0)}^{(\hvar,2)} &=& - \frac{i Z \alpha}{4 \pi} \, \Gamma_{S} \,
\Bigg\{\frac{2}{\epsilon^2} + 
\frac{2 \left(2 \ln \left(\frac{\mu}{2
m_{\mu}}\right)+i \pi \right)}{\epsilon} \nonumber \\
&& \hspace{18mm} +\frac{1}{6} \bigg[24
\ln^2\left(\frac{2 m_{\mu}}{\mu}\right)-24 i \pi  \ln
\left(\frac{2 m_{\mu}}{\mu}\right) -7 \pi^2\bigg] \Bigg\}.
\eea

\smallsection{Collinear region}
Assuming $l \sim l^{\c}$ and expanding the propagators in $\lambda$, we find
\bea
\label{eq:diag-b-col-first}
i \mathcal{M}_{(b,l^0)}^{\c} &=& \dfrac{Z e^2 V_{(b,l^0)}}{2 (2 \pi)^d} \mu^{2 \epsilon}\int d n_+l \int d^{d-2} l_{\perp} \int dn_-l \dfrac{1}{l_{\perp}^2 + n_-l \, n_+l + i 0} \nonumber \\
&& \hspace{-8mm} \times \dfrac{1}{l_{\perp}^2 + n_-l \, n_+l - n_+p' \, n_-l - n_-p' \, n_+l + i 0} \, \,  \dfrac{1}{n_-k' \, n_+l + i 0} \, .
\eea
We start by performing the $n_-l$ integral using residues. The poles are
\ali{
P_{(b),1}^{\c} = \dfrac{n_+l \, n_- p' - l_{\perp}^2 - i0}{n_+l - n_+p'},
\qquad \qquad
P_{(b),2}^{\c} = -\dfrac{l_{\perp}^2 + i0}{n_+l}.
}
The integral is only non-vanishing if $0 < n_+l < n_+p'$. In that case, $P_{(b),1}^{\c}$ is located above the real axis; we pick it so that we close the contour in $+i \infty$. Then,
\bea
\label{eq:diag-b-col-first-2}
i \mathcal{M}_{(b,l^0)}^{\c} &=& -4 m_{\mu} M_N \Gamma_{S}  \dfrac{Z e^2 \mu^{2 \epsilon}}{2 (2 \pi)^d} \int d n_+l
\nonumber \\
&& \hspace{-8mm} \times \int d^{d-2} l_{\perp}  \dfrac{2 \pi i}{(n_-k \, n_+l-i0) \left(n_+l^2 \, n_-p' - n_+p' \, l_{\perp}^2+i0 \right)}.
\eea
Finally, proceeding in a similar way as before, we find
\bea
i \mathcal{M}_{(b,l^0)}^{\c} &=& 
- \frac{i Z\alpha }{4 \pi} \, \Gamma_{S} \,
\Bigg\{-\frac{1}{\epsilon
^2} + \frac{2 \ln
\left(\frac{m_e}{\mu}\right)}{\epsilon}-2 \ln^2\left(\frac{\mu}{m_e}\right)-\frac{\pi^2}{12}
\Bigg\}.
\eea

\smallsection{Total}
As predicted, the original amplitude is equal to the sum of the hard and collinear regions,
\be
i \mathcal{M}_{(b,l^0)} = i \mathcal{M}_{(b,l^0)}^{(\hvar,1)} + i \mathcal{M}_{(b,l^0)}^{(\hvar,2)} + i \mathcal{M}_{(b,l^0)}^{\c}.
\ee

\subsubsection{Diagram (c)}

In this case, the amplitude is decomposed into hard and collinear regions,
\ali{
i \mathcal{M}_{(c,l^0)} = i \mathcal{M}_{(b,l^0)}^{\h} + i \mathcal{M}_{(c,l^0)}^{\c},
}
with
\bs
\ali{
\label{eq:McL0}
i \mathcal{M}_{(c,l^0)} &= \frac{i \alpha}{2 \pi} \, \Gamma_{S} 
\Bigg\{
\frac{1}{\epsilon} \ln \left(\frac{2 m_{\mu}}{m_e}\right) + \ln \left(\frac{2 m_{\mu}}{m_e}\right) \ln \left(\frac{\mu^2}{2 m_e m_{\mu}}\right)+\frac{\pi^2}{12}
\Bigg\}, \\
\label{eq:MchL0}
i \mathcal{M}_{(c,l^0)}^{\h}
&=
- \frac{i \alpha}{4 \pi} \, \Gamma_{S}  
\Bigg\{\frac{1}{\epsilon ^2}
-\frac{2}{\epsilon } \ln \left(\frac{2 m_{\mu}}{\mu}\right) + 2 \ln^2\left(\frac{2 m_{\mu}}{\mu}\right) - \frac{\pi^2}{12}\Bigg\}, \\
\label{eq:iMccolL0}
i \mathcal{M}_{(c,l^0)}^{\c}
&=
\frac{i \alpha}{4 \pi}  \, \Gamma_{S}  
\Bigg\{\frac{1}{\epsilon^2} + 
\frac{2}{\epsilon} \ln \left(\frac{\mu}{m_e}\right)
+ 2 \ln^2\left(\frac{\mu}{m_e}\right)+\frac{\pi^2}{12}
\Bigg\}.
}
\es

\subsubsection{Diagram (d)}

This case is characterized by a single region, the hard-nuclear one. we have
\ali{
i \mathcal{M}_{(d,l^0)} = i \mathcal{M}_{(d,l^0)}^{\sh},
}
with
\ali{
\label{eq:MdL0}
i \mathcal{M}_{(d,l^0)}^{\sh} = i \frac{Z^2 \alpha}{2 \pi} \, \Gamma_{S} 
\left[
\frac{1}{\epsilon} + \ln \left( \frac{\mu^2}{M_N^2} \right)
\right].
}

\subsubsection{Diagram (e)}

A single region, the hard one, also characterizes diagram (e),
\ali{
i \mathcal{M}_{(e,l^0)} = i \mathcal{M}_{(e,l^0)}^{\h},
}
with
\ali{
\label{eq:MeL0}
i \mathcal{M}_{(e,l^0)}^{\h} = - i \frac{Z \alpha}{2 \pi} \, \, \Gamma_{S} \, \left[ \frac{1}{\epsilon} + \ln \left(\frac{\mu^2}{m_{\mu}^2}\right) -2\right].
}

\subsubsection{Diagram (f)}

Diagram (f) is decomposed into hard and collinear regions,
\ali{
i \mathcal{M}_{(f,l^0)} = i \mathcal{M}_{(f,l^0)}^{\h} + i \mathcal{M}_{(f,l^0)}^{\c},
}
with
\bs
\ali{
i \mathcal{M}_{(f,l^0)} &= - i \frac{Z \alpha}{2 \pi} \Gamma_{S}
\Bigg\{ \frac{1}{\epsilon} \ln \left( \frac{2 m_{\mu}}{m_e} \right) - \ln \left( \frac{2 m_{\mu}}{m_e}\right) \ln \left( \frac{2 m_{\mu} m_e}{\mu^2}\right) - \frac{\pi^2}{6} \Bigg\}, \\
\label{eq:MghL0}
i \mathcal{M}_{(f,l^0)}^{\h}
&= - \frac{i Z \alpha}{4 \pi} \, \Gamma_{S} \,
\Bigg\{ - \frac{1}{\epsilon^2} - \frac{2}{\epsilon} \ln \left( \frac{\mu}{2 m_{\mu}} \right) - 2 \ln^2\left(\frac{\mu}{2 m_{\mu}}\right) - \frac{5 \pi^2}{12}\Bigg\}, \\
\label{eq:iMfcolL0}
i \mathcal{M}_{(f,l^0)}^{\c}
&=
- \frac{i Z \alpha}{4 \pi} \, \Gamma_{S} \,
\Bigg\{\frac{1}{\epsilon^2} + \frac{2}{\epsilon} \ln \left( \frac{\mu}{m_e} \right) + 2 \ln^2\left(\frac{\mu}{m_e}\right) + \frac{\pi^2}{12}\Bigg\}.
}
\es

\subsection{Real corrections}

As discussed in section \ref{sec:Framework}, the real corrections follow from the emission of photons from the muon and the electron (recall figure \ref{fig:real-diags-full}). As usual, we start by discussing the calculation in the full theory, and only then do we consider the regions. Ultimately, we prove that the sum of the results of the regions leads to the full theory result.

A word of caution related to terminology is in order. We are interested in calculating what is usually known as the cross section for the ``soft'' real emission according to the Low-Burnett-Kroll (LBK) theorem \cite{Low:1958sn, Burnett:1967km, Weinberg:1965nx, Jackiw:1968zza, Larkoski:2014bxa, Beneke:2021umj}.%
\fn{Collinear emission is not considered, as the electron is taken to be massive.}
This concerns the real emission of photons with low energy, and \textit{not} the soft mode defined in table \ref{table:modes}. As we will see, the cross section is reproduced by two regions: the soft and the soft-collinear, corresponding to modes defined table \ref{table:modes}. 

Let us then start by performing the calculation using the full theory. Here, the differential cross section for the soft (in the sense of low energy) real emission diagrams can be written as \cite{Denner:2019vbn}
\be
d \sigma_{\text{real}}=\delta_{\text{real}} \, d \sigma_{\mathrm{LO}},
\ee
where $d \sigma^{\mathrm{LO}}$ is the leading order cross section without photon emission, and 
\be
\delta_{\text{real}} =  \sum_{\substack{n, n^{\prime}}} \delta_{\text{real},nn'} = -\frac{\alpha}{2 \pi} \sum_{\substack{n, n^{\prime}}} \sigma_n Q_n \sigma_{n^{\prime}} Q_{n^{\prime}} I_{n n^{\prime}} ,
\ee
where the sum over $n$ and $n'$ is made over the relevant charged particles. For a particle $n$, $Q_n$ is its charge and $\sigma_n$ a sign ($-1$ for an outgoing particle and +1 for an incoming one). Furthermore,
\bea
\label{eq:Inn}
I_{n n^{\prime}} &=& \frac{(2 \pi \mu)^{4-d}}{2 \pi} \int_{l_0<\Delta E} \frac{d^{d-1} \vec{l}}{l_0} \frac{p_n \cdot p_{n^{\prime}}}{p_n \cdot l \, \, p_{n^{\prime}} \cdot l} \bigg|_{l_0=|\vec{l}|} .
\label{eq:real-emission-integrals}
\eea
We find
\bs
\bea
\label{eq:full-ee-real}
\delta_{\text{real},ee} &=& -\dfrac{\alpha}{2 \pi} \bigg\{ -\frac{1}{\epsilon} + 2 \ln \left( \frac{\Delta E \, m_e}{m_{\mu} \mu} \right) \bigg\}, \\
\label{eq:full-mumu-real}
\delta_{\text{real},\mu\mu} &=& -\dfrac{\alpha}{2 \pi} \bigg\{ -\frac{1}{\epsilon} + 2 \ln \left( \frac{2 \Delta E}{\mu}\right) - 2\bigg\}, \\
\delta_{\text{real},e\mu} = \delta_{\text{real},\mu e} &=& \dfrac{\alpha}{2 \pi} \bigg\{ -\frac{1}{\epsilon} \ln \left(\frac{2 m_{\mu}}{m_e} \right) + \ln \left(\frac{2 m_{\mu}}{m_e} \right) \ln \left(\frac{2 \, \Delta E^2 m_e}{m_{\mu} \, \mu^2} \right)  - \frac{\pi^2}{6}\bigg\}, \hspace{10mm} 
\eea
\es
implying
\be
\label{eq:IR-real-full}
\delta_{\text{real}} = \dfrac{\alpha}{\pi}
\Bigg\{
\frac{1}{\epsilon}-\frac{1}{\epsilon} \ln\left(\dfrac{2m_{\mu}}{m_e}\right)
+ \ln \left(\dfrac{m_{\mu} \mu^2}{2 \Delta E^2 m_e}\right)
+ \ln \left(\dfrac{2 m_{\mu}}{m_e}\right) \ln \left(\dfrac{2 \Delta E^2 m_e}{m_{\mu} \mu^2}\right) - \dfrac{\pi^2}{6} + 1
\Bigg\}.
\ee

\smallsection{Soft region}
The soft region of $\delta_{\rm real}$, $\delta^{\s}_{\textrm{real}}$, can be calculated by
a) considering the definition of the soft function $\mathcal{S}$ in eq. \ref{eq:S-function} (ignoring the Wilson lines for the nucleus),
b) setting $| \mathcal{X}^{\s} \rangle = |A(k)\rangle$ and
c) expanding one of the Wilson lines of each matrix element to the first non-trivial order.
As before, we write
\be
\delta^{\s}_{\textrm{real}} = \delta^{\s}_{\textrm{real},ee} + \delta^{\s}_{\textrm{real},\mu\mu} + \delta^{\s}_{\textrm{real},e\mu} +
\delta^{\s}_{\textrm{real},\mu e},
\ee
with
\bs
\ali{
\label{eq:S-ee-real}
\delta^{\s}_{\textrm{real},ee} &= 0, \\
\delta^{\s}_{\textrm{real},\mu\mu} &= - \frac{\alpha}{2 \pi} \bigg\{ -\frac{1}{\epsilon} + 2 \ln \left( \frac{2 \Delta E}{\mu}\right) - 2\bigg\}, \\
\delta^{\s}_{\textrm{real},\mu e} = \delta^{\s}_{\textrm{real},e\mu} &= \frac{\alpha}{2\pi} \bigg\{\frac{1}{2 \epsilon^2} + \frac{1}{\epsilon} \ln \left(\frac{\mu}{2 \Delta E}\right)  + \ln^2\left(\frac{\mu}{2 \Delta E}\right) - \frac{\pi^2}{8}\bigg\},
}
\es
so that
\be
\label{eq:EFT-III-soft-real}
\delta^{\s}_{\textrm{real}} = \frac{\alpha}{2 \pi } \Bigg\{ \frac{1}{\epsilon^2} + \frac{2}{\epsilon} \bigg[ \frac{1}{2} + \ln \left(\frac{\mu }{2 \Delta E}\right) \bigg] + 2 \ln^2 \left(\dfrac{2 \Delta E}{\mu}\right) - 2 \ln \left(\dfrac{2 \Delta E}{\mu}\right) + 2 - \dfrac{\pi^2}{4}\Bigg\}.
\ee
We note that $\delta^{\s}_{\textrm{real}} = \left[ \int_0^{\Delta E} d\mathcal{E}  \,\mathcal{S}(\mathcal{E}) \right]_{\rm{NLO}}$.

\smallsection{Soft-collinear region}
Following an equivalent procedure for the soft-collinear region, we write
\be
\delta^{\sc}_{\textrm{real}} = \delta^{\sc}_{\textrm{real},ee} + \delta^{\sc}_{\textrm{real},\mu\mu} + \delta^{\sc}_{\textrm{real},e\mu} +
\delta^{\sc}_{\textrm{real},\mu e},
\ee
with
\ali{
\label{eq:IV-ee-real}
\delta^{\sc}_{\textrm{real},ee} &= -\frac{\alpha}{2 \pi} \Bigg\{-\frac{1}{\epsilon} + 2 \ln\left( \frac{\Delta E \, m_e}{m_{\mu} \, \mu}\right) \Bigg\}, \\
\delta^{\s}_{\textrm{real},\mu\mu} &= 0, \\
\delta^{\sc}_{\textrm{real},\mu e} = \delta^{\sc}_{\textrm{real},e\mu} &= \frac{\alpha}{2 \pi} \Bigg\{-\frac{1}{2 \epsilon^2} + \frac{1}{\epsilon} \ln \left( \frac{\Delta E m_e}{m_{\mu} \, \mu}\right) - \ln^2 \Bigg(\frac{\Delta E m_e}{m_{\mu} \, \mu} \Bigg) - \frac{\pi^2}{24}  \Bigg\},
}
which leads to
\be
\label{eq:S-IV}
\delta^{\sc}_{\textrm{real}} = \frac{\alpha}{\pi} \Bigg\{ -\frac{1}{2 \epsilon^2} + \frac{1}{\epsilon} \left[ \frac{1}{2} + \ln \left( \frac{\Delta E m_e}{m_{\mu} \, \mu}\right) \right] - \ln^2 \Bigg(\frac{\Delta E m_e}{m_{\mu} \, \mu} \Bigg) - \ln\left( \frac{\Delta E \, m_e}{m_{\mu} \, \mu}\right) - \frac{\pi^2}{24}  \Bigg\}.
\ee

\smallsection{Total}
As predicted, the full theory result is the sum of the soft and the soft-collinear regions:
\ali{
\delta_{\textrm{real}} = \delta^{\s}_{\textrm{real}} + \delta^{\sc}_{\textrm{real}}.
}

 \bibliographystyle{JHEP}
 \bibliography{biblio.bib}

\end{document}